\documentclass[journal,draftcls,onecolumn,12pt,twoside]{IEEEtran}
\usepackage{mathrsfs}
\usepackage{amsfonts}
\usepackage{amsthm}
\theoremstyle{remark}

\newtheorem{theorem}{Theorem}
\newtheorem{lemma}{Lemma}

\newtheorem{remark}{Remark}
\usepackage{algorithm}
\usepackage{color}
\usepackage{cite}

\usepackage{url}

\newcommand{\Z}{\mathbb{Z}}
\newcommand{\defas}{:=}
\newcommand{\abs}[1]{\left| #1 \right|}

\ifCLASSINFOpdf
   \usepackage[pdftex]{graphicx}
\else
   \usepackage[dvips]{graphicx}
   \graphicspath{{../eps/}}
   \DeclareGraphicsExtensions{.eps}
\fi

\usepackage[cmex10]{amsmath}
\usepackage{algorithmic}
\begin{document}
\title{Rate Distortion for Lossy In-network Linear Function Computation and Consensus: Distortion Accumulation and Sequential Reverse Water-Filling}

\author{
\IEEEauthorblockN{Yaoqing Yang, Pulkit Grover and Soummya Kar}\\
% \IEEEauthorblockA{Department of Electrical and Computer Engineering, Carnegie Mellon University, Pittsburgh, PA, 15213}

\thanks{A preliminary version of this work was presented in part at the 53rd Annual Allerton Conference on Communication, Control and Computing, 2015. This work is supported in part by the National Science Foundation under grants CCF-1513936, by ECCS-1343324 and CCF-1350314 (NSF CAREER) for Pulkit Grover, and by Systems on Nanoscale Information fabriCs (SONIC), one of the six SRC STARnet Centers, sponsored by MARCO and DARPA.

Y. Yang, P. Grover and S. Kar are with the Department of Electrical and Computer Engineering, Carnegie Mellon University, Pittsburgh, PA, 15213, USA. Email: \{yyaoqing,pgrover,soummyak\}@andrew.cmu.edu.}

}%


%\author{\IEEEauthorblockN{Michael Shell\IEEEauthorrefmark{1}
%Telephone: (800) 555--1212, Fax: (888) 555--1212}


\maketitle
\begin{abstract}
  We consider the problem of distributed lossy linear function computation in a tree network. We examine two cases: (i) data aggregation (only one sink node computes) and (ii) consensus (all nodes compute the same function). By quantifying the accumulation of information loss in distributed computing, we obtain fundamental limits on network computation rate as a function of incremental distortions (and hence incremental loss of information) along the edges of the network. The above characterization, based on quantifying distortion accumulation, offers an improvement over classical cut-set type techniques which are based on overall distortions instead of incremental distortions. This quantification of information loss qualitatively resembles information dissipation in cascaded channels~\cite{Poly_ArXiv_14}. Surprisingly, this accumulation effect of distortion happens even at infinite blocklength. Combining this observation with an inequality on the dominance of mean-square quantities over relative-entropy quantities, we obtain outer bounds on the rate distortion function that are tighter than classical cut-set bounds by a difference which can be arbitrarily large in both data aggregation and consensus. We also obtain inner bounds on the optimal rate using random Gaussian coding, which differ from the outer bounds by $\mathcal{O}(\sqrt{D})$, where $D$ is the overall distortion. The obtained inner and outer bounds can provide insights on rate (bit) allocations for both the data aggregation problem and the consensus problem. We show that for tree networks, the rate allocation results have a mathematical structure similar to classical reverse water-filling for parallel Gaussian sources.
\end{abstract}


% \fi
\IEEEpeerreviewmaketitle

\section{Introduction}\label{Intro}
The phenomenon of information dissipation~\cite{Cal_ISIT_15,Anan_ISIT_14,Poly_ArXiv_14,Eva_TIT_99,Rag_ISIT_13,Erk_TIT_98} has been of increasing interest recently from an information-theoretic viewpoint. These results characterize and quantify the gradual loss of information as it is transmitted through cascaded noisy channels. This study has also yielded data processing inequalities that are stronger than those used classically~\cite{Cal_ISIT_15,Rag_ISIT_13}.

The dissipation of information cannot be quantified easily using classical information-theoretic tools that rely on the law of large numbers, because the dissipation of information is often due to finite-length of codewords and power constraints on the channel inputs~\cite{Poly_ArXiv_14}. In many classical network information theory problems, such as relay networks, the dissipation of information is not observed because it can be suppressed by use of asymptotically infinite blocklengths~\cite{Poly_ArXiv_14,Dig_TIT_11}\footnote{In \cite{Dig_TIT_11}, it is shown that cut set bounds are order-optimal in an arbitrary wireless network with a single source and a single destination.}. However, information dissipation does happen in many problems of communications and computation. For example, in~\cite{Eva_TIT_99}, Evans and Schulman obtain bounds on the information dissipation in noisy circuits, and in~\cite{Poly_ArXiv_14}, Polyanskiy and Wu examine a similar problem in cascaded AWGN channels with power-constrained inputs. Our earlier works~\cite{Yang_All_14,Yang_Arx_15} show that under some conditions, error-correcting codes can be used to overcome information dissipation and achieve reliable linear computation using unreliable circuit components. In many of these works~\cite{Cal_ISIT_15,Anan_ISIT_14,Poly_ArXiv_14,Eva_TIT_99,Rag_ISIT_13,Erk_TIT_98}, quantifying dissipation of information requires use of tools that go beyond those commonly used in classical information theory, e.g., cut-set techniques and the data processing inequality.

Does the information dissipation problem exist in lossy noiseless networks? For lossy compression and communication of a single source over a noiseless line network, information can be preserved by repeatedly transmitting the same codeword from one end to the other. However, in this paper, we show that in distributed lossy \emph{computation}, information does dissipate. We first study the problem of lossily computing a weighted sum of independent Gaussian sources over a tree network at an arbitrarily determined sink node. We prove that distortion must accumulate, and hence information, if measured in the way of mean-square distortion, must dissipate, along the way from leaves to the sink node due to repeated lossy quantization of distributed data scattered in the network. In contrast with dissipation results in channel coding~\cite{Poly_ArXiv_14,Dig_TIT_11}, this information loss, measured in mean-square distortion, happens even at infinite blocklength. Moreover, by quantifying ``incremental distortion'', i.e., incremental information loss on each link of the tree network, we derive an information-theoretic outer bound on the rate distortion function that is tighter than classical cut-set bounds obtained for this problem in the work of Cuff, Su and El Gamal~\cite{Cuf_ISIT_09}. Using the same technique, we improve the classical outer bound on the sum rate of network consensus (all nodes compute the same linear function) for tree networks from $\mathcal{O}\left( n{\log_2 }\frac{1}{{{n}^{3/2}}D} \right)$ (see~\cite[Proposition 4]{Su_TIT_10}\footnote{Note that the original bound in~\cite{Su_TIT_10} has a normalization term $\frac{1}{n}$. The bound in \cite{Su_TIT_10} is useful only if $D=O(n^{-\frac{3}{2}})$. Our outer bound is useful for all $D$.}) to $\mathcal{O}\left( n{\log_2 }\frac{1}{D} \right)$, where $n$ is the number of nodes in the tree network and $D$ is the required overall distortion. In Remark~\ref{comparison_remark}, we provide the intuition underlying the difference between our bound and the cut-set bound for lossy in-network computation. \textcolor{black}{Note that although our definition of information loss (measured in terms of distortion accumulation) is different from that of \cite{Poly_ArXiv_14}, this definition provides a new perspective in this line of study.}

A crucial step in our derivation is to bound the difference in differential entropies of two distributions, where we use the dominance of the mean-square quantities over the quantities based on relative entropy (see Eq.~\eqref{diff_in_entro}). This inequality was used by Raginsky and Sason in~\cite{Rag_FaT_13} (credited to Wu~\cite{Wu_HWI}) as a means of proving a weak version of the ``HWI inequality''~\cite{Otto_JFA_00} (H, W and I stand for divergence, Wasserstein distance and Fisher information distance respectively), which has deep connections with log-Sobolev type inequalities~\cite{Rag_FaT_13}.

In Section~\ref{main_sec} and Section~\ref{Gaussian_sec}, we provide information-theoretic bounds on the rate distortion function for linear function computation in a tree network, where the function is computed at an arbitrarily predetermined sink node. For simplicity, we restrict our attention to independent Gaussian sources. In Section~\ref{consensus_sec}, we extend our results to the problem of network consensus, in which all nodes compute the same linear function. In both cases, the difference between the inner and outer bounds is shown to approach zero in the high-resolution (i.e., zero distortion) limit. Note that in~\cite[Section V]{Cuf_ISIT_09}, the authors show a constant difference between their lower and the inner bounds in the Gaussian case. Using our improved outer bound, we can upper-bound the difference by $\mathcal{O}(D^{1/2})$, where $D$ is the required distortion. Therefore, the inner bound and the outer bound match in the asymptotic zero-distortion limit. In the special case of a line network, we show that the rate distortion function is very similar to the reverse water-filling result for parallel Gaussian sources~\cite[Theorem 10.3.3]{Cover_Wiley_06}.

\textcolor{black}{The inner bound obtained in this paper is based on random Gaussian codebooks. The main difficulty here is to bound the overall distortion for random coding in linear function computation. In order to compute the overall distortion, we quantify a non-trivial equivalence between random-coding-based estimates and MMSE estimates. Relying on the distortion accumulation result for MMSE estimates, we equivalently obtain the distortion accumulation result for Gaussian random codebooks, and hence obtain the overall distortion. This equivalence between random coding and MMSE is easy to obtain for point-to-point channels, but hard for network function computation, due to information loss about the exact source distribution after successive quantization. The key technique is to bound this information loss using bounds on associated KL-divergences, and hence to show the equivalence between network computation and point-to-point communications. (\textcolor{black}{See also Remark~\ref{why_MMSE} for details on why our analysis is conceptually different from classical techniques such as Wyner-Ziv coding and why such new proof techniques are needed.})}

\textcolor{black}{We briefly summarize the main technical contributions of this paper:
\begin{itemize}
  \item we analyze the distortion accumulation effect associated with the incremental distortion, and use this to provide an outer bound on the rate-distortion function for linear function computation;
  \item we provide an inner bound that matches with the outer bound in the zero distortion limit using Gaussian random codebooks; we also quantify the equivalence between random coding and MMSE estimates for linear function computation;
  \item we extend the results from linear function computation to the problem of network consensus.
\end{itemize}}

\subsection{Related Works}
Problems of in-network linear function computing have been extensively studied for the goal of distributed data aggregation and distributed signal processing~\cite{Gir_CM_06,Dim_PI_10}.

From an information-theoretic and in particular rate-distortion viewpoint, the in-network computing problem is often studied from the perspective of distributed source coding for source reconstruction or function computation. The network structures considered include multi-encoder networks (CEO-type function computing problems)~\cite{Kri_TIT_09,Wag_TIT_08,Wag_TIT_11}, Gaussian multiple-access networks~\cite{Sou_TIT_12}, three-node relay networks~\cite{Sef_ISIT_12}, line or tree networks~\cite{Son_TIT_13,App_TIT_11,Vis_ISIT_10,Sef_ITW_13,Mis_ISIT_13} or even general networks in lossless settings~\cite{Kow_TIT_12,Kan_JSAC_13}. Among these works, \cite{Mis_ISIT_13} considers the problem of lossy computation in a line network, which is most closely related to our work  (ours is lossy computation in a tree network). However, the result in~\cite{Mis_ISIT_13} only characterizes the limit $\underset{R\to \infty }{\mathop{\lim }}\,-\frac{\log D}{R}$, where $R$ and $D$ are respectively the overall rate and the overall distortion.

Our work is also closely related to~\cite{Su_TIT_10,Cuf_ISIT_09,Aya_TIT_10,Xu_ISIT_14}, where outer bounds based on cut-set techniques~\cite{Gamal_Camb_11} are obtained on the rate, or on the computation time, that is required to meet certain fidelity requirements on linear function computation. Our work is especially inspired by the works by Su, Cuff and El Gamal~\cite{Su_TIT_10,Cuf_ISIT_09}. However, we show that many outer bounds in~\cite{Su_TIT_10,Cuf_ISIT_09} can be significantly tightened with information-dissipation-inspired techniques beyond the cut-set bounds (see, for example,~\cite{Rag_FaT_13}). Many recent works improve on cut-set bounds in certain instances in network information theory, such as the sum capacity of a multi-cast deterministic network~\cite{Sho_ISIT_14} and the capacity region of a multi-cast noisy network~\cite{Kam_All_14}. However, the above-mentioned references do not consider noiseless lossy in-network computation.

Some previous works on information-theoretic distributed computing also rely on random-coding-based techniques to provide inner bounds \cite{Cuf_ISIT_09,Sef_ITW_13}. The achievable schemes in \cite{Su_TIT_10} utilize Gaussian test channels, which also implicitly require random coding arguments. However, we find it hard to directly analyze the random coding schemes for distributed lossy computing with Gaussian sources, especially for computing the overall mean-square error of the consensus value, \textcolor{black}{because we may need to obtain a non-trivial generalization of the ``Markov Lemma''~\cite[Lecture Notes 13]{Gamal_Camb_11} to Gaussian sources (see Remark~\ref{why_MMSE} for details). However, this generalization may be cumbersome and not directly related to the main result, the outer bound obtained using distortion accumulation, in this paper.} To overcome this difficulty, we show a non-trivial equivalence between the estimate based on Gaussian random coding and the estimate based on MMSE: in the limit of infinite block-length, the MMSE estimate of a Gaussian source given the codeword generated by Gaussian random coding is just the codeword itself, which means that the analysis for MMSE is also applicable in the analysis of the random-coding scheme. Further, for MMSE estimates, we have shown in Section~\ref{main_sec} that the incremental error (incremental distortion) at different stages of the distributed computation scheme are uncorrelated with each other. Thus, using this property of MMSE estimates, we are able to complete the computation of the overall distortion for our proposed scheme based on random coding.

Our work is organized as follows: Section \ref{model} provides the model and the problem formulation of distributed lossy function computation; Section~\ref{main_sec} provides the main results of this paper, which contain the result on distortion accumulation and the information-theoretic outer bound on the rate-distortion function for distributed lossy computation; Section~\ref{Gaussian_sec} provides the inner bound using Gaussian random codebooks and using the equivalence between random coding and MMSE; Section~\ref{consensus_sec} generalizes the outer and inner bounds to the problem of distributed lossy network consensus; Section~\ref{conclude_sec} concludes the paper. Proofs of various intermediate results are often relegated to the appendices.

\subsection{Notation and Preliminary Results}
Vectors are written in bold font, e.g., $\mathbf{x}$ and $\mathbf{y}$. Sets are written in calligraphic letters, such as $\mathcal{S}$. Scalar random variables are written in uppercase letters, e.g., $U$ and $V$. Quantities that measure mean-square distortions are denoted by $D$ or $d$ with subscripts and superscripts. A Gaussian distribution with mean $\boldsymbol{\mu}$ and covariance $\bf{\Sigma}$ is denoted by $\mathcal{N}(\boldsymbol{\mu},\bf{\Sigma})$. The all-zero vector with length $N$ is denoted by $\mathbf{0}_N$, and the $N\times N$ identity matrix is denoted by $\mathbf{I}_N$.

\begin{figure}
\centering
\includegraphics[scale=0.6]{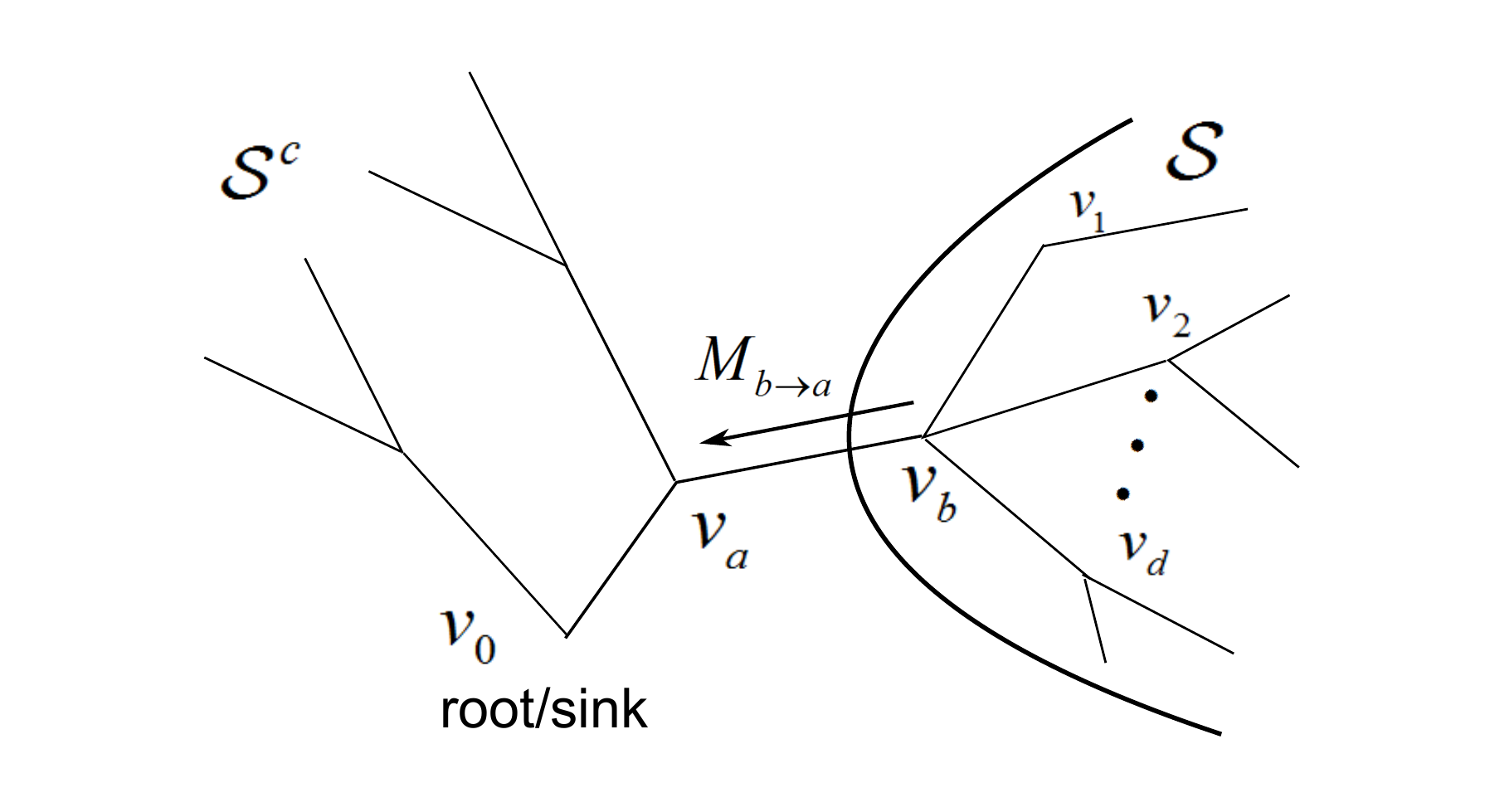}
\caption{This is an illustration of linear function computation considered in this paper. The goal is to compute a weighted sum of distributed Gaussian sources over a tree-network. The notation $M_{b\to a}$ denotes the set of bits transmitted from $v_b$ to $v_a$. The set $\mathcal{S}$ in this figure can also be written as $\mathcal{S}_b$, which denotes the set that contains $v_b$ and all its descendants in the network.}
\label{cut_set_figure}
\end{figure}

The calligraphic letter $\mathcal{T}=(\mathcal{V},\mathcal{E})$ is used to represent a tree graph with a node set $\mathcal{V}=\{v_i\}_{i=0}^n$ with cardinality $n+1$ and an edge set $\mathcal{E}$. In this paper, an edge is always undirected\footnote{Although we consider an undirected tree graph, we specify a unique root node, which makes the subsequent definitions on descendants and children valid.}. The neighborhood $\mathcal{N}(v_i)$ of a node $v_i$ is defined as all the nodes that are connected with $v_i$. A root node $v_0$ is specified for the tree graph. Since in a tree graph, each node has a unique path to the root node, for an arbitrary node $v_i\neq v_0$, a unique parent node which is the {neighboring node of} $v_i$ on the path from $v_i$ to $v_0$ can be determined, which is denoted as $v_{\text{PN}(i)}$. The \emph{children} of $v_i$ are defined as the set of nodes $\{v_j\in \mathcal{V}\mid v_i=v_{PN(j)}\}$. The \emph{descendants} of $v_i$ are defined as the set of nodes that includes all nodes $v_j$ that have $v_i$ on the unique path from $v_j$ to the root $v_0$. The set ${\mathcal{S}_i}$ is used to denote the set that is constituted by node $v_i$ and all the descendants of $v_i$. As shown in Fig.~\ref{cut_set_figure}, the set $\mathcal{S}$ is constituted by a node $v_b$ and its descendants. Thus, in Fig.~\ref{cut_set_figure}, $\mathcal{S}={\mathcal{S}_b}$ and $v_a=v_{\text{PN}(b)}$. When there is no ambiguity, we use $v_1$, $v_2,\ldots v_d$ to denote the children of a particular node $v_b$.

We will obtain scaling bounds on the communication rate. Throughout this paper, we rely on the family of ``big-O'' notation~\cite{knuth}. The notation $f_1(N)=\mathcal{O}(f_2(N))$ and $f_1(N)=\Omega(f_2(N))$ respectively mean that $f_1(N)/f_2(N)\le C_1$ and $f_1(N)/f_2(N)\ge C_2$ for two positive constants $C_1$, $C_2$ and sufficiently large $N$. By $f_1(N)=\Theta(f_2(N))$ we mean that $f_1(N)=\mathcal{O}(f_2(N))$ and $f_1(N)=\Omega(f_2(N))$.

We will use some results on mean-square error estimation. First, we state the orthogonality principle and the statisticians' Pythagoras theorem, which we will use frequently in this paper.
\begin{lemma}\label{Pyth_lmm}(Pythagoras theorem, \cite[Theorem 9.4]{Will_Cam_91}, \cite[Section 8.1]{Sch_MA_91}) For a random (vector) variable $X$ such that $\mathbb{E}[X^\top X]<\infty$ and a $\sigma$-algebra $\mathcal{G}$, the conditional expectation $\mathbb{E}[X|\mathcal{G}]$ is a version of the orthogonal projection of $X$ onto the probability space $\mathcal{L}^2(\Omega,\mathcal{G},\mathbf{P})$: for all $\mathcal{G}$-measurable (vector) functions $Y$, it holds that $Y\perp \left(X-\mathbb{E}[X|\mathcal{G}]\right)$, or equivalently
\begin{equation}
  \mathbb{E}\left[Y\left(X-\mathbb{E}[X|\mathcal{G}]\right)^\top\right]=0.
\end{equation}
\end{lemma}
Second, we provide a lemma that describes the relationship between the Kullback-Leibler divergence and the mean-square error under Gaussian smoothing.
\begin{lemma}\label{Transportation_Inequality_lmm}(\cite{Wu_HWI}\cite[Lemma 3.4.2]{Rag_FaT_13})
Let $\mathbf{x}$ and $\mathbf{y}$ be a pair of $N$-dimensional real-valued random vectors, and let $\mathbf{z}\sim\mathcal{N}(\mathbf{0}_N,\mathbf{I}_N)$ be independent of $(\mathbf{x}, \mathbf{y})$. Then, for any $t>0$,
\begin{equation}\label{Transportation_Inequality}
  D\left( {P_{\mathbf{x}+\sqrt{t}\mathbf{z}}}||{P_{\mathbf{y}+\sqrt{t}\mathbf{z}}} \right)\le \frac{1}{2t}\mathbb{E}\left[ {{\left\| \mathbf{x}-\mathbf{y} \right\|}_2^2 } \right].
\end{equation}
\end{lemma}
\begin{IEEEproof}
See page 116 of~\cite{Rag_FaT_13}. The proof follows from~\cite{Wu_HWI}. However, the proof in~\cite{Rag_FaT_13} is presented for the case when the vector length $N=1$. Thus, we include the complete proof for general $N$ in Appendix~\ref{PofLm2}.
\end{IEEEproof}

\section{System Model and Problem Formulation}\label{model}
We consider a linear function computation problem in a tree network $\mathcal{T}=(\mathcal{V},\mathcal{E})$. Suppose each node $v_i\in \mathcal{V}$ observes an independent random vector ${\mathbf{x}_i}\sim \mathcal{N}(\mathbf{0}_N,\mathbf{I}_N)$. We assume that each edge in $\mathcal{E}$ is a noiseless bidirectional link, through which bits can be sent. The objective is to obtain a weighted sum $\mathbf{y}=\sum\limits_{i=1}^n{w_i{\mathbf{x}_i}}$ at the pre-assigned sink node ${v_0}$, which is taken to be the root node. In Section~\ref{consensus_sec}, we will also consider an extension of the problem where the weighted sum is computed at all nodes.

Throughout the paper, we assume time is slotted. In each time slot, we assume that only one node transmits along only one edge. We follow the notion of distributed computation scheme introduced in \cite{Su_TIT_10}. By a \emph{distributed computation scheme}, we denote a five-tuple $(T,\mathscr{S},\mathscr{G},\mathbf{v},\mathbf{e})$ described in the following. We use $T$ to denote the total number of time slots, $\mathscr{S}$ to denote a sequence of real mappings $\mathscr{S}=\{f_t\}_{t=1}^T$, and $\mathscr{G}$ to denote a sequence of encoding mappings $\mathscr{G}=\{g_t\}_{t=1}^T$. We use $\mathbf{v}=[v(1),v(2),\ldots v(T)]$ to denote a vector of node indices and $\mathbf{e}$ to denote a vector of edge indices $\mathbf{e}=[e(1),e(2),\ldots e(T)]$, such that at each time slot $t$, the transmitting node $v(t)$ computes the mapping $f_{t}$ (whose arguments are to be made precise below) and transmits an encoded version $g_t(f_t)$ to one of its neighbors through the edge $e(t)$. The only assumption that we make about the encoding mappings is that each mapping $g_t$ outputs a binary sequence of a finite length. The arguments of $f_t$ may consist of all the information available at the transmitting node $v(t)$ up to time $t$, including its observation $\mathbf{x}_{v(t)}$, randomly generated data, and information obtained from its neighborhood up to time $t$. Note that the total number of time slots $T$ can be greater than number of vertices $n$ in general, i.e., nodes may be allowed to transmit multiple times. For an arbitrary link $v_i\to v_j$, define $M_{i\to j}$ as all the bits transmitted on the link $v_i\to v_j$ (see Fig.~\ref{cut_set_figure}). Denote by $R_{i\to j}$ the number of bits in $M_{i\to j}$ normalized by $N$. Note that $R_{i\to j}$ is the (normalized) total number of bits transmitted possibly over multiple time slots to node $v_j$. Also note that $R_{i\to j}>0$ only if $v_i$ and $v_j$ are connected. By sum rate $R$, we mean the total number of bits communicated in the distributed computation scheme normalized by $N$. Since we only consider tree graphs,
\begin{equation}\label{total_rate}
\begin{split}
  R=&\frac{1}{N}\sum\limits_{i=1}^n(NR_{i\to \text{PN}(i)}+NR_{\text{PN}(i)\to i})=\sum\limits_{i=1}^n(R_{i\to \text{PN}(i)}+R_{\text{PN}(i)\to i}).
\end{split}
\end{equation}
We only consider oblivious distributed computation schemes, i.e., the five-tuple $(T,\mathscr{S},\mathscr{G},\mathbf{v},\mathbf{e})$ is fixed and does not change with inputs. Further, we assume that a scheme terminates in finite time, i.e., $T < \infty$. A scheme must be feasible, i.e., all arguments of $f_t$ should be available in $v(t)$ before time $t$. Denote by $\mathcal{F}$ the set of all feasible oblivious distributed computation schemes (five-tuples). Although a feasible scheme is general, in that it allows a given edge $e$ to be active at multiple (non-consecutive) slots, our inner bound scheme is based on a sequential scheduling, where each node transmits to its parent node only once.

Since the goal is to compute $\mathbf{y}=\sum\limits_{i=1}^n{w_i{\mathbf{x}_i}}$ at the sink node $v_0$, without loss of generality, we assume $v(T)=v_0$ and the output of the mapping $f(T)$ computed at $v(T)$ is the final estimate $\widehat{\mathbf{y}}$. Denote by $D$ the overall (normalized) mean-square distortion
\begin{equation}\label{Y_dist}
  D=\frac{1}{N}\mathbb{E}\left[\left\|\mathbf{y}-\widehat{\mathbf{y}}\right\|_2^2\right].
\end{equation}
The objective is to compute the minimum value of the {sum rate} $R$ (defined in~\eqref{total_rate}) such that the overall distortion is smaller than $D^\text{tar}$.
\begin{equation}\label{RD_op_problem}
\begin{split}
&\min_{(T,\mathscr{S},\mathscr{G},\mathbf{v},\mathbf{e})\in\mathcal{F}}{\;\;\;\;}R,\\
&{\;\;\;\;\;\;}\text{s.t.}{\;}D\le D^\text{tar}.
\end{split}
\end{equation}

In what follows, we define some quantities associated with the ``incremental distortion'' that we mentioned in Section~\ref{Intro}. For an arbitrary set $\mathcal{S}\subset \mathcal{V}$, define ${\mathbf{y}_\mathcal{S}}=\sum\limits_{v_j\in \mathcal{S}}{{{w}_{j}}{\mathbf{x}_j}}$ as the partial sum in $\mathcal{S}$. We use $\sigma_\mathcal{S}^2=\sum\limits_{v_j\in \mathcal{S}}{w_{j}^2 }$ to denote the variance of each entry of $\mathbf{y}_\mathcal{S}$. Suppose at the final time slot $T$, all the available information (observations of random variables) at a node $v_i\in \mathcal{V}$ is $I_i$. Denote by $\widehat{\mathbf{y}}^\text{mmse}_{\mathcal{S},i}$ the MMSE estimate of ${\mathbf{y}_\mathcal{S}}$ at any node $v_i$, given the information $I_i$, which can be written as
\begin{equation}\label{MMSE_estimator}
  {\widehat{\mathbf{y}}^\text{mmse}_{\mathcal{S},i}}=\mathbb{E}\left[ {\mathbf{y}_\mathcal{S}}|{{I}_i} \right].
\end{equation}
\textcolor{black}{For an arbitrary (non-sink) node $v_i$ and its parent node $v_{\text{PN}(i)}$, denote by $D_i^\text{Tx}$ and ${D_i^\text{Rx}}$ the MMSE distortions of estimating ${\mathbf{y}_{\mathcal{S}_i}}$, respectively at $v_i$ and $v_{\text{PN}(i)}$, where, recall, $\mathcal{S}_i$ denotes the set of descendants of node $v_i$ (including itself). The information about $\mathbf{y}_{\mathcal{S}_i}$ should be transmitted from $v_i$ to its parent $v_{\text{PN}(i)}$. Therefore, the superscript $^\text{Tx}$ means that the distortion is defined for the transmitting node $v_i$, and the superscript $^\text{Rx}$ means the receiving node $v_{\text{PN}(i)}$. }Define $D_i^\text{Inc}$ to be the mean-square difference between the two estimates $\widehat{\mathbf{y}}^\text{mmse}_{\mathcal{S}_i,i}$ and $\widehat{\mathbf{y}}^\text{mmse}_{\mathcal{S}_i,\text{PN}(i)}$. Thus,
\begin{align}
  &D_i^\text{Tx}=\frac{1}{N}\mathbb{E}\left[ {{\left\| {\mathbf{y}_{\mathcal{S}_i}}-{\widehat{\mathbf{y}}^\text{mmse}_{\mathcal{S}_i,i}} \right\|_2^2 }} \right],\label{MMSE_i}\\
  &{D_i^\text{Rx}}=\frac{1}{N}\mathbb{E}\left[ {{\left\| {\mathbf{y}_{\mathcal{S}_i}}-{\widehat{\mathbf{y}}^\text{mmse}_{\mathcal{S}_i,\text{PN}(i)}} \right\|}_2^2 } \right],\label{MMSE_father}\\
  &{D_i^\text{Inc}}=\frac{1}{N}\mathbb{E}\left[ {{\left\| {\widehat{\mathbf{y}}^\text{mmse}_{\mathcal{S}_i,\text{PN}(i)}}-{\widehat{\mathbf{y}}^\text{mmse}_{\mathcal{S}_i,i}} \right\|}_2^2 } \right].\label{new_distortion}
\end{align}
Denote the MMSE distortion in estimating $\mathbf{y}=\sum\limits_{i=1}^n{w_i{\mathbf{x}_i}}$ at ${v_0}$ by $D_0^\text{mmse}$. Because for the same distributed computation scheme, the overall distortion $D$ cannot be less than $D_0^\text{mmse}$, the overall distortion with MMSE estimate at the sink $v_0$,
\begin{equation}\label{lb_d0}
  D\ge D_0^\text{mmse}.
\end{equation}
In Section~\ref{Dist_acc_section}, we will show that $D_i^\text{Inc}={D_i^\text{Rx}}-D_i^\text{Tx}$ (for all feasible distributed computation schemes) and the overall MMSE distortion $D_0^\text{mmse}$ can be written as the summation of $D_i^\text{Inc}$ on all links. Therefore, we call $D_i^\text{Inc}$ the incremental distortion.

\section{Main Results: Outer Bounds Based on Incremental Distortion}\label{main_sec}
\subsection{Distortion Accumulation}\label{Dist_acc_section}
Our first result shows that the overall MMSE distortion can be written as the summation of the distortion on all the tree links. It asserts that the distortion for in-network computing must accumulate along the way from all the leaves to the sink node.

\begin{theorem}[Distortion Accumulation]\label{Distortion_add_up}
For any feasible distributed computation scheme (see the model of Section~\ref{model}) and for each node $v_i\in \mathcal{V}\setminus \{v_0\}$, the incremental distortion $D_i^\text{Inc}$ and the MMSE distortions $D_i^\text{Tx}$ and $D_i^\text{Rx}$ satisfy
\begin{equation}\label{Pyth_all}
  {D_i^\text{Rx}}=D_i^\text{Tx}+{D_i^\text{Inc}}.
\end{equation}
\textcolor{black}{Thus, we also have
\begin{equation}\label{Tr_Distortion}
  D_i^\text{Tx}=\sum\limits_{v_j\in \mathcal{S}_i\setminus\{v_i\}}{{D_j^\text{Inc}}},
\end{equation}}
\begin{equation}\label{Total_Distortion}
  D_0^\text{mmse}=\sum\limits_{i=1}^n{{D_i^\text{Inc}}}.
\end{equation}
\end{theorem}
\begin{IEEEproof}
See Appendix~\ref{add_distortion_app}.
\end{IEEEproof}
\textcolor{black}{\begin{remark}\label{induction_on_tree}
In some of the proofs in this paper, we adopt an `induction method in the tree network', which we often briefly refer to as \emph{induction in the tree}. The idea is that, to prove that some property $P$ holds for each node $v_i\in \mathcal{V}$, firstly, we prove that $P$ holds at all leaves. Secondly, we prove that, for an arbitrary node $v_b$, if $P$ holds at $v_b$, then $P$ also holds at its parent-node $v_a$. It is obvious that these two arguments lead to the conclusion that $P$ holds for all nodes in the tree network.
\end{remark}}
{\begin{remark}Note that the distortion accumulation effect does not happen in classical relay networks that can be understood quite well using deterministic abstractions. However, our result shows that it is unclear if similar abstractions can be made to obtain insight on in-network computation. Coming up with such abstractions is a fruitful direction of research in rate-limited and/or noisy computing.\end{remark}}

\subsection{Rate Distortion Outer Bound}
Our second result provides an outer bound on the rate distortion function for linear computation over a tree network using incremental distortions.
\begin{theorem}[Incremental-Distortion-Based Outer Bound]\label{main_thm}
For the model of Section~\ref{model}, given a feasible distributed computation scheme, the sum rate is lower-bounded by
\begin{equation}\label{rate_distortion_lower_bound}
\begin{split}
  R&\ge\frac{1}{2}\sum\limits_{i=1}^n \left[
  {\log_2 }\frac{\sigma _{{\mathcal{S}_i}}^2 }{{D_i^\text{Inc}}}
  -\frac{D_i^\text{Tx}}{2w_i^2 }
  -\frac{{\log_2 }e}{2\sigma _{\mathcal{S}_i}^2 }\sqrt{2D_i^\text{Tx}\left( 4\sigma_{\mathcal{S}_i}^2 +D_i^\text{Tx}\right)} \right]\\
   &=\frac{1}{2}\sum\limits_{i=1}^n{\left[ {\log_2 }\frac{\sigma _{{\mathcal{S}_i}}^2 }{D_i^\text{Rx}-D_i^\text{Tx}}-\mathcal{O}\left((D_i^\text{Tx})^{1/2}\right) \right]},
\end{split}
\end{equation}
where $w_i$ is the weight of the observation $\mathbf{x}_i$, $\mathcal{S}_i$ is the node set that contains node $v_i$ and its descendants, $\sigma_{\mathcal{S}_i}^2$ is the variance of each entry of the partial sum ${\mathbf{y}_{\mathcal{S}_i}}=\sum\limits_{v_j\in \mathcal{S}_i}{{{w}_{j}}{\mathbf{x}_j}}$, $D_i^\text{Tx}$ and $D_i^\text{Inc}$ are the MMSE distortion and the incremental distortion at the node $v_i$, which are respectively defined in~\eqref{MMSE_i} and~\eqref{new_distortion}. %\footnote{{Since $D_i^\text{Tx}$ and $D_i^\text{Inc}$ both depend on the distributed computation scheme, the bound obtained is not fundamental. Specifically, the dependence on the distributed computation scheme is through the $D_i^\text{Tx}$'s and the $D_i^\text{Inc}$'s only. See Remark~\ref{remark1} and Corollary~\ref{corollary_1}. } }.
By optimizing over the incremental distortions $D_i^\text{Inc}$, one obtains the following scheme-independent bound stated in an optimization form
\begin{equation}\label{min_opti}
\begin{split}
&\mathop{\min}\limits_{D_i^\text{Inc},1\le i\le n} \frac{1}{2}\sum\limits_{i=1}^n \left[
  {\log_2 }\frac{\sigma _{{\mathcal{S}_i}}^2 }{{D_i^\text{Inc}}}
  -\frac{D_i^\text{Tx}}{2w_i^2 }
  -\frac{{\log_2 }e}{2\sigma _{\mathcal{S}_i}^2 }\sqrt{2D_i^\text{Tx}\left( 4\sigma_{\mathcal{S}_i}^2 +D_i^\text{Tx}\right)} \right],\\
  &{\;\;\;\;\;\;\;\;\;\;\;\;\;\;\;\;\;\;\;}\text{s.t.} \left\{ \begin{matrix}
   D_i^\text{Tx}=\sum\limits_{v_j\in \mathcal{S}_i\setminus\{v_i\}}{{D_j^\text{Inc}}},\forall i\neq 0,  \\
   \sum\limits_{i=1}^n{{D_i^\text{Inc}}}=D_0^\text{mmse}\le D.
\end{matrix} \right.
\end{split}
\end{equation}
Define the function $\psi_i(\cdot)$ as
\begin{equation}
\psi_i(x)=\frac{x}{2w_i^2 }
  +\frac{{\log_2 }e}{2\sigma _{\mathcal{S}_i}^2 }\sqrt{2x\left( 4\sigma_{\mathcal{S}_i}^2 +x\right)}.
\end{equation}
Then, a lower bound on $R$ can be obtained from the optimization in \eqref{min_opti}:
\begin{equation}\label{111111111}
\begin{split}
R\ge \frac{1}{2}\,{\log_2 }\frac{\prod\limits_{i=1}^n{\sigma_{\mathcal{S}_i}^2}}{{{\left( {{D}}/n \right)}^n}}-\frac{1}{2}\sum_{i=1}^n \psi_i(D),
\end{split}
\end{equation}
which means that in the limit of small distortion $D$, the optimization problem \eqref{min_opti} provides the following lower bound in order sense
\begin{equation}\label{minimize_lower_bound}
\begin{split}
  R\ge\frac{1}{2}\,{\log_2 }\frac{\prod\limits_{i=1}^n{\sigma_{\mathcal{S}_i}^2}}{{{\left( {{D}}/n \right)}^n}}-n\mathcal{O}(D^{1/2}).
\end{split}
\end{equation}
\end{theorem}
\textit{Proof Sketch:} The complete proof is in Appendix~\ref{lower_bound_proof}. The first step is to prove, on an arbitrary link $v_b\to v_a$ towards the root (see Fig.~\ref{cut_set_figure}), $NR_{b\to a}\ge h({\widehat{\mathbf{y}}^\text{mmse}_{\mathcal{S},b}})-\frac{N}{2}{\log_2 }2\pi e{D_b^\text{Inc}}$, where $h(\cdot)$ denotes differential entropy, and hence the rate $R_{b\to a}$ is related to the incremental distortion $D_b^\text{Inc}$.

Then, we prove that $h({\widehat{\mathbf{y}}^\text{mmse}_{\mathcal{S},b}})>h({\mathbf{y}_\mathcal{S}})-\mathcal{O}\left(N(D_b^\text{Tx})^{1/2}\right)$, using inequality~\eqref{Transportation_Inequality}. Thus, using $h({\mathbf{y}_\mathcal{S}})=\frac{N}{2}\log 2\pi e \sigma _{{\mathcal{S}_b}}^2 $ (note that $\mathcal{S}$ and $\mathcal{S}_b$ here denote the same set), we get $R_{b\to a}\ge\frac{1}{2}{\log_2 }\frac{\sigma _{{\mathcal{S}_b}}^2 }{D_b^\text{Inc}}-\mathcal{O}\left((D_b^\text{Tx})^{1/2}\right)$. Inequality \eqref{rate_distortion_lower_bound} can be obtained by summing over all links towards the root. The optimization form obtained in \eqref{min_opti} only requires the minimization of the scheme-dependent bound over the choices of $D_i^\text{Inc}$. The proof of the last inequality~\eqref{111111111} and its order-sense form \eqref{minimize_lower_bound} can be obtained by lower-bounding the optimization problem \eqref{min_opti}.\qed

This outer bound is obtained when all incremental distortions are equal, which is very similar to the reverse water-filling solution for the parallel Gaussian lossy source coding problem~\cite[Theorem 10.3.3]{Cover_Wiley_06} in the limit of large rate (zero distortion). We will prove that this rate (in the small distortion regime) is also achievable using Gaussian random codebooks (see Section~\ref{Gaussian_sec}). To achieve the optimal sum rate, the rate on the link $v_i\to v_{\text{PN}(i)}$ should be approximately equal to $\frac{1}{2}\log_2\frac{\sigma_{\mathcal{S}_i}^2}{D/n}$, where $\sigma_{\mathcal{S}_i}^2$ is the variance of each entry of the partial sum $\mathbf{y}_{\mathcal{S}_i}$.
\subsection{Comparison With the Cut-Set Bound}\label{comparison_section}
Using the classical cut-set bound technique~\cite[Thm. 1]{Su_TIT_10}, we can obtain another bound different from the one in Theorem~\ref{main_thm}. This bound {is in the same mathematical form as the sum rate expression in~\cite[Sec. V-A.3]{Cuf_ISIT_09}}.
\begin{theorem}[Cut-Set Outer Bound]\label{cut_set_thm}
For the model of Section~\ref{model}, the sum rate is lower-bounded by
\begin{equation}\label{abbas_cutset_bound}
  R\ge \frac{1}{2}\sum\limits_{i=1}^n{{\log_2 }\frac{\sigma _{{{\mathcal{S}}_i}}^2 }{{D_i^\text{Rx}}}}.
\end{equation}
\end{theorem}
\begin{IEEEproof}
See Appendix~\ref{cut_set_proof}.
\end{IEEEproof}
Denote by $R_1$ the outer bound obtained by the classical cut-set bound (Theorem~\ref{cut_set_thm}) and by $R_2$ the outer bound obtained by Theorem~\ref{main_thm}. From~\eqref{rate_distortion_lower_bound} and~\eqref{abbas_cutset_bound}
\begin{equation}\label{comparison}
\begin{split}
  \Delta_R:={R_{2}}-{R_{1}}
  %=\frac{1}{2}\sum\limits_{i=1}^n{\left[ {\log_2 }\frac{\sigma_{\mathcal{S}_i}^2}{{D_i^\text{Inc}}}-\mathcal{O}\left((D_i^\text{Tx})^{1/2}\right) \right]}-\frac{1}{2}\sum\limits_{i=1}^n{{\log_2 }\frac{\sigma _{{{\mathcal{S}}_i}}^2 }{{D_i^\text{Rx}}}}\\
  =\frac{1}{2}\sum\limits_{i=1}^n{\left[ {\log_2 }\frac{{D_i^\text{Rx}}}{{D_i^\text{Rx}}-D_i^\text{Tx}}-\mathcal{O}\left((D_i^\text{Tx})^{1/2}\right) \right]}.
\end{split}
\end{equation}
In order to illustrate the improvement on the outer bound $R_2$, we consider the case when $\mathcal{T}=(\mathcal{V},\mathcal{E})$ is a line network, connected as $v_0\leftrightarrow v_1 \leftrightarrow \ldots \leftrightarrow v_n$. Then,
\begin{equation}\label{line_net_1}
  {\widehat{\mathbf{y}}^\text{mmse}_{\mathcal{S}_{i-1},i-1}}\overset{(a)}{\mathop{=}}\,{\widehat{\mathbf{y}}^\text{mmse}_{\mathcal{S}_{i-1},\text{PN}(i)}}
  \overset{(b)}{\mathop{=}}\,{\widehat{\mathbf{y}}^\text{mmse}_{\mathcal{S}_i,\text{PN}(i)}}+{{w}_{i-1}}{\mathbf{x}_{i-1}},
\end{equation}
where $(a)$ holds because $v_{i-1}$ is the parent-node of $v_i$, and $(b)$ follows from ${\mathbf{y}_{\mathcal{S}_{i-1}}}={\mathbf{y}_{\mathcal{S}_i}}+{{w}_{i-1}}{\mathbf{x}_{i-1}}$. Therefore, $ {\mathbf{y}_{\mathcal{S}_{i-1}}}-{\widehat{\mathbf{y}}^\text{mmse}_{\mathcal{S}_{i-1},i-1}} =  {\mathbf{y}_{\mathcal{S}_i}}-{\widehat{\mathbf{y}}^\text{mmse}_{\mathcal{S}_i,\text{PN}(i)}} $. Using~\eqref{MMSE_i},~\eqref{MMSE_father}, we obtain ${D_{i-1}^\text{Tx}}={D_i^\text{Rx}}$. Thus,~\eqref{comparison} changes to
\begin{equation}\label{comparison_1}
\begin{split}
  \Delta_R=\frac{1}{2}\sum\limits_{i=1}^n{\left[ {\log_2 }\frac{{D_{i-1}^\text{Tx}}}{{D_{i-1}^\text{Tx}}-D_i^\text{Tx}}-\mathcal{O}\left((D_i^\text{Tx})^{1/2}\right) \right]},
\end{split}
\end{equation}
where $0=D_n^\text{Tx}<D_{n-1}^\text{Tx}<\ldots<D_1^\text{Tx}<D_0^\text{mmse}\le D$.

Then, we consider a typical choice of $D_i^\text{Tx}$, which minimizes the rate outer bound. In~\eqref{minimize_lower_bound}, we can show that, when $D$ is required to be small enough, the way to minimize the RHS of~\eqref{rate_distortion_lower_bound} is to make $D_i^\text{Rx}-D_i^\text{Tx}$ to be a constant for all $i$. This strategy yields a lower bound on the minimum possible rate. In the case of a line network, this strategy becomes $D_i^\text{Tx}=\frac{n-i}{n}D_0^\text{mmse}, \forall i$. Then
\begin{equation}\label{comparison_2}
\begin{split}
  \Delta_R=\sum\limits_{i=1}^n{\left[ \frac{{\log_2 }(n-i+1)}{2}-\mathcal{O}\left((D_i^\text{Tx})^{1/2}\right) \right]}\approx &\frac{1}{2}\log_2 (n!)
  =\Theta(n\log_2 n),
\end{split}
\end{equation}
when the overall distortion $D$ is small, i.e., the gap between the two bounds can be arbitrarily large.
\begin{remark}\label{comparison_remark}
Here, we point out the intuition underlying the difference between the proofs of the incremental-distortion-based bound (Theorem~\ref{main_thm}) and the cut-set bound (Theorem~\ref{cut_set_thm}). The classical proofs of cut-set bounds for lossy computation often rely on the following key steps (see Appendix~\ref{cut_set_proof}, as well as the proofs of \cite[Theorem III.1]{Aya_TIT_10}) and \cite[Proposition 4]{Su_TIT_10}):
\begin{equation}
\begin{split}
&\text{Rate}\ge I\left(\text{Computed Result};\text{True Result}\right)\\
&\ge h(\text{True Result})-h(\text{True Result}|\text{Computed Result}),
\end{split}
\end{equation}
where $h(\text{True Result}|\text{Computed Result})$ can be upper-bounded by a function of overall distortion and the expression $h(\text{True Result})$ can be obtained explicitly. However, the proof of the incremental-distortion-based bound is based on the following key steps (see Appendix~\ref{lower_bound_proof}):
\begin{equation}
\begin{split}
&{\;\;\;\;}\text{Rate on Link }e=(v_1,v_2)\\
&\ge I\left(\text{Computed Result 1};\text{Computed Result 2}\right)\\
&\ge h(\text{Computed Result 1})-h(\text{Computed Result 1}|\text{Computed Result 2}),
\end{split}
\end{equation}
where ``Computed Result 1'' denotes the MMSE estimate at the parent-node $v_1$ on link $e=(v_1,v_2)$ and ``Computed Result 2'' denotes the MMSE estimate at the child-node $v_2$ on link $e$. The term $h(\text{Computed Result 1}|\text{Computed Result 2})$ leads to a function of incremental distortion between two estimates, which yields a tighter bound than cut-set bounds for lossy in-network computing. However, the distribution of ``Computed Result 1'', the MMSE estimate, is unknown, and hence $h(\text{Computed Result 1})$ cannot be obtained directly. To solve this problem, we lower-bound $h(\text{Computed Result 1})$ by upper-bounding the difference between $h(\text{Computed Result 1})$ and $h(\text{True Result})$, using the inequality in Lemma~\ref{Transportation_Inequality_lmm}.
\end{remark}

\section{Achievable Rates with Random Gaussian Codebooks}\label{Gaussian_sec}

In this section, we use random Gaussian codebooks to give an incremental-distortion based sum rate inner bound. The main achievable result in this paper is as follows.
\begin{theorem}[Inner Bound]\label{Gaussian_code_thm}
\textcolor{black}{Using random Gaussian codebooks, we can find a distributed computation scheme, such that the sum rate $R$ is upper-bounded by
\begin{equation}\label{Gaussian_code_rate}
    R\le\frac{1}{2}\sum\limits_{i=1}^n{\log_2 }\frac{\sigma_{\mathcal{S}_i}^2}{d_i}+n\delta_N,
\end{equation}
where $\lim_{N\to\infty} \delta_N=0$ is a parameter defined in \eqref{r_up_12}, and $d_i$'s are tunable distortion parameters, and $\sigma _\mathcal{S}^2 =\sum\limits_{v_j\in \mathcal{S}}{w_{j}^2 }$. Further, the overall distortion $D$ satisfies
\begin{equation}\label{Gaussian_code_distortion}
    D\le \sum\limits_{i=1}^n d_i+\epsilon_N,
\end{equation}
where $\lim_{N\to\infty} \epsilon_N =0$ is a parameter defined in~\eqref{555}\footnote{The parameter $\delta_N$ is used for providing a slight excess rate of the rate defined by mutual information in \eqref{r_up_12}, and the parameter $\epsilon_N$ upper-bounds the deviation of the overall sum distortion $D$ from the summation of the tunable distortion parameters $\sum\limits_{i=1}^n d_i$. Note that here $N$ denotes the code length of the random Gaussian codebooks.}.} The limit sum rate $\lim_{N\to\infty}R$ exists, and can be upper-bounded by
\begin{equation}\label{minimize_upper_bound}
  \lim_{N\to\infty}R\le \frac{1}{2}\,{\log_2 }\frac{\prod\limits_{i=1}^n{\sigma_{\mathcal{S}_i}^2}}{{{\left( {{D}}/n \right)}^n}}.
\end{equation}
\end{theorem}
\begin{IEEEproof}
See Section~\ref{Gaussian_code_alg}.
\end{IEEEproof}
%\begin{remark}
%Notice that the inner bound also depends on the specific choice of a set of parameters $d_i,i=1,2,\ldots n$. To obtain the optimal rate-distortion function $R(D)$, one can optimize the inner bound with the best choice of $d_i$. The optimized inner bound when the code length $N\to \infty$ is shown in Corollary~\ref{corollary_2}, which, together with the lower bound \eqref{minimize_lower_bound} in Theorem~\ref{main_thm}, shows the tightness of the inner and outer bounds.
%\end{remark}
We rely on typicality-based arguments to prove the inner bound. Therefore, before we elaborate on the main distributed computation scheme in Section~\ref{Gaussian_code_alg}, we first review some notation and techniques on typicality.

\subsection{Notation on Typicality-Based Coding}\label{test_channel}
We first define some random variables, the pdfs of which we will use in the distributed computation scheme. (We will clarify the absolute continuity and hence existence of densities with respect to the appropriate Lebesgue measure of the various random objects used in our proofs.) At each node $v_i$, we define an \emph{estimate random variable} $U^\text{TC}_i$ and a \emph{description random variable} $V^\text{TC}_i$. The superscript $^\text{TC}$ represents the Gaussian test channel, which we will use to define these scalar random variables. Denote the variance of $U^\text{TC}_i$ by $\widehat{\sigma}_i^2$. The estimate random variables $U^\text{TC}_i$'s are defined from the leaves to the root $v_0$ in the tree. For an arbitrary leaf $v_l$, define
\begin{equation}\label{leaf_estimator}
  {U^\text{TC}_l}=w_lX_l,
\end{equation}
where $X_l\sim\mathcal{N}(0,1)$ is a scalar random variable, and $w_l$ is the weight at node $v_l$ in the weighted sum $\mathbf{y}=\sum\limits_{i=1}^n{w_i{\mathbf{x}_i}}$. For non-leaf nodes, without loss of generality, we use $v_1$, $v_2$, \ldots $v_d$ to denote the children of an arbitrary node $v_b$ (see Fig. \ref{cut_set_figure}). Suppose the description random variables $\{V_i^\text{TC}\}_{i=1}^d$ at the children of node $v_b$ have been defined (\textcolor{black}{the formal definitions of the description random variables are provided later in equation \eqref{tree_test_channel}}). Then, define the estimate random variable for the non-leaf node $v_b$ as
\begin{equation}\label{non_leaf_estimator}
  {U^\text{TC}_b}=\sum\limits_{k=1}^{d}V^\text{TC}_k+w_bX_b,
\end{equation}
where $X_b\sim\mathcal{N}(0,1)$ is a scalar random variable, and $w_b$ is the weight at $v_b$. At each node $v_i$, the description random variable $V^\text{TC}_i$ is now defined based on the estimate random variable using a Gaussian test channel
\begin{equation}\label{tree_test_channel}
  U^\text{TC}_i=V^\text{TC}_i+Z_i,
\end{equation}
where \textcolor{black}{$Z_i\sim\mathcal{N}(0,d_i)$ is independent of $V_i^\text{TC}$ and $d_i$ is a variable that will be chosen later. From the definition of Gaussian test channels, $\text{var}[V^\text{TC}_i]=\widehat{\sigma }_{i}^2 -d_i$. Readers are referred to Appendix~\ref{Gaussian_rate_calculation} for details on the definition of Gaussian test channels.} Then, using~\eqref{tree_test_channel}, we have that
\begin{equation}\label{var_induction}
  \widehat{\sigma }_b^2 =\sum\limits_{k=1}^{d}\text{var}[V^\text{TC}_i]+w_b^2=\sum\limits_{k=1}^{d}{(\widehat{\sigma }_{k}^2 -d_k)}+w_b^2 .
\end{equation}
\textcolor{black}{Note that the estimate random variables and the description random variables are both defined from leaves to the root. However, we have different definitions of the estimate random variables for leaves and non-leaf nodes (\eqref{leaf_estimator} and \eqref{non_leaf_estimator}) but the same definition of description random variables}. Note that the Gaussian test channel \eqref{tree_test_channel} and the definitions in~\eqref{leaf_estimator} and~\eqref{non_leaf_estimator} involve linear transformations. Therefore, all estimate random variables $U^\text{TC}_i$'s and description random variables $V^\text{TC}_i$'s are scalar Gaussian random variables with zero mean. We will not directly use the random variables $U^\text{TC}_i$ and $V^\text{TC}_i$ in the achievability proof (because they are scalars and cannot be directly used for coding). However, we use the pdfs of these random variables. We use $\phi_{U^\text{TC}_i}$ and $\phi_{V^\text{TC}_i}$ to denote the pdfs of $U^\text{TC}_i$ and $V^\text{TC}_i$. We also use joint pdfs, where the meanings are always clear from the context. Note that the variance of $U^\text{TC}_i$ and $V^\text{TC}_i$ are tunable, since the parameter $d_i$, which is related to the variance of the added Gaussian noise $Z_i$, is a tuning parameter.
\begin{remark}\label{remark5}
In fact, the way in which we define the description random variables and estimate random variables in Section~\ref{test_channel} essentially implies the basic idea of our distributed computation scheme. Although we consider block computation in the entire paper, we can view these description random variables and estimate random variables as the `typical' intermediate results during the computation. In particular, the estimate random variable $U^\text{TC}_i$ represents the typical properties of the estimate $\widehat{\mathbf{s}}_i$ of the partial sum $\mathbf{y}_{\mathcal{S}_i}$ at the node $v_i$ (by representing the typical properties, we mean the typical sets that the estimate $\widehat{\mathbf{s}}_i$ belongs to are defined based on the distributions of the estimate random variable $U^\text{TC}_i$), while the description random variable $V^\text{TC}_i$ represents the typical properties of the descriptions $\widehat{\mathbf{r}}_i$. Note that the messages to be further transmitted from the node $v_i$ to its parent node is the description sequence $\widehat{\mathbf{r}}_i$. The estimate ${U^\text{TC}_0}$ represents the properties of the estimate of $Y$ at the sink $v_0$. Based on this intuition, we can provide an intuitive explanation of the formula in Theorem~\ref{Gaussian_code_thm}: suppose $U^\text{TC}_i$ and $V^\text{TC}_i$ are length-$N$ vectors (this is of course technically incorrect, and we only try to provide some intuition on Theorem~\ref{Gaussian_code_thm} here), then, since $U^\text{TC}_i$ and $V^\text{TC}_i$ are all Gaussian, it can be proved that $V^\text{TC}_i$ is just the MMSE estimate ${\widehat{\mathbf{y}}^\text{mmse}_{\mathcal{S}_i,\text{PN}(i)}}=\mathbb{E}\left[ {\mathbf{y}_{\mathcal{S}_i}}|{{I}_{\text{PN}(i)}}\right]= \mathbb{E}\left[ {\mathbf{y}_{\mathcal{S}_i}}|V^\text{TC}_i\right]$ of the required partial sum $\mathbf{y}_{\mathcal{S}_i}$ at node $v_{\text{PN}(i)}$, the parent node of $v_i$. Then, we can apply the distortion accumulation result (\eqref{Total_Distortion} in Theorem \ref{Distortion_add_up}) to $U^\text{TC}_i$ and $V^\text{TC}_i$, and obtain $D= \sum\limits_{i=1}^n d_i$, since $d_i=\mathbb{E}[(U^\text{TC}_i)^2-(V^\text{TC}_i)^2]$ is the counterpart of the incremental distortion $D_i^\text{Inc}$. In Section \ref{Gaussian_code_alg}, we will formalize this intuitive argument using Gaussian random codes.
\end{remark}

Denote by $q_U$, $q_V$ and $q_{U,V}$ the $N$-fold product distribution of the scalar distributions $\phi_{U_i^\text{TC}}$, $\phi_{V_i^\text{TC}}$ and $\phi_{U_i^\text{TC},V_i^\text{TC}}$. Denote by $\mathcal{T}_{U,\varepsilon}^N$ and $\mathcal{T}_{V,\varepsilon}^N$ the two sets of $N$-length sequences ${s^N}$ and $r^N$ that are respectively typical with respect to ${\phi_U^\text{TC}}$ and ${\phi_V^\text{TC}}$. Denote by $\mathcal{J}_{\varepsilon }^{2N}$ the set of all $2N$-length sequences $\left( {s^N},{r^N} \right)$ that are jointly typical with respect to ${\phi}_{U^\text{TC},V^\text{TC}}$. Denote by $\mathcal{T}_{V,\varepsilon}^N(s^N)$ the set of sequences ${r^N}$ that are jointly typical with a particular typical sequence ${s^N}$. The formal definitions of these typical sets are provided in the following equations (\textcolor{black}{note that we will use a general definition of typical sets from \cite{Jeon_Arx_15}, and we will show that the definitions below are special cases of the general definition}):
\begin{equation}\label{typi1}
\mathcal{T}_{U,\varepsilon}^N=\left\{s^N:\left|-\frac{1}{N}\log q_U(s^N)-h(U_i^\text{TC})\right|< \varepsilon_N,\left|\frac{1}{N}||s^N||_2^2-\hat{\sigma}^2_i\right|<\varepsilon_N\right\},
\end{equation}

\begin{equation}\label{typi2}
\mathcal{T}_{V,\varepsilon}^N=\left\{r^N:\left|-\frac{1}{N}\log q_V(r^N)-h(V_i^\text{TC})\right|< \varepsilon_N,\left|\frac{1}{N}||r^N||_2^2-(\hat{\sigma}^2_i-d_i)\right|<\varepsilon_N\right\},
\end{equation}

\begin{equation}\label{joint_typicality}
\begin{split}
\mathcal{J}_{\epsilon }^{2N}=\left\{(s^N,r^N):s^N\in \mathcal{T}_{U,\varepsilon}^N, r^N\in \mathcal{T}_{V,\varepsilon}^N,\left|-\frac{1}{N}\log q_{U,V}(s^N,r^N)-h(U_i^\text{TC},V_i^\text{TC})\right|< \varepsilon_N,\right.\\
\left.\left|\frac{1}{N}||s^N-r^N||_2^2-d_i\right|<\varepsilon_N\right\},
\end{split}
\end{equation}

\begin{equation}\label{typi4}
\mathcal{T}_{V,\varepsilon}^N(s^N)=\left\{r^N:(s^N,r^N)\in \mathcal{J}_{\epsilon }^{2N}\right\}.
\end{equation}

\subsection{Applying Gaussian Codes in Function Computing}\label{Gaussian_code_alg}
The illustrative explanation in Remark~\ref{remark5} relies on Gaussian test channels, {which is a heuristic to provide insights into the design of the achievability strategy}. In this part, we rigorously prove the achievability using explicit random Gaussian codebooks.

Note that all computations are block computations. According to the system model, each node $v_i$ has a random vector ${\mathbf{x}_i}$, where each coordinate is generated by $\mathcal{N}(0,1)$. The sink $v_0$ has the goal to compute the weighted sum $\mathbf{y}=\sum\limits_{i=1}^n{w_i{\mathbf{x}_i}}$. Recall that ${\mathbf{y}_\mathcal{S}}=\sum\limits_{v_j\in \mathcal{S}}{{{w}_{j}}{\mathbf{x}_j}}$ and $\sigma _\mathcal{S}^2 =\sum\limits_{v_j\in \mathcal{S}}{w_{j}^2 }$.

Before the computation starts, each node $v_i$ generates a codebook\footnote{Notice that the rate of this code should be $\log_2 (2^{NR_i}+1)\approx R_i$. However, when $N\to\infty$ (which is the case considered in this section), the code rate converges to $R_i$. In other words, a single codeword $\mathbf{c}_i(0)$ has asymptotically no effect on the coding rate.} $\mathcal{C}_i=\{\mathbf{c}_i(w):\; w \in \{0,1,\ldots 2^{NR_i}\}\}$, where each codeword is generated i.i.d. according to distribution $p_{V^\text{TC}_i}$. The rate is chosen such that
\begin{equation}\label{r_up}
 R_i=I(U^\text{TC}_i;V^\text{TC}_i)+\delta_N=\frac{1}{2}\log \frac{\widehat{\sigma}_i^2}{d_i}+\delta_N,
\end{equation}
where $U^\text{TC}_i$ and $V^\text{TC}_i$ are scalar test-channel random variables defined in Section~\ref{test_channel} and $\underset{N\to \infty }{\mathop{\lim }}\delta_N=0$. We claim that, for each node $v_i\in\mathcal{V}$,
\begin{equation}\label{variance_less}
  \widehat{\sigma }_i^2 \le \sigma _{{{\mathcal{S}}_i}}^2 .
\end{equation}
\begin{IEEEproof}
See Appendix~\ref{pf_variance_less}.
\end{IEEEproof}
This leads to
\begin{equation}\label{r_up_12}
 R_i\le\frac{1}{2}\log \frac{\sigma_{\mathcal{S}_i}^2}{d_i}+\delta_N.
\end{equation}
Summing up~\eqref{r_up_12} over all links, we obtain the first inequality \eqref{Gaussian_code_rate} in Theorem~\ref{Gaussian_code_thm}.

The codebook $\mathcal{C}_i$ is revealed to $v_i$'s parent-node $v_{\text{PN}(i)}$. At the beginning of the distributed computation scheme, each leaf $v_l$ uses $w_l \mathbf{x}_l$ as the estimate $\widehat{\mathbf{s}}_l$. During the distributed computation scheme, as shown in Fig. \ref{cut_set_figure}, each non-leaf node $v_b$, upon receiving description indices $M_{1b}, M_{2b},\ldots  M_{db}$ from the $d$ children $v_1,\ldots  v_d$, decodes these description indices, computes the sum of these descriptions and the data vector generated at $v_b$ as follows
\begin{equation}\label{U_b_hat}
  \widehat{\mathbf{s}}_b=\sum\limits_{k=1}^{d}{\mathbf{c}_k(M_{k\to b})}+w_b\mathbf{x}_b,
\end{equation}
and re-encodes $\widehat{\mathbf{s}}_b$ into a new description index $M_{b\to a}\in [1:2^{NR_b}]$ and sends the description index to the parent-node $v_a$ using rate $R_b$. We denote the reconstructed description by $\widehat{\mathbf{r}}_b=\mathbf{c}_b(M_{b\to a})$. The decoding and encoding at the node $v_b$ are defined as follows. Note that the leaves only encode and the root $v_0$ only decodes.
\begin{itemize}
  \item \textbf{Decoding: }In each codebook $\mathcal{C}_k,k=1,\ldots d$, use the codeword $\mathbf{c}_k(M_{k\to b})$ as the description $\widehat{\mathbf{r}}_k$. If $v_b=v_0$ is the root, it computes the sum of all codewords $\mathbf{c}_k(M_{k\to 0})$ as the estimate of $\mathbf{y}$:
      \begin{equation}
        \widehat{\mathbf{y}}=\mathop{\sum}\limits_{v_k\in \mathcal{N}(v_0)} \mathbf{c}_k(M_{k\to 0})=\mathop{\sum}\limits_{v_k\in \mathcal{N}(v_0)} \widehat{\mathbf{r}}_k.
      \end{equation}
  \item \textbf{Encoding: }Find a codeword $\mathbf{c}_b(M_{b\to a}) \in \mathcal{C}_b\setminus \{\mathbf{c}_b(0)\}$ such that the two vectors $\widehat{\mathbf{s}}_b=\sum\limits_{k=1}^{d}{\mathbf{c}_k(M_{k\to b})}+w_b\mathbf{x}_b$ and $\widehat{\mathbf{r}}_b=\mathbf{c}_b(M_{b\to a})$ are jointly typical with respect to the test-channel distribution $\phi_{U^\text{TC}_b,V^\text{TC}_b}$ (in $\mathcal{J}_{\epsilon }^{2N}$). If there are more than one codewords that satisfy this condition, arbitrarily choose one of them. However, if $\widehat{\mathbf{s}}_b=\sum\limits_{k=1}^{d}{\mathbf{c}_k(M_{k\to b})}+w_b\mathbf{x}_b$ is not typical with respect to the test-channel distribution $\phi_{U^\text{TC}_b}$ (not in $\mathcal{T}_{U,\varepsilon}^N$), or if there is no codeword in $\mathcal{C}_b\setminus \{\mathbf{c}_b(0)\}$ that satisfies the joint typicality condition, send description index $M_{b\to a}=0$ (note that this means the index of the $0$-th random codeword $\mathbf{c}_b(0)$, instead of a vector $\mathbf{0}^N$).
\end{itemize}

Since all codebooks $\mathcal{C}_k,k=1,\ldots d,$ have been revealed to $v_b$, the decoding is always successful, in that the decoding process is simply the mapping from the description index $M_{k\to b}$ to the description $\mathbf{c}_k(M_{k\to b})$. However, the encoding may fail. In this case, the description index $M_{b\to a}=0$ is sent and this description index is decoded to a predetermined random sequence $\mathbf{c}_b(0)$ on the receiver side. Note that the rate $R_i$ is the same as $R_{i\to \text{PN}(i)}$ in~\eqref{total_rate}, and the notation $R_i$ is used here for simplicity. We still use the notation $R_{i\to \text{PN}(i)}$ for the results on network consensus, where each node may have to send descriptions to different nodes, and $R_{i\to \text{PN}(i)}$ can usefully indicate that the direction of information transmission is from the node $v_i$ to its parent node $v_{\text{PN}(i)}$.

\subsection{The Proof of Theorem~\ref{Gaussian_code_thm}: Analysis of the Gaussian Random Codes}
In this part, we analyze the expected distortion of the Gaussian random codes. \textcolor{black}{Note that, unless specifically clarified, all results in this part are stated for the random coding ensemble, i.e., the expectation $\mathbb{E}[\cdot]$ and the probability $\Pr(\cdot)$ are taken over random data sampling, codeword selection and random codebook generation. The result in Theorem~\ref{Gaussian_code_thm} holds for at least one code in this random coding ensemble.}

The following Lemma~\ref{covering_lemma} states that the estimate $\widehat{\mathbf{s}}_b$ and the description $\widehat{\mathbf{r}}_b$ are jointly typical for all $b$ with high probability.
\begin{lemma}[Covering Lemma for Lossy In-network Linear Function Computing]\label{covering_lemma}
For the encoding and decoding schemes as described in this section, denote by $E_i=1$ the event that the encoding at the node $v_i$ is not successful. Then
\begin{equation}\label{encoding_error}
 \underset{N\to \infty }{\mathop{\lim }}\,\underset{1\le i\le n}{\mathop{\sup }}\,\Pr ({E_i}=1)=0,
\end{equation}
where the probability is taken over random data sampling and random codebook generation.
\end{lemma}
\begin{IEEEproof}
See Appendix~\ref{pf_covering}.
\end{IEEEproof}
In Lemma~\ref{var_bounded}, we provide bounds on the variances of $\widehat{\mathbf{s}}_b$ and $\widehat{\mathbf{r}}_b$. Note that the inequalities in Lemma~\ref{var_bounded} do not trivially follow from the typicality of $\widehat{\mathbf{s}}_b$ and $\widehat{\mathbf{r}}_b$ because the typicality of $\widehat{\mathbf{s}}_b$ only ensures that $\frac{1}{N}\left\|\widehat{\mathbf{s}}_b\right\|_2^2-\widehat{\sigma}_b^2$ converges to zero in probability, while \eqref{var_bounded_eqn_1} requires convergence in mean value to zero. This is a standard issue. In Appendix~\ref{app_var_bounded}, we use a standard technique to overcome this issue. The key idea is that, for non-typical case (when encoding fails), we send a predetermined random sequence, on the variance of which we can provide a bound.
\begin{lemma}\label{var_bounded}
At each node $v_b$, the description $\widehat{\mathbf{r}}_b=\mathbf{c}_b(M_{b\to a})$ and the estimate $\widehat{\mathbf{s}}_b$ defined in~\eqref{U_b_hat} satisfy
\begin{equation}\label{var_bounded_eqn_1}
  \left|\mathbb{E}\left[\frac{1}{N}\left\|\widehat{\mathbf{s}}_b\right\|_2^2\right]-\widehat{\sigma}_b^2\right|<\varepsilon_N,
\end{equation}
\begin{equation}\label{var_bounded_eqn_2}
  \left|\mathbb{E}\left[\frac{1}{N}\left\|\widehat{\mathbf{r}}_b\right\|_2^2\right]-(\widehat{\sigma}_b^2-d_b)\right|<\varepsilon_N,
\end{equation}
\begin{equation}\label{typical_small_distortion_1}
  \left|\mathbb{E}\left[ \frac{1}{N}{{\left\| \widehat{\mathbf{r}}_b-\widehat{\mathbf{s}}_b \right\|}_2^2 } \right]-d_b \right|<\varepsilon_N,
\end{equation}
where $\lim_{N\to \infty}\varepsilon_N=0$.
\end{lemma}
\begin{IEEEproof}
See Appendix~\ref{app_var_bounded}.
\end{IEEEproof}
\begin{lemma}\label{ent_bounded}
At each node $v_b$, the description $\widehat{\mathbf{r}}_b=\mathbf{c}_b(M_{b\to a})$ and the estimate $\widehat{\mathbf{s}}_b$ defined in~\eqref{U_b_hat} satisfy
\begin{equation}\label{ent_bounded_eqn_1}
 h(\widehat{\mathbf{s}}_b)>\frac{N}{2}\log_{2}2\pi e \widehat{\sigma}^2_b-N\beta_N,
\end{equation}
\begin{equation}\label{ent_bounded_eqn_2}
  h(\widehat{\mathbf{r}}_b)>\frac{N}{2}\log_{2}2\pi e (\widehat{\sigma}^2_b-d_b)-N\beta_N,
\end{equation}
where $\lim_{N\to \infty}\beta_N=0$, $h(\cdot)$ is the differential entropy function, and \textcolor{black}{the random vectors $\widehat{\mathbf{s}}_b$ and $\widehat{\mathbf{r}}_b$ are defined in the probability space that contains the random codebook generation\footnote{\textcolor{black}{To define differential entropy for the two random vectors $\widehat{\mathbf{s}}_b$ and $\widehat{\mathbf{r}}_b$, we need to first define the densities (with respect to the Lebesgue measure) of the two random vectors. The estimate $\widehat{\mathbf{s}}_b$ is certainly absolutely continuous, because it is smoothed by the Gaussian random variable $\mathbf{x}_b$ (see \eqref{U_b_hat}). However, conditioned on a specific instance of the Gaussian codebooks, the random vector $\widehat{\mathbf{r}}_b$ has a finite support, and the (conditional) differential entropy of $\widehat{\mathbf{r}}_b$ is $-\infty$. To overcome this difficulty, we cast the analysis on the unconditional distribution of $\widehat{\mathbf{r}}_b$, i.e., taking into account the code generation randomness. In this way, $\widehat{\mathbf{r}}_b$ is also absolutely continuous.}}.}
\end{lemma}
\begin{IEEEproof}
See Appendix~\ref{ent_var_bounded}.
\end{IEEEproof}

Lemma~\ref{ent_bounded} indicates that $\widehat{\mathbf{s}}_b$ and $\widehat{\mathbf{r}}_b$ are close to Gaussian-distributed random variables in differential entropy sense. We will use Lemma~\ref{var_bounded} and Lemma~\ref{ent_bounded} to show a non-trivial relationship between the Gaussian-code-based distortion $d_i$ and the MMSE-based incremental distortion $D_i^\text{Inc}$. This relationship is characterized in Lemma~\ref{distortion_close_lemma}. \textcolor{black}{The proof is based on an observation that, when the true distribution of the source is close (in the sense of differential entropy) to the expected distribution, the estimate based on random coding can provide a distortion that is approximately equal to the MMSE estimate.}
\begin{remark}\label{why_MMSE}
\textcolor{black}{If we try to directly obtain the overall distortion bound in~\eqref{Gaussian_code_distortion} using some classical coding schemes such as Wyner-Ziv coding \cite[Chapter 15.9]{Cover_Wiley_06}, we have to prove that the incremental errors (the term $d_i$) due to successive quantizations along the network are `approximately uncorrelated' (so that $d_i$ for different $i$ can be summed up to obtain the bound on the overall distortion~\eqref{Gaussian_code_distortion}). While we do not pursue this direction, the above may be achieved by obtaining a non-trivial generalization of the ``Markov Lemma''~\cite[Lecture Notes 13]{Gamal_Camb_11} to Gaussian sources. To bypass this difficulty, we directly relate the Gaussian-code-based distortion $d_i$ and the MMSE-based distortion $D_i^\text{Tx}$, which simultaneously shows some nontrivial connections between Gaussian random codes and MMSE estimates. This is why the proof of the inner bound is conceptually different from existing literature.}
\end{remark}

Recall that the MMSE estimate of the sum $\mathbf{y}_{\mathcal{S}_i}$ at the node $v_j$ is denoted by
$\widehat{\mathbf{y}}^\text{mmse}_{\mathcal{S}_i,j}=\mathbb{E}_{\mathcal{C}_i}\left[ \mathbf{y}_{\mathcal{S}_i}|{{I}_{j}} \right]$, where ${{I}_{j}}$, as before, denotes the information available to the node $v_j$. Define $D_i^\text{Tx}$, $D_i^\text{Rx}$ and $D_i^\text{Inc}$ similar to~\eqref{MMSE_i},~\eqref{MMSE_father} and~\eqref{new_distortion}. That is, $D_i^\text{Tx}=\mathbb{E}\left[ \frac{1}{N}\left\| \mathbf{y}_{\mathcal{S}_i}-\widehat{\mathbf{y}}^\text{mmse}_{\mathcal{S}_i,i} \right\|_{2}^2  \right]$, ${D_i^\text{Rx}}=\mathbb{E}\left[ \frac{1}{N}\left\| \mathbf{y}_{\mathcal{S}_i}-\widehat{\mathbf{y}}^\text{mmse}_{\mathcal{S}_i,\text{PN}(i)} \right\|_{2}^2  \right]$ and ${D_i^\text{Inc}}=\frac{1}{N}\mathbb{E}\left[ {{\left\| {\widehat{\mathbf{y}}^\text{mmse}_{\mathcal{S}_i,\text{PN}(i)}}-{\widehat{\mathbf{y}}^\text{mmse}_{\mathcal{S}_i,i}} \right\|}_2^2 } \right]$. Notice that the inner $\mathbb{E}[\cdot]$ (for the MMSE estimate $\widehat{\mathbf{y}}^\text{mmse}_{\mathcal{S}_i,j}=\mathbb{E}_{\mathcal{C}_i}\left[ \mathbf{y}_{\mathcal{S}_i}|{{I}_{j}} \right]$) is for a given codebook $\mathcal{C}_i$ at $v_i$, because both $v_i$ and its parent $v_{\text{PN}(i)}$ know the codebook $\mathcal{C}_i$ (see the codebook construction in Section~\ref{Gaussian_code_alg}). However, the outer $\mathbb{E}[\cdot]$ (for $D_i^\text{Tx}=\mathbb{E}\left[ \frac{1}{N}\left\| \mathbf{y}_{\mathcal{S}_i}-\widehat{\mathbf{y}}^\text{mmse}_{\mathcal{S}_i,i} \right\|_{2}^2  \right]$) is still taken over both the codeword selection and the random codebook generation. In this subsection, the quantities $D_i^\text{Tx}$, $D_i^\text{Rx}$ and $D_i^\text{Inc}$ are all averaged over the random codebook ensemble.
\begin{lemma}\label{distortion_close_lemma}
For an arbitrary node $v_i$
\begin{equation}\label{distortion_close}
\sqrt{d_i -{\varepsilon }_{N}}-{{\eta }_{N}}\le\sqrt{D_i^\text{Inc}}\le \sqrt{d_i +{\varepsilon }_{N}}+{{\eta }_{N}},
\end{equation}
where $\underset{N\to \infty }{\mathop{\lim }}\,{{\eta }_{N}}=0$ and $\varepsilon_N$ is the same as in~\eqref{typical_small_distortion_1}. Further, the mean-square difference between the MMSE estimate $\widehat{\mathbf{y}}^\text{mmse}_{\mathcal{S}_i,i}$ and the estimate $\widehat{\mathbf{s}}_i$ based on Gaussian random codes satisfies
\begin{equation}\label{mmse_close_333}
  \frac{1}{N}\mathbb{E}\left[\left\|\widehat{\mathbf{s}}_i-\widehat{\mathbf{y}}^\text{mmse}_{\mathcal{S}_i,i}\right\|_2^2\right]\le \Delta_N,
\end{equation}
where $\underset{N\to \infty }{\mathop{\lim }}\Delta_N=0$.
\end{lemma}
\begin{IEEEproof}
See Appendix~\ref{pf_distortion}.
\end{IEEEproof}

Since the distributed computation scheme using Gaussian random codes in Theorem~\ref{Gaussian_code_thm} (see Section \ref{Gaussian_code_alg}) satisfies the model in Section~\ref{model}, the distortion accumulation result in Theorem~\ref{Distortion_add_up} holds, i.e.,
\begin{equation}\label{333}
  \frac{1}{N}\mathbb{E}\left[ {{\left\| {\mathbf{y}}-{{\widehat{\mathbf{y}}}_{\mathcal{S}_0,0}^\text{mmse}} \right\|_2^2 }} \right]=\sum\limits_{i=1}^n{{D_i^\text{Inc}}},
\end{equation}
where $\mathbf{y}$ is the overall weighted sum, ${\widehat{\mathbf{y}}}_{\mathcal{S}_0,0}$ is the MMSE estimate of $\mathbf{y}$ at the sink $v_0$, and all expectation operations are taken over the random codebook ensemble. Using \eqref{mmse_close_333} in Lemma~\ref{distortion_close_lemma}, we have that
\begin{equation}\label{3332}
  \frac{1}{N}\mathbb{E}\left[\left\|\widehat{\mathbf{y}}-\widehat{\mathbf{y}}^\text{mmse}_{\mathcal{S}_0,0}\right\|_2^2\right]\le \Delta_N,
\end{equation}
where $\lim_{N\to \infty}\Delta_N=0$ and $\widehat{\mathbf{y}}$ is the estimate of the overall sum $\mathbf{y}$ at the sink using random Gaussian code. From Lemma~\ref{Pyth_lmm}, we have that
\begin{equation}
  \mathbb{E}\left[\left\|\widehat{\mathbf{y}}-\mathbf{y}\right\|_2^2\right]=\mathbb{E}\left[ {{\left\| {\mathbf{y}}-{{\widehat{\mathbf{y}}}_{\mathcal{S}_0,0}^\text{mmse}} \right\|_2^2 }} \right]+\mathbb{E}\left[\left\|\widehat{\mathbf{y}}-\widehat{\mathbf{y}}^\text{mmse}_{\mathcal{S}_0,0}\right\|_2^2\right].
\end{equation}
Plugging in~\eqref{333}, \eqref{3332} and using the triangle inequality, we get
\begin{equation}\label{444}
  D=\frac{1}{N}\mathbb{E}\left[\left\|\widehat{\mathbf{y}}-\mathbf{y}\right\|_2^2\right]\le\sum\limits_{i=1}^n{{D_i^\text{Inc}}}+ \Delta_N.
\end{equation}
Using \eqref{distortion_close} in Lemma~\ref{distortion_close_lemma}, we get
\begin{equation}
\begin{split}
  D\le& \sum\limits_{i=1}^n{{{\left( \sqrt{{{d}_i} +{{\varepsilon }_{N}}}+{{\eta }_{N}} \right)}^2 }}\\
  =&\sum\limits_{i=1}^n{{{d}_i}}+\sum\limits_{i=1}^n{{{\varepsilon }_{N}}+\eta _{N}^2 +2{{\eta }_{N}}\sqrt{{{d}_i}+{{\varepsilon }_{N}}}}.
\end{split}
\end{equation}
By defining $\epsilon_N=\sum\limits_{i=1}^n{{{\varepsilon }_{N}}+\eta _{N}^2 +2{{\eta }_{N}}\sqrt{{{d}_i} +{{\varepsilon }_{N}}}}$, we get
\begin{equation}\label{555}
  D\le \sum\limits_{i=1}^n{d_i }+{{\epsilon }_{N}},
\end{equation}
where $\lim_{N\to \infty}\epsilon_N=0$. Finally, noticing that \eqref{555} holds for the random code ensemble, we can find at least one code in the ensemble such that the distortion bound \eqref{Gaussian_code_distortion} holds.

Since we can tune the distortion parameter $d_i$ directly, we can set $d_1=d_2=\ldots=d_n=d$. Then, in the limit of large $N$, $D=nd$, which means that $d_1=d_2=\ldots=d_n=D/n$. Thus, we can obtain the minimized achievable result $R=\frac{1}{2}\,{\log_2 }\frac{\prod\limits_{i=1}^n{\sigma_{\mathcal{S}_i}^2}}{{{\left( {{D}}/n \right)}^n}}$, which is~\eqref{minimize_upper_bound} in Theorem~\ref{Gaussian_code_thm}.

\section{Extension to Network Consensus}\label{consensus_sec}
The results in the preceding sections can be extended to the case when each node in the network $\mathcal{T}$ wants to obtain an estimate of $\mathbf{y}=\sum\limits_{i=1}^n{w_i{\mathbf{x}_i}}$. Note that the network consensus problem considered in this paper is a generalization of average consensus, which is the case where $w_i=\frac{1}{n},\forall i$. The generalized definition in this paper is similar with the general form of distributed averaging in \cite{Dim_PI_10,Aya_TIT_10}.

Define ${{\mathcal{S}}_{i\to j}}\subset\mathcal{V}$ as the set that contains node $v_i$ and all its descendants when neighboring node $v_j$ is defined to be the parent-node of $v_i$. As in~\eqref{MMSE_i}-\eqref{new_distortion}, define
\begin{equation}\label{con_Dij}
  {D_{i\to j}^\text{Tx}}=\frac{1}{N}\mathbb{E}\left[ {{\left\| {\mathbf{y}_{{\mathcal{S}}_{i\to j}}}-{\widehat{\mathbf{y}}^\text{mmse}_{{\mathcal{S}}_{i\to j},i}} \right\|}_2^2 } \right],
\end{equation}
\begin{equation}\label{con_TDij}
  {D_{i\to j}^\text{Rx}}=\frac{1}{N}\mathbb{E}\left[ {{\left\| {\mathbf{y}_{{\mathcal{S}}_{i\to j}}}-{\widehat{\mathbf{y}}^\text{mmse}_{{\mathcal{S}}_{i\to j},j}} \right\|}_2^2 } \right],
\end{equation}
and
\begin{equation}\label{con_DDij}
  D_{i\to j}^\text{Inc}=\frac{1}{N}\mathbb{E}\left[ {{\left\| {\widehat{\mathbf{y}}^\text{mmse}_{{\mathcal{S}}_{i\to j},i}}-{\widehat{\mathbf{y}}^\text{mmse}_{{\mathcal{S}}_{i\to j},j}} \right\|}_2^2 } \right],
\end{equation}
where ${\widehat{\mathbf{y}}^\text{mmse}_{{\mathcal{S}}_{i\to j},i}}$ and ${\widehat{\mathbf{y}}^\text{mmse}_{{\mathcal{S}}_{i\to j},j}}$ are defined by~\eqref{MMSE_estimator}, i.e., the MMSE estimates of ${\mathbf{y}_{{\mathcal{S}}_{i\to j}}}$ with information at $v_i$ or at $v_j$. Since each node $v_i$ makes an estimate of the weighted sum $\mathbf{y}$, for a given distributed computation scheme, we define the overall distortion of the MMSE estimate $\widehat{\mathbf{y}}^\text{mmse}_i$ of $\mathbf{y}$ at the node $v_i$ as
\begin{equation}\label{D_i_mmse}
  D_i^{\text{mmse}}=\frac{1}{N}\mathbb{E}\left[\left\|\widehat{\mathbf{y}}^\text{mmse}_i-\mathbf{y}\right\|_2^2\right].
\end{equation}
For the same distributed computation scheme, define the overall distortion of the estimate $\widehat{\mathbf{y}}_i$ of $\mathbf{y}$ at the node $v_i$ as
\begin{equation}\label{D_i_total}
  D_i^{\text{Total}}=\frac{1}{N}\mathbb{E}\left[\left\|\widehat{\mathbf{y}}_i-\mathbf{y}\right\|_2^2\right].
\end{equation}
Then, we have that $D_i^{\text{Total}}\ge D_i^{\text{mmse}}$. For a feasible and oblivious distributed computation scheme $(T,\mathscr{S},\mathscr{G},\mathbf{v},\mathbf{e})\in\mathcal{F}$ (see the distributed computation model in Section~\ref{model}), the sum rate $R$ is defined in the same way as in the problem of linear function computation:
\begin{equation}\label{total_rate_consensus}
  R=\sum\limits_{i=1}^n{\sum\limits_{v_j\in \mathcal{N}(i)}{{R_{i\to j}}}}.\\
\end{equation}
The distortion is defined as the sum distortion
\begin{equation}\label{total_distortion_consensus}
  D=\sum\limits_{i=1}^n D_i^{\text{Total}}.\\
\end{equation}
Thus, the problem to be considered is
\begin{equation}\label{RD_op_problem_consensus}
\begin{split}
&\min_{(T,\mathscr{S},\mathscr{G},\mathbf{v},\mathbf{e})\in\mathcal{F}}{\;\;\;\;}R,\\
&{\;\;\;\;\;\;}\text{s.t.}{\;}D\le D^\text{tar}.
\end{split}
\end{equation}

We define ${{\overrightarrow{\mathcal{T}}}_{k}}$ as the edge set of the directed tree towards the root $v_k$. The set ${{\overrightarrow{\mathcal{T}}}_{k}}$ can be written as
\[\begin{split}
  &{{\overrightarrow{\mathcal{T}}}_{k}}=\{\text{directed edges } (v_i,v_j):{\;\;} (v_i,v_j)\in \mathcal{E},\text{ and }v_j\\
  &\text{ is the parent node of }v_i\text{ when } v_k\text{ is }\text{defined as the root}\}.
\end{split}\]
In all, we define $n$ different directed edge sets of directed trees towards $n$ different roots. These directed trees are all defined based on the original tree $\mathcal{T}$. The only difference is that the edges are directed. We use $(i,j)\in{{\overrightarrow{\mathcal{T}}}_{k}}$ to represent that the ordered pair $(i,j)$ is a directed edge in the directed edge set ${{\overrightarrow{\mathcal{T}}}_{k}}$.

\begin{theorem}[Distortion Accumulation for Network Consensus]\label{Distortion_consensus_add_up}
For the network consensus problem, the overall distortion of estimating $Y$ at the node $v_k$ satisfies
\begin{equation}\label{Total_Distortion_consensus}
 {D_k^{\text{mmse}}}=\sum\limits_{(i,j)\in {{\overrightarrow{\mathcal{T}}}_{k}}}{D_{i\to j}^\text{Inc}},
\end{equation}
where ${{\overrightarrow{\mathcal{T}}}_{k}}$ is the directed edge set of the directed tree towards the root $v_k$, and $D_{i\to j}^\text{Inc}$ is as defined in~\eqref{con_DDij}.
\end{theorem}
\begin{IEEEproof}
See Appendix~\ref{consensus_add_up_proof}.
\end{IEEEproof}

\subsection{Inner and Outer Bounds Based on Incremental-Distortion}
Recall that $\sigma _\mathcal{S}^2 $ is the variance of $Y_\mathcal{S}$. The counterpart of Theorem~\ref{main_thm} is stated as follows.
\begin{theorem}[Incremental-Distortion-Based Outer Bound for Network Consensus]\label{main_thm_2}
For the network consensus problem, given a feasible distributed computation scheme, the sum rate is lower-bounded by
\begin{equation}\label{consensus_lower_bound}
\begin{split}
  & R=\sum\limits_{i=1}^n{\sum\limits_{v_j\in \mathcal{N}(i)}{{R_{i\to j}}}}\\
  &\ge \frac{1}{2}\sum\limits_{i=1}^n{\sum\limits_{v_j\in \mathcal{N}(i)}{\left[ {\log_2 }\frac{\sigma _{{{\mathcal{S}}_{i\to j}}}^2 }{D_{i\to j}^\text{Inc}}
  -\frac{{D_{i\to j}^\text{Tx}}}{2w_i^2 }
  -\frac{{\log_2 }e}{2\sigma _{\mathcal{S}_{i\to j}}^2 }\sqrt{2{D_{i\to j}^\text{Tx}}\left( 4\sigma _{\mathcal{S}_{i\to j}}^2 +{D_{i\to j}^\text{Tx}} \right)}
   \right]}} \\
 & =\frac{1}{2}\sum\limits_{i=1}^n{\sum\limits_{v_j\in \mathcal{N}(i)}{\left[ {\log_2 }\frac{\sigma _{{{\mathcal{S}}_{i\to j}}}^2 }{{D_{i\to j}^\text{Rx}}-{D_{i\to j}^\text{Tx}}}-\mathcal{O}\left((D_{i\to j}^\text{Tx})^{1/2}\right) \right]}},
\end{split}
\end{equation}
where $\sigma^2_{\mathcal{S}_{i\to j}}$ is the variance of each entry of the partial sum $\mathbf{y}_{\mathcal{S}_{i\to j}}=\sum\limits_{v_k\in \mathcal{S}_{i\to j}}{{{w}_{k}}{\mathbf{x}_k}}$, and $D_{i\to j}^\text{Tx}$, $D_{i\to j}^\text{Rx}$ and $D_{i\to j}^\text{Inc}$ are respectively defined in~\eqref{con_Dij},~\eqref{con_TDij} and~\eqref{con_DDij}. By optimizing over the incremental distortions $D_{i\to j}^\text{Inc}$, one obtains the following scheme-independent bound stated in an optimization form
\begin{equation}
\begin{split}
&\mathop{\min}\limits_{D_{i\to j}^\text{Inc},\forall (i,j)\in \mathcal{E}} \frac{1}{2}\sum\limits_{i=1}^n \sum\limits_{v_j\in \mathcal{N}(i)}\left[
  {\log_2 }\frac{\sigma _{{\mathcal{S}_{i\to j}}}^2 }{{D_{i\to j}^\text{Inc}}}
  -\frac{D_{i\to j}^\text{Tx}}{2w_i^2 }
  -\frac{{\log_2 }e}{2\sigma _{\mathcal{S}_{i\to j}}^2 }\sqrt{2D_{i\to j}^\text{Tx}\left( 4\sigma_{\mathcal{S}_{i\to j}}^2 +D_{i\to j}^\text{Tx}\right)} \right],\\
  &{\;\;\;\;\;\;\;\;\;\;\;\;\;\;\;\;\;\;\;}\text{s.t.} \left\{ \begin{matrix}
   D_{i\to j}^\text{Tx}=\sum\limits_{v_k\in \mathcal{S}_{i\to j}\setminus\{v_i\},(k,l)\in \vec{\mathcal{T}}_j}{{D_{k\to l}^\text{Inc}}},\forall (i,j)\in\mathcal{E},  \\
   D\ge {D_k^{\text{mmse}}}=\sum\limits_{(i,j)\in {{\overrightarrow{\mathcal{T}}}_{k}}}{D_{i\to j}^\text{Inc}}.
\end{matrix} \right.
\end{split}
\end{equation}
\end{theorem}
\begin{IEEEproof}
See Appendix~\ref{consensus_lower_bound_proof}.
\end{IEEEproof}
Then, we present an achievable result using Gaussian codes to show that the outer bound in Theorem~\ref{main_thm_2} is tight in the low distortion regime.

\begin{theorem}[Inner Bound for Network Consensus]\label{Gaussian_code_thm_2}
Using Gaussian random codebooks, we can find a distributed computation scheme, such that the sum rate $R$ satisfies
\begin{equation}\label{consensus_upper_bound}
 R\le \frac{1}{2}\sum\limits_{i=1}^n{\sum\limits_{v_j\in \mathcal{N}(i)}{{\log_2 }\frac{\sigma_{\mathcal{S}_{i\to j}}^2}{d_{i\to j}}}}+(2n-2)\delta_N,
\end{equation}
where $\lim_{N\to\infty}\delta_N=0$, and the $d_{i\to j}$'s are distortion parameters. Further, the overall distortion $D$ in all nodes $v_i$ satisfies
\begin{equation}\label{distortion_upper_bound_consensus}
  D<\sum_{k=1}^n\sum\limits_{(i,j)\in {{\overrightarrow{\mathcal{T}}}_{k}}}{d_{i\to j}}+n\epsilon_N,
\end{equation}
where $\lim_{N\to \infty}\epsilon_N=0$.
\end{theorem}
\begin{IEEEproof}
See Appendix~\ref{Gaussian_thm2_proof}.
\end{IEEEproof}
If we ignore the small gap between the inner bound~\eqref{consensus_upper_bound} and the outer bound~\eqref{consensus_lower_bound} when the resolution level $D$ is fine enough, the optimal rate can be obtained by solving the following convex optimization problem:
\begin{equation}\label{convex_optimization}
\begin{split}
  &\underset{D_{i\to j}^\text{Inc},\forall (i,j)\in \mathcal{E}}{\mathop{\min }} \frac{1}{2}\sum\limits_{i=1}^n{\sum\limits_{v_j\in \mathcal{N}(i)}{{\log_2 }\frac{\sigma _{{{\mathcal{S}}_{i\to j}}}^2 }{D_{i\to j}^\text{Inc}}}},\\
  &s.t.{\;\;\;\;}\sum_{k=1}^n\sum\limits_{(i,j)\in {{\overrightarrow{\mathcal{T}}}_{k}}}{D_{i\to j}^\text{Inc}}\le D.
\end{split}
\end{equation}
\begin{remark}
The rate distortion outer bound in~\eqref{consensus_lower_bound} depends on the distributed computation scheme. Using convex optimization techniques, we can minimize over all incremental distortions $D_{i\to j}^\text{Inc}$ with the linear constrains specified by \eqref{Total_Distortion_consensus} to obtain a fundamental outer bound on the rate distortion function of distributed consensus. The outer bound is essentially obtained by rate allocation in the network. If the $\mathcal{O}(D^{1/2})$ gap between the inner and outer bound is neglected, the rate (measured in number of bits) allocated to the link $v_i\to v_j$ is $\frac{1}{2}{{\log_2 }\frac{\sigma _{{{\mathcal{S}}_{i\to j}}}^2 }{D_{i\to j}^\text{Inc}}}$.

We consider a special case when $w_i=\frac{1}{n},\forall i$. This is the classical case of lossy distributed network consensus with the same distortion requirement at all nodes~\cite{Su_TIT_10}. We again consider the line network as shown in Section~\ref{comparison_section}. In this case, it can be shown that the optimal solution is
\[D_{i\to j}^\text{Inc}=\frac{D}{2(n-1)},\forall (i,j)\in \mathcal{E},\]
if all $\mathcal{O}\left((D_{i\to j}^\text{Tx})^{1/2}\right)$ terms are neglected, in the limit of zero-distortion (high resolution)\footnote{We can neglect the $\mathcal{O}\left((D_{i\to j}^\text{Tx})^{1/2}\right)$ terms, because in the zero-distortion limit, $\log \frac{1}{D_{i\to j}^\text{Inc}}> \log \frac{1}{D_{i\to j}^\text{Tx}}>> (D_{i\to j}^\text{Tx})^{1/2}$.}. Similar with the data-aggregation case, this solution for network consensus is also very similar to the reverse water-filling solution for parallel Gaussian lossy source coding problem~\cite[Theorem 10.3.3]{Cover_Wiley_06} in the limit of large rate (zero distortion). This solution yields a sum rate of $\mathcal{O}\left( n{\log_2 }\frac{1}{D} \right)$. The classical outer bound~\cite[Prop. 4]{Su_TIT_10} about the distributed network consensus in a tree network is $\mathcal{O}\left( n{\log_2 }\frac{1}{{{n}^{3/2}}D} \right)$. This means that our result is certainly tighter than the classical result in a line network in the zero-distortion limit. Moreover, this $\mathcal{O}(n\log n)$ gap is also consistent with the $\log (n!)$ gap in Section~\ref{comparison_section}.
\end{remark}

\section{Conclusions}\label{conclude_sec}
In this paper, we have considered the lossy linear function computation problem in a Gaussian tree network. Our results show that the phenomenon of information dissipation exists in this problem, and by quantifying the information dissipation, we obtain an information-theoretic outer bound on the rate distortion function that is tighter than classical cut-set bounds for lossy linear function computing for both data aggregation and network consensus problems. The results also show that linear Gaussian codes can achieve within $\mathcal{O}(\sqrt D)$ of the obtained outer bound, which means that our outer bound is tight when the required distortion is small (high resolution scenario). A meaningful future work is to investigate tighter outer bound for all values of $D$, and investigate compression algorithms, e.g., lattice codes~\cite{Nazer_TIT_07,Kri_TIT_09}, that achieve the outer bound for all values of $D$. Another research topic of interest is the study of deterministic abstractions that account for the distortion accumulation effect. Since our work focuses on a special case of noiseless networks, it may prove useful in initiating this direction of research. \textcolor{black}{It is also interesting to investigate the generalization of the distortion accumulation effect and the inequalities developed in this paper to other computation and inference problems in networks, especially in networks with cycles and in the case when data is not stored at all nodes \cite{Dim_PI_10,San_SPM_14}. One can obtain loose upper bounds for simple non-tree networks. For instance, for an achievable distortion bound in non-tree networks, a simple extension could be to the case of a directed acyclic network with only one source node with message $\mathbf{x}$, and only two paths to the sink node. In this case, if the mean-square error on one path is $D_1$ and the mean-square error on the other path is $D_2$, an achievable (if suboptimal) variance of estimating the source message using these two messages is  $\min\{D_1,D_2\}$. This is achieved by either choosing the first message or the second message, and the equality is achieved when the two messages are the same. Therefore, one can obtain (loose) upper bounds on the accumulation of distortion using our achievability results. However, because of obvious looseness in the bound, we may not achieve an asymptotically tight result, as we obtained in Theorem \ref{Distortion_add_up}. Another interesting direction is the possible extension of distortion accumulation to non-Gaussian sources using the Wasserstein distance as a distance metric \cite{7467523}, although we suspect that a simple form of distortion accumulation may not be easily obtained.}

\appendices
\section{Proof of Lemma~\ref{Transportation_Inequality_lmm}}\label{PofLm2}
Using the chain rule for divergence \cite[Theorem 2.5.3]{Cover_Wiley_06}, we can expand the divergence $D({P_{\mathbf{x},\mathbf{y},\mathbf{x}+\sqrt{t}\mathbf{z}}}\parallel {P_{\mathbf{x},\mathbf{y},\mathbf{y}+\sqrt{t}\mathbf{z}}})$ in two ways:
\begin{equation}\label{expand_der1}
\begin{split}
  &D({P_{\mathbf{x},\mathbf{y},\mathbf{x}+\sqrt{t}\mathbf{z}}}\parallel {P_{\mathbf{x},\mathbf{y},\mathbf{y}+\sqrt{t}\mathbf{z}}})\\
  =&D({P_{\mathbf{x}+\sqrt{t}\mathbf{z}}}\parallel {P_{\mathbf{y}+\sqrt{t}\mathbf{z}}})+D({P_{\mathbf{x},\mathbf{y}|\mathbf{x}+\sqrt{t}\mathbf{z}}}\parallel {P_{\mathbf{x},\mathbf{y}|\mathbf{y}+\sqrt{t}\mathbf{z}}})\\
  \ge& D({P_{\mathbf{x}+\sqrt{t}\mathbf{z}}}\parallel {P_{\mathbf{y}+\sqrt{t}\mathbf{z}}}),
\end{split}
\end{equation}
{and
\begin{equation}\label{expand_der2}
\begin{split}
  &D({P_{\mathbf{x},\mathbf{y},\mathbf{x}+\sqrt{t}\mathbf{z}}}\parallel {P_{\mathbf{x},\mathbf{y},\mathbf{y}+\sqrt{t}\mathbf{z}}})\\
  =&D({P_{\mathbf{x},\mathbf{y}}}\parallel {P_{\mathbf{x},\mathbf{y}}})+D({P_{\mathbf{x}+\sqrt{t}\mathbf{z}|\mathbf{x},\mathbf{y}}}\parallel {P_{\mathbf{y}+\sqrt{t}\mathbf{z}|\mathbf{x},\mathbf{y}}})\\
  =&D({P_{\mathbf{x}+\sqrt{t}\mathbf{z}|\mathbf{x},\mathbf{y}}}\parallel {P_{\mathbf{y}+\sqrt{t}\mathbf{z}|\mathbf{x},\mathbf{y}}})\\
  =&\int_{d{{x}^N}}P({{x}^N})d{{x}^N}
  \int_{d{{y}^N}}P({{y}^N}|{x}^N)\times\\
  &{\;\;\;\;\;\;\;\;\;\;\;\;\;\;\;\;\;\;\;\;}D\left( P\left( {{x}^N}+\sqrt{t}{{z}^N}|{{x}^N},{{y}^N} \right)\left\| P\left( {{y}^N}+\sqrt{t}{{z}^N}|{{x}^N},{{y}^N} \right) \right. \right)d{{y}^N}\\
  =&\mathbb{E}\left[ D\left( P\left( {\mathbf{x}}+\sqrt{t}{{z}^N}|{\mathbf{x}},{\mathbf{y}} \right)\right.\right.\left.\left.\left\| P\left( {\mathbf{y}}+\sqrt{t}{{z}^N}|{\mathbf{x}},{\mathbf{y}} \right) \right. \right) \right]\\
  =&\mathbb{E}\left[ D\left( \left. \mathcal{N}({\mathbf{x}},t{{\mathbf{I}}_{N}}) \right\|\mathcal{N}({\mathbf{y}},t{{\mathbf{I}}_{N}}) \right) \right]\\
  \overset{(a)}{=}&\mathbb{E}\left[ \mathbb{E}\left[ \left. \frac{1}{2t}{{\left( {\mathbf{x}}-{\mathbf{y}} \right)}^{\top }}\left( {\mathbf{x}}-{\mathbf{y}} \right) \right|{\mathbf{x}},{\mathbf{y}} \right] \right]\\
  =&\frac{1}{2t}\mathbb{E}\left[ \left\| {\mathbf{x}}-{\mathbf{y}} \right\|_{2}^2  \right],
\end{split}
\end{equation}}
where (a) follows from a known result (see, e.g., \cite[Pg. 13]{Duc_Ber_07}) that the KL-divergence between two $N$-dimensional multivariate normal distributions $\mathcal{N}(\boldsymbol{\mu}_0,\mathbf{\Sigma}_0)$ and $\mathcal{N}({{\boldsymbol{\mu }}_{1}},{{\mathbf{\Sigma }}_{1}})$ is
\begin{equation}\label{Gaussian_div}
\begin{split}
  &D\left( \left. \mathcal{N}({{\boldsymbol{\mu}}_0},{{\mathbf{\Sigma }}_0}) \right\|\mathcal{N}({{\boldsymbol{\mu}}_{1}},{{\mathbf{\Sigma }}_{1}}) \right)\\
  =&\frac{1}{2}\left(\text{tr}\left( \mathbf{\Sigma }_{1}^{-1}{{\mathbf{\Sigma }}_0} \right)-N+{{\left( {{\boldsymbol{\mu}}_{1}}-{{\boldsymbol{\mu}}_0} \right)}^{\top }}\mathbf{\Sigma }_{1}^{-1}\left( {{\boldsymbol{\mu}}_{1}}-{{\boldsymbol{\mu}}_0} \right)+\ln \left( \frac{\det {{\mathbf{\Sigma }}_{1}}}{\det {{\mathbf{\Sigma }}_0}} \right) \right).
\end{split}
\end{equation}
Combining~\eqref{expand_der1} and~\eqref{expand_der2}, we obtain Lemma~\ref{Transportation_Inequality_lmm}.

\section{Proofs for Section~\ref{main_sec}}
\subsection{Proof of Theorem~\ref{Distortion_add_up}}\label{add_distortion_app}
We first examine the change of distortion on an arbitrary link $v_b\to v_a$ as shown in Fig.~\ref{cut_set_figure}. Then, we prove this theorem by summing up all distortion on all links. By definition, we have
\begin{equation}\label{MMSE_b}
  {\widehat{\mathbf{y}}^\text{mmse}_{\mathcal{S},b}}=\mathbb{E}\left[ {\mathbf{y}_\mathcal{S}}|I_b \right],
\end{equation}
where ${I}_b$ denotes all available information at the node $v_b$. Similarly, we have
\begin{equation}\label{MMSE_a}
  {\widehat{\mathbf{y}}^\text{mmse}_{\mathcal{S},a}}=\mathbb{E}\left[ {\mathbf{y}_\mathcal{S}}|{{I}_a} \right].
\end{equation}
However, since the only information available at $v_a$ to estimate $\mathbf{y}_\mathcal{S}$ is $M_{b\to a}$, because the data $\mathbf{x}_i$'s are uncorrelated, we have that
\begin{equation}\label{MMSE_a_2}
  {\widehat{\mathbf{y}}^\text{mmse}_{\mathcal{S},a}}=\mathbb{E}\left[ {\mathbf{y}_\mathcal{S}}|M_{b\to a} \right].
\end{equation}

It is certain that $M_{b\to a}$, the message bits transmitted from node $v_b$ to node $v_a$, must be a function of all the available information in $v_b$. This means that $\sigma \left( M_{b\to a} \right)\subset \sigma \left( I_b \right)$, where $\sigma(\cdot)$ denotes the $\sigma$-algebra generated by the argument and $\sigma \left( I_b \right)$ denotes all the available information including the observations of all random variables at node $v_b$. Since ${\widehat{\mathbf{y}}^\text{mmse}_{\mathcal{S},b}}$, the conditional expectation estimate of $\mathbf{y}_\mathcal{S}$ given all the available information in $v_b$, is the projection of ${\mathbf{y}_\mathcal{S}}$ onto $\sigma \left( I_b \right)$ and $\widehat{\mathbf{y}}^\text{mmse}_{\mathcal{S},a}$ is the projection of ${\mathbf{y}_\mathcal{S}}$ onto $\sigma \left( M_{b\to a} \right)\subset \sigma \left( I_b \right)$, we have that ${\widehat{\mathbf{y}}^\text{mmse}_{\mathcal{S},b}}-\widehat{\mathbf{y}}^\text{mmse}_{\mathcal{S},a}$ is $\sigma \left( I_b \right)$-measurable. Therefore, using the orthogonality principle (Lemma~1), we can show that
\begin{equation}\label{orthogonal}
  \left({\widehat{\mathbf{y}}^\text{mmse}_{\mathcal{S},b}}-\widehat{\mathbf{y}}^\text{mmse}_{\mathcal{S},a}\right)\bot \left({\widehat{\mathbf{y}}^\text{mmse}_{\mathcal{S},b}}-{\mathbf{y}_\mathcal{S}}\right),
\end{equation}
where the LHS is $\sigma \left( I_b \right)$-measurable, and the RHS is the projection error of the conditional expectation estimate ${\widehat{\mathbf{y}}^\text{mmse}_{\mathcal{S},b}}$ (Lemma 1 basically says that the projection error $\mathbb{E}[X|\mathcal{G}]-X$ between the original vector $X$ and the projection (conditional expectation) $\mathbb{E}[X|\mathcal{G}]$ is uncorrelated of the sigma-algebra $\mathcal{G}$, i.e., all $\mathcal{G}$-measurable random variables). Therefore, using Pythagoras theorem and the observation that $\mathbb{E}[{\widehat{\mathbf{y}}^\text{mmse}_{\mathcal{S},b}}]=\mathbb{E}[\widehat{\mathbf{y}}^\text{mmse}_{\mathcal{S},a}]=\mathbb{E}[{\mathbf{y}_\mathcal{S}}]=\mathbf{0}_N$, we get
\begin{equation}\label{Pyth}
  {D_b^\text{Rx}}=D_b^\text{Tx}+{D_b^\text{Inc}},
\end{equation}
where, recall that $D_b^\text{Tx}=\frac{1}{N}\mathbb{E}\left[ {{\left\| {\mathbf{y}_\mathcal{S}}-{\widehat{\mathbf{y}}^\text{mmse}_{\mathcal{S},b}} \right\|}_2^2 } \right]$, ${D_b^\text{Rx}}=\frac{1}{N}\mathbb{E}\left[ {{\left\| {\mathbf{y}_\mathcal{S}}-{\widehat{\mathbf{y}}^\text{mmse}_{\mathcal{S},a}} \right\|}_2^2 } \right]$, and ${D_b^\text{Inc}}=\frac{1}{N}\mathbb{E}\left[ {{\left\| {\widehat{\mathbf{y}}^\text{mmse}_{\mathcal{S},b}}-{\widehat{\mathbf{y}}^\text{mmse}_{\mathcal{S},a}} \right\|}_2^2 } \right]$. Since the link $v_b\to v_a$ is arbitrarily chosen, equation~\eqref{Pyth} can be generalized to all nodes, and hence~\eqref{Pyth_all} is proved.

Now, we show that the distortion $D_b^\text{Tx}$ can be written as the sum of the distortions from the children of $v_b$. Without loss of generality, suppose the node $v_b$ has $d$ children $v_1,v_2,\ldots v_d$, as shown in Fig.~\ref{cut_set_figure}. By definition, we have
\begin{equation}\label{Sum_decompose}
  {\mathbf{y}_\mathcal{S}}=\sum\limits_{k=1}^{d}{\mathbf{y}_{\mathcal{S}_k}}+w_b{\mathbf{x}_b}.
\end{equation}
By the definition of MMSE estimator, we have that
\begin{equation}\label{MMSE_decompose}
\begin{split}
  {\widehat{\mathbf{y}}^\text{mmse}_{\mathcal{S},b}}&=\mathbb{E}\left[ \mathbf{y}_\mathcal{S}|I_b \right]
  =\mathbb{E}\left[ \sum\limits_{k=1}^{d}{\mathbf{y}_{\mathcal{S}_k}}+w_b{\mathbf{x}_b}|I_b \right]
  =\sum\limits_{k=1}^{d}{\widehat{\mathbf{y}}^\text{mmse}_{\mathcal{S}_k,b}}+w_b{\mathbf{x}_b}.
  \end{split}
\end{equation}
Therefore, we have
\begin{equation}\label{MMSE_decompose_2}
\begin{split}
  D_b^\text{Tx}=&\frac{1}{N}\mathbb{E}\left[ {{\left\| {\mathbf{y}_\mathcal{S}}-{\widehat{\mathbf{y}}^\text{mmse}_{\mathcal{S},b}} \right\|}_2^2 } \right]
  \overset{(a)}{=}\frac{1}{N}\sum\limits_{k=1}^{d}{\mathbb{E}\left[ {{\left\| {\mathbf{y}_{\mathcal{S}_k}}-{\widehat{\mathbf{y}}^\text{mmse}_{\mathcal{S}_k,b}} \right\|}_2^2 } \right]}=\sum\limits_{k=1}^{d} D_k^\text{Rx},
\end{split}
\end{equation}
where (a) holds because different estimates $\widehat{\mathbf{y}}^\text{mmse}_{\mathcal{S}_k,b}$ on different links $v_k\to v_b$ are independent of each other.

Combining~\eqref{MMSE_decompose_2} with~\eqref{Pyth}, we have that
\begin{equation}\label{MMSE_decompose_3}
\begin{split}
D_b^\text{Tx}
%&=\sum\limits_{k=1}^{d}\left\{ \mathbb{E}\left[ {{\left\| \mathbf{y}_{\mathcal{S}_k}-{\widehat{\mathbf{y}}^\text{mmse}_{\mathcal{S}_k,k}} \right\|}_2^2 } \right]+\mathbb{E}\left[ {{\left\| {\widehat{\mathbf{y}}^\text{mmse}_{\mathcal{S}_k,b}}-{\widehat{\mathbf{y}}^\text{mmse}_{\mathcal{S}_k,k}} \right\|}_2^2 } \right] \right\}\\
=\sum\limits_{k=1}^{d}{\left( {D_k^\text{Tx}}+{D_k^\text{Inc}} \right)}.
\end{split}
\end{equation}
\textcolor{black}{Using~\eqref{MMSE_decompose_3}, we can prove \eqref{Tr_Distortion} using induction in the tree (see Remark~\ref{induction_on_tree}). Equation \eqref{Total_Distortion} is obtained by carrying out the induction in the tree until the sink node $v_0$.}
\subsection{Proof of Theorem~\ref{main_thm}}\label{lower_bound_proof}
We still consider the specific set $\mathcal{S}$ as shown in Fig.~\ref{cut_set_figure}. On the link $v_b\to v_a$, we have that
\begin{equation}\label{cut_set_bound_ineq}
\begin{split}
  & N{R_{b\to a}}\overset{(a)}{\mathop{\ge }}\,H(M_{b\to a})\\
  & \ge I(M_{b\to a};{\widehat{\mathbf{y}}^\text{mmse}_{\mathcal{S},b}})\\
  & \overset{(b)}{=}\,I(M_{b\to a},{\widehat{\mathbf{y}}^\text{mmse}_{\mathcal{S},a}};{\widehat{\mathbf{y}}^\text{mmse}_{\mathcal{S},b}})\\
  & \overset{(c)}{=}\,I({\widehat{\mathbf{y}}^\text{mmse}_{\mathcal{S},a}};{\widehat{\mathbf{y}}^\text{mmse}_{\mathcal{S},b}})+I(M_{b\to a};{\widehat{\mathbf{y}}^\text{mmse}_{\mathcal{S},b}}|{\widehat{\mathbf{y}}^\text{mmse}_{\mathcal{S},a}}) \\
 & \ge I({\widehat{\mathbf{y}}^\text{mmse}_{\mathcal{S},a}};{\widehat{\mathbf{y}}^\text{mmse}_{\mathcal{S},b}})\\
 &=h({\widehat{\mathbf{y}}^\text{mmse}_{\mathcal{S},b}})-h({\widehat{\mathbf{y}}^\text{mmse}_{\mathcal{S},b}}|{\widehat{\mathbf{y}}^\text{mmse}_{\mathcal{S},a}})\\
 & =h({\widehat{\mathbf{y}}^\text{mmse}_{\mathcal{S},b}})-h({\widehat{\mathbf{y}}^\text{mmse}_{\mathcal{S},b}}-{\widehat{\mathbf{y}}^\text{mmse}_{\mathcal{S},a}}|{\widehat{\mathbf{y}}^\text{mmse}_{\mathcal{S},a}}) \\
 & \ge h({\widehat{\mathbf{y}}^\text{mmse}_{\mathcal{S},b}})-h({\widehat{\mathbf{y}}^\text{mmse}_{\mathcal{S},b}}-{\widehat{\mathbf{y}}^\text{mmse}_{\mathcal{S},a}})\\
 &\overset{(d)}{\mathop{\ge }}\,h({\widehat{\mathbf{y}}^\text{mmse}_{\mathcal{S},b}})-\frac{N}{2}{\log_2 }2\pi e{D_b^\text{Inc}},
\end{split}
\end{equation}
where
\begin{description}
  \item[$(a)$] holds because $M_{b\to a}$ is a binary information sequence;
  \item[$(b)$] holds because $\widehat{\mathbf{y}}^\text{mmse}_{\mathcal{S},a}$ is a function of $M_{b\to a}$;
  \item[$(c)$] follows from the chain rule for mutual information;
  \item[$(d)$] holds because the entropy-maximizing distribution under variance constraint is Gaussian.
\end{description}

Now we only need to lower-bound $h({\widehat{\mathbf{y}}^\text{mmse}_{\mathcal{S},b}})$. We know that
\begin{align}
  &{\widehat{\mathbf{y}}^\text{mmse}_{\mathcal{S},b}}={\widehat{\mathbf{y}}^\text{mmse}_{\mathcal{S}\setminus\{b\},b}}+w_b{\mathbf{x}_b}\label{HWI_hatY}\\
  &{\mathbf{y}_\mathcal{S}}={\mathbf{y}_{\mathcal{S}\setminus\{b\}}}+w_b{\mathbf{x}_b}\label{HWI_Y}.
\end{align}
Suppose ${\widehat{\mathbf{y}}^\text{mmse}_{\mathcal{S},b}}\sim r\left( x^N \right)$ and ${\mathbf{y}_\mathcal{S}}\sim s\left( x^N \right)$. {Observe that~\eqref{HWI_hatY} and~\eqref{HWI_Y} are in the form of the random variables in Lemma~\ref{Transportation_Inequality_lmm} with $t=w_b^2$ and $\mathbf{z}=\mathbf{x}_b\sim \mathcal{N}(\mathbf{0}_N,\mathbf{I}_N)$.} Then, using Lemma~\ref{Transportation_Inequality_lmm}, we have that
\begin{equation}\label{KL_bound}
  \begin{split}
D\left( r||s \right)\le& \frac{1}{2w_b^2 }\mathbb{E}\left[ {{\left\| {\mathbf{y}_{\mathcal{S}\setminus\{b\}}}-{\widehat{\mathbf{y}}^\text{mmse}_{\mathcal{S}\setminus\{b\},b}} \right\|}_2^2 } \right]=\frac{1}{2w_b^2 }\mathbb{E}\left[ {{\left\| {\mathbf{y}_\mathcal{S}}-{\widehat{\mathbf{y}}^\text{mmse}_{\mathcal{S},b}} \right\|}_2^2 } \right]=\frac{{ND_b^\text{Tx}}}{2w_b^2 }.
\end{split}
\end{equation}

By definition, we have that
\begin{equation}\label{q_form}
  {\mathbf{y}_\mathcal{S}}\sim s\left( x^N \right)=\frac{1}{\left(\sqrt{2\pi} {{\sigma }_\mathcal{S}}\right)^N}\exp \left( -\frac{{\left\|x^N\right\|_2^2 }}{2\sigma _\mathcal{S}^2 } \right).
\end{equation}
Therefore,
\begin{equation}\label{q_entropy}
  h({\mathbf{y}_\mathcal{S}})=\frac{N}{2}{\log_2 }2\pi e\sigma _\mathcal{S}^2 .
\end{equation}
The difference between $h({\widehat{\mathbf{y}}^\text{mmse}_{\mathcal{S},b}})$ and $h({\mathbf{y}_\mathcal{S}})$ is
\begin{equation}\label{entropy_to_MMSE}
\begin{split}
   h({\widehat{\mathbf{y}}^\text{mmse}_{\mathcal{S},b}})-h({\mathbf{y}_\mathcal{S}})
  &=-\int_{x^N\in R^N}{r\log rdx^N}+\int_{x^N\in R^N}{s\log sdx^N} \\
 & =-\int_{x^N\in R^N}{r\log \frac{r}{s}dx^N}+\int_{x^N\in R^N}{(s-r)\log sdx^N} \\
 & \overset{(a)}{=}-D\left( r||s \right)+{\log_2 }e\int_{x^N\in R^N}{(s-r)\left( -\frac{{\left\|{x}^N\right\|_2^2 }}{2\sigma _\mathcal{S}^2 } \right)dx^N} \\
 & =-D\left( r||s \right)+\frac{{\log_2 }e}{2\sigma _\mathcal{S}^2 }\mathbb{E}\left[ \left\|\widehat{\mathbf{y}}^\text{mmse}_{\mathcal{S},b}\right\|_2^2 -\left\|\mathbf{y}_\mathcal{S}\right\|_2^2  \right],
\end{split}
\end{equation}
{where we used~\eqref{q_form} in step (a)}. The second term of the RHS can be bounded by
\begin{equation}\label{der11}
\begin{split}
  \left| \mathbb{E}\left[ \left\|\widehat{\mathbf{y}}^\text{mmse}_{\mathcal{S},b}\right\|_2^2 -\left\|\mathbf{y}_\mathcal{S}\right\|_2^2  \right] \right|
 &=\left| \mathbb{E}\left[ \left( {\widehat{\mathbf{y}}^\text{mmse}_{\mathcal{S},b}}-{\mathbf{y}_\mathcal{S}} \right)^\top\left( {\widehat{\mathbf{y}}^\text{mmse}_{\mathcal{S},b}}+{\mathbf{y}_\mathcal{S}} \right) \right] \right|\\
 &\le \sqrt{\mathbb{E}\left[ {{\left\| {\widehat{\mathbf{y}}^\text{mmse}_{\mathcal{S},b}}-{\mathbf{y}_\mathcal{S}} \right\|}_2^2 } \right]\mathbb{E}\left[ {{\left\| {\widehat{\mathbf{y}}^\text{mmse}_{\mathcal{S},b}}+{\mathbf{y}_\mathcal{S}} \right\|}_2^2 } \right]} \\
 & =\sqrt{{ND_b^\text{Tx}}\mathbb{E}\left[ {{\left\| {\widehat{\mathbf{y}}^\text{mmse}_{\mathcal{S},b}}-{\mathbf{y}_\mathcal{S}}+2{\mathbf{y}_\mathcal{S}} \right\|}_2^2 } \right]}\\
 &\le \sqrt{ND_b^\text{Tx}\cdot 2\left\{ \mathbb{E}\left[ 4\left\|\mathbf{y}_\mathcal{S}\right\|_2^2  \right]+\mathbb{E}\left[ {{\left\| {\widehat{\mathbf{y}}^\text{mmse}_{\mathcal{S},b}}-{\mathbf{y}_\mathcal{S}} \right\|}_2^2 } \right] \right\}}\\
 &=\sqrt{2ND_b^\text{Tx}\left( 4N\sigma _\mathcal{S}^2 +ND_b^\text{Tx} \right)}. \\
\end{split}
\end{equation}
Therefore, {combining~\eqref{KL_bound} and \eqref{q_entropy}-\eqref{der11}}, we get
\begin{equation}\label{diff_in_entro}
\begin{split}
&h({\widehat{\mathbf{y}}^\text{mmse}_{\mathcal{S},b}})\ge h({\mathbf{y}_\mathcal{S}})-\frac{ND_b^\text{Tx}}{2w_b^2 }-\frac{N{\log_2 }e}{2\sigma _\mathcal{S}^2 }\sqrt{2D_b^\text{Tx}\left( 4\sigma _\mathcal{S}^2 +D_b^\text{Tx} \right)}\\
&=\frac{N}{2}{\log_2 }2\pi e\sigma _\mathcal{S}^2 -\frac{ND_b^\text{Tx}}{2w_b^2 }-\frac{N{\log_2 }e}{2\sigma _\mathcal{S}^2 }\sqrt{2D_b^\text{Tx}\left( 4\sigma _\mathcal{S}^2 +D_b^\text{Tx} \right)}.
\end{split}
\end{equation}
Plugging the above inequality into~\eqref{cut_set_bound_ineq}, we get
\begin{equation}\label{good_bound}
\begin{split}
  {R_{b\to a}}\ge& \frac{1}{2}{\log_2 }\frac{\sigma _\mathcal{S}^2 }{{D_b^\text{Inc}}} -\frac{D_b^\text{Tx}}{2w_b^2 }
  -\frac{{\log_2 }e}{2\sigma _\mathcal{S}^2 }\sqrt{2D_b^\text{Tx}\left( 4\sigma_\mathcal{S}^2 +D_b^\text{Tx}\right)}\\
   =&\frac{1}{2}{\log_2 }\frac{\sigma_\mathcal{S}^2 }{{D_b^\text{Inc}}}-\mathcal{O}\left((D_b^\text{Tx})^{1/2}\right),
\end{split}
\end{equation}
in the limit of small $D_b^\text{Tx}$. Summing~\eqref{good_bound} over all links, we get
\begin{equation}\label{proof_end}
\begin{split}
  \sum\limits_{i=1}^n{{R_{i\to \text{PN}(i)}}}
  &\ge \frac{1}{2}\sum\limits_{i=1}^n \left[
  {\log_2 }\frac{\sigma _{{\mathcal{S}_i}}^2 }{{D_i^\text{Inc}}}
  -\frac{D_i^\text{Tx}}{2w_i^2 }
  -\frac{{\log_2 }e}{2\sigma _{\mathcal{S}_i}^2 }\sqrt{2D_i^\text{Tx}\left( 4\sigma_{\mathcal{S}_i}^2 +D_i^\text{Tx}\right)} \right].\\
  & =\frac{1}{2}\sum\limits_{i=1}^n{\left[ {\log_2 }\frac{\sigma _{{\mathcal{S}_i}}^2 }{D_i^\text{Rx}-D_i^\text{Tx}}-\mathcal{O}\left((D_i^\text{Tx})^{1/2}\right) \right]},
\end{split}
\end{equation}
in the limit of small $D_i^\text{Tx},\forall i$. The last equality in \eqref{proof_end} can be obtained using ${D_i^\text{Rx}}=D_i^\text{Tx}+{D_i^\text{Inc}}$ (see the distortion accumulation equation~\eqref{Pyth_all}).
The optimization bound shown in \eqref{min_opti} is basically the same bound \eqref{proof_end} stated in an optimization form over the choices of the incremental distortions $D_i^\text{Inc}$. Now, we prove that the solution of the optimization satisfies \eqref{111111111}, which finally leads to the order-sense bound \eqref{minimize_lower_bound}. When the constraints in~\eqref{min_opti} are satisfied,
\begin{equation}
  \begin{split}
    R&\ge \frac{1}{2}\sum\limits_{i=1}^n{\left[ {\log_2 }\frac{\sigma_{\mathcal{S}_i}^2}{{D_i^\text{Inc}}}-\psi_i\left(D_i^\text{Tx}\right) \right]}\\
    &\overset{(a)}{\mathop{\ge }}\frac{1}{2}\,{\log_2 }\frac{\prod\limits_{i=1}^n{\sigma_{\mathcal{S}_i}^2}}{\prod\limits_{i=1}^n{{D_i^\text{Inc}}}}-\frac{1}{2}\sum_{i=1}^n\psi_i\left(D_0^\text{mmse}\right)\\
    &\overset{(b)}{\mathop{\ge }}\frac{1}{2}\,{\log_2 }\frac{\prod\limits_{i=1}^n{\sigma_{\mathcal{S}_i}^2}}{{{\left( {D_0^\text{mmse}}/n \right)}^n}}-\frac{1}{2}\sum_{i=1}^n\psi_i\left(D_0^\text{mmse}\right)\\
    &\overset{(c)}{\mathop{\ge }}\frac{1}{2}\,{\log_2 }\frac{\prod\limits_{i=1}^n{\sigma_{\mathcal{S}_i}^2}}{{{\left( {{D}}/n \right)}^n}}-\frac{1}{2}\sum_{i=1}^n\psi_i\left(D\right),
  \end{split}
\end{equation}
where $(a)$ holds because $D_i^\text{Tx}<D_0^\text{mmse}$ (which can be easily seen by comparing~\eqref{Tr_Distortion} and \eqref{Total_Distortion}) and the functions $\psi_i(\cdot)$, $i=1,\ldots n$ are monotone, $(b)$ follows from the constraint $D_0^\text{mmse}=\sum_{i=1}^n D_i^\text{Inc}$ in~\eqref{Total_Distortion} and the fact that the arithmetic mean is greater or equal to the geometric mean, and $(c)$ follows from the inequality $D_0^\text{mmse}\le D$ in~\eqref{lb_d0}. Further, using the fact that $\psi_i(D)=\frac{D}{2w_i^2 }
  +\frac{{\log_2 }e}{2\sigma _{\mathcal{S}_i}^2 }\sqrt{2D\left( 4\sigma_{\mathcal{S}_i}^2 +D\right)}=\mathcal{O}(\sqrt{D})$, we obtain the lower bound \eqref{minimize_lower_bound}.
\subsection{Proof of Theorem~\ref{cut_set_thm}}\label{cut_set_proof}
We still look at a specific set $\mathcal{S}$ as shown in Fig.~\ref{cut_set_figure}. Then, we have
\begin{equation}\label{cut_set_bound_ineq_small}
  \begin{split}
  & N{R_{b\to a}}\ge H(M_{b\to a}) \\
 & \ge I(M_{b\to a};{\mathbf{y}_\mathcal{S}}) \\
 & =h({\mathbf{y}_\mathcal{S}})-h({\mathbf{y}_\mathcal{S}}|M_{b\to a}) \\
 & =h({\mathbf{y}_\mathcal{S}})-h({\mathbf{y}_\mathcal{S}}|M_{b\to a},{\widehat{\mathbf{y}}^\text{mmse}_{\mathbf{S},a}}) \\
 & =h({\mathbf{y}_\mathcal{S}})-h({\mathbf{y}_\mathcal{S}}-{\widehat{\mathbf{y}}^\text{mmse}_{\mathbf{S},a}}|{\widehat{\mathbf{y}}^\text{mmse}_{\mathbf{S},a}},M_{b\to a}) \\
 & \ge h({\mathbf{y}_\mathcal{S}})-h({\mathbf{y}_\mathcal{S}}-{\widehat{\mathbf{y}}^\text{mmse}_{\mathbf{S},a}}) \\
 & \ge \frac{N}{2}{\log_2 }2\pi e\sigma _{\mathcal{S}}^2 -\frac{N}{2}{\log_2 }2\pi e{D_b^\text{Rx}}=\frac{N}{2}{\log_2 }\frac{\sigma _{\mathcal{S}}^2 }{{D_b^\text{Rx}}}.
\end{split}
\end{equation}
Summing~\eqref{cut_set_bound_ineq_small} over all links, we get the outer bound~\eqref{abbas_cutset_bound}.

\section{Proofs for Section~\ref{Gaussian_sec}}
\subsection{A Review on Gaussian Test Channels}\label{Gaussian_rate_calculation}
First, we elaborate on the details of Gaussian test channels. Suppose a transmitter has a source $X\sim\mathcal{N}(0,P)$ and wishes to send an approximate description $\widehat{X}$ to a receiver with distortion $D$. Then
\begin{equation}\label{Gaussian_rd}
  R(D) = \min_{p(\widehat{x}|x): \mathbb{E}\left[ (X-\widehat{X})^2 \right]\le D} I(X;\widehat{X}) = \frac{1}{2}\log\frac{P}{D},\forall P\ge D.
\end{equation}
The ``test channel'' in this case is the inverse Gaussian channel
\begin{equation}\label{inverse_Gaussian}
  X=\widehat{X}+Z,
\end{equation}
where $Z\sim\mathcal{N}(0,D)$ is an additive noise independent of $\widehat{X}$ (see \cite[Theorem 10.3.2]{Cover_Wiley_06}). The test channel is useful for understanding orthogonality properties of {codewords in random codebooks}. To achieve the rate in~\eqref{Gaussian_rd}, we can use a random code $\{\widehat{\mathbf{c}}(w):\; w \in \{1,2,\ldots 2^{NR}\}\}$ with joint typicality encoding and decoding, where each codeword $\widehat{\mathbf{c}}(w)$ is generated i.i.d. with each entry distributed as $\mathcal{N}(0,P-D)$. When $N\to\infty$, the rate~\eqref{Gaussian_rd} is asymptotically achieved.

\subsection{Proof of \eqref{variance_less}}\label{pf_variance_less}
We use induction in the tree (see Remark~\ref{induction_on_tree}) to prove \eqref{variance_less}. For an arbitrary leaf $v_l$, we know that $\widehat{\sigma }_l^2=\sigma_{\mathcal{S}_l}^2=w_l^2$. For an arbitrary non-leaf node $v_b$, we have that (see \eqref{var_induction})
\begin{equation}\label{777}
  \widehat{\sigma }_b^2 =\sum\limits_{k=1}^{d}{(\widehat{\sigma }_{k}^2-d_b)}+w_b^2 .
\end{equation}
By definition, we have
\begin{equation}
  {\mathbf{y}_\mathcal{S}}=\sum\limits_{k=1}^{d}{\mathbf{y}_{\mathcal{S}_k}}+w_b{\mathbf{x}_b},
\end{equation}
which means
\begin{equation}\label{888}
  \sigma_{\mathcal{S}_b}^2=\sum\limits_{k=1}^{d}\sigma_{\mathcal{S}_k}^2+w_bw_b^2,
\end{equation}
Comparing \eqref{777} and \eqref{888}, we know that, if \eqref{variance_less} holds at all children of $v_b$, it also holds at $v_b$. Thus, by induction in the tree, we can show that~\eqref{variance_less} is true.

\subsection{Proof of Lemma~\ref{covering_lemma}}\label{pf_covering}
The key idea is to use the generalized covering lemma~\cite[pg. 70]{Gamal_Camb_11}, which is rephrased as follows.
\begin{lemma}\label{gamal_lemma}
Suppose $\mathbf{x}$ is an arbitrary sequence and $\underset{N\to \infty }{\mathop{\lim }}\Pr(\mathbf{x}\notin\mathcal{T}_\epsilon^{(N)}(p_X))=0$. Let $\widehat{\mathbf{x}}(m),m\in \mathcal{A}$, where $|\mathcal{A}|>2^{nR}$, be random sequences independent from $\mathbf{x}$, each distributed according to $p_{\widehat{X}}$. Then, there exists $\delta (\varepsilon )\to 0$ as $\varepsilon \to 0$ such that
\[\underset{N\to \infty }{\mathop{\lim }}\,\Pr \left((\mathbf{x},\widehat{\mathbf{x}}(m))\notin \mathcal{T}_{\epsilon }^{(N)}({p_{X,\widehat{X}}}),\forall m\in \mathcal{A}\right)=0,\]
if $R>I(X;\widehat{X})+\delta (\varepsilon )$.
\end{lemma}
The covering lemma follows directly from the conditional typicality lemma and the joint typicality lemma, e.g., see~\cite{Gamal_Camb_11} and other recent works such as~\cite{App_TIT_11}\cite{Vis_ISIT_10}\cite{Cuf_ISIT_09} on distributed source coding and computing. The original covering lemma is stated for discrete sources with a finite distortion measure and strong typicality. The generalized version to abstract sources with infinite measure has been obtained in~\cite{Jeon_ISIT_14,Jeon_Arx_15}. We first present this version of the generalized covering lemma.

\begin{lemma}[Generalized covering lemma in \cite{Jeon_Arx_15}, page 18, Theorem IV.5]
	Let $R\geq0$ be a nonnegative real number such that
	$R>I(\mathbf{y};\mathbf{z})$. Then, there exists a $\mu_{YZ}$-typicality criterion
	$\mathcal{W}_{0}$ and a positive number $c>0$ such that, for any $\mu_{YZ}$-typicality criterion
	$\mathcal{W}\leq\mathcal{W}_{0}$, there exists a $\mu_{Y}$-typicality criterion
	$\mathcal{V}$ so that we have the following for all sufficiently large $n\in\Z^{+}$:
	
	Let $I_{N}$ be a finite set with $|I_{N}|\geq2^{NR}$.
	Let $\mathbf{y}^{N}$ be a random variable taking values in
	$Y^{N}$, and for each $m\in I_{N}$, let $\mathbf{z}^{N}(m)$ be a random
	variable taking values in $Z^{N}$. Assume that for $m,m'\in I_{N}$ with
	$m\neq m'$, $(\mathbf{z}^{N}(m),\mathbf{z}^{N}(m'))$ follows a
	distribution $\mu_{Z}^{N}\times\mu_{Z}^{N}$. Then we have
	\begin{displaymath}
		\Pr\left(\mathbf{y}^{N}
		\in\mathcal{T}_{\mathcal{V}}^{(N)}(\mu_{Y})\ \textrm{and}\
		(\mathbf{y}^{N},\mathbf{z}^{N}(m))
		\notin\mathcal{T}_{\mathcal{W}}^{(N)}(\mu_{YZ})\quad
		\textrm{for all $m\in I_{N}$}\right)\leq2^{-cN}.
	\end{displaymath}
\end{lemma}

The generalization of the classical typicality is stated using a typicality definition called ``$\mu$-typicality'', where $\mu$ is a probability measure, or a probability density function on a general alphabet $X$ on which the typical sets are defined. The alphabet can be discrete or continuous (including a $d$-dimensional space). The key definition of typicality involves a triple $\mathcal{U}=(\mathscr{F},\epsilon,\mathscr{N})$ called a ``typicality criterion''. In this triple, the set $\mathscr{F}$ is called the ``typicality requirements'' and is composed of a finite set of $\mu$-integrable functions $\{f_1,f_2,\ldots f_M\}$ (i.e., the expectation of these functions $\mathbb{E}_\mu[f]$ with respect to $\mu$ is finite) that are used for defining the typical set. For example, choosing $\mathscr{F}$ to be a single point set containing the log-likelihood function corresponds to weak-typicality, or containing the indicator functions corresponds to strong typicality (see the book by Cover and Thomas \cite[Section 13.6]{Cover_Wiley_06}). The constant $\epsilon$ in definition of $\mathcal{W}$ is the usual small value for bounding the difference between empirical mean and true expectation of the functions in  $\mathscr{F}$,  \textcolor{black}{such as the constant $\varepsilon_N$ in equations \eqref{typi1} to \eqref{typi4}}. The set $\mathscr{N}$ is a $\mu$-null set ($0$-measure set with respect to the measure $\mu$) which is only used for some special cases. When the density function $\mu$ and the typicality requirement set $\mathscr{F}=\{f_1,f_2,\ldots f_k\}$ is defined, the typical set is defined as (see Definition II.2 in \cite{Jeon_Arx_15})
\begin{eqnarray}\label{typ_set}
\mathcal{T}_{\mathcal{U}}^{(n)}(\mu)&\defas&
		\Bigg\{x^{N}\in(X\setminus \mathscr{N})^{N}:\abs{\frac{1}{N}\sum_{i=1}^{N}f(x_{i})
		-\int f\,d\mu}\leq\epsilon\quad
		\textrm{for all $f\in\mathscr{F}$}\Bigg\}.
\end{eqnarray}

\textcolor{black}{In the definitions of typical sets (\eqref{typi1} to \eqref{typi4} in this response), the set $\mathscr{F}$ is defined using appropriate functions. For example, in \eqref{typi1}, $\mathscr{F}$ consists of the functions $\log q_U(\cdot)$ and $\left\|\cdot\right\|^2$}. Note that $\mathbb{E}_{q_U}[\log q_U(x)]$ and $\mathbb{E}_{q_U}[\left\|x\right\|^2]$ are the differential entropy of $q_U$ and the second moment of $q_U$, which are always finite for Gaussian distributions\footnote{Note that for scalar Gaussian distributions $\mathbb{E}_{q_U}[\log q_U(x)]$ and $\mathbb{E}_{q_U}[\left\|x\right\|^2]$ are linear functions of each other. However, this does not matter because a typicality requirement can definitely include the same functions.}. Therefore, \eqref{typi1} to \eqref{typi4} are all valid typicality criteria. Here are some key properties of $\mu$-typicality that may help further explain the concept:
\begin{itemize}
\item To define a typical set, you have to first specify the density function $\mu$ and the typicality requirements $\mathscr{F}$. The expectation $\mathbb{E}_\mu [f]=\int f\,d\mu$ for $f\in\mathscr{F}$ must be respect to $\mu$, rather than any other distribution.
\item If the sequence $x^N$ is distributed according to $\mu^N$, it automatically belongs to any valid typicality criterion defined for $\mu$ with high probability, because of the law of large numbers.
    \end{itemize}

Now we explain the generalized covering lemma. In the first sentence of the lemma (``Let $R\ge 0$ be a nonnegative real number [...]''), the variable $R$ is the coding rate so $R>I(\mathbf{y};\mathbf{z})$ simply means the coding rate has to be bigger than the mutual information\footnote{Note that in the original paper \cite{Jeon_Arx_15} the generalized covering lemma has an auxiliary random variable $\mathbf{x}$ that is useful in multi-terminal source coding problems. For our problem with independent sources at different nodes, the auxiliary variable $\mathbf{x}$ in the generalized covering lemma does not exist so we just removed all terms related to the random variable $\mathbf{x}$.}. The mutual information is defined as (see the end of page 16 in \cite{Jeon_Arx_15}) $I(\mathbf{y};\mathbf{z})=D(\mu_{YZ}||\mu_{Y}\mu_{Z})$.

At first, we skip the sentence ``Then, there exists a $\mu_{YZ}$-typicality criterion [...]'' (we will explain it later) and look at the middle part of the generalized covering lemma (``let $I_N$ be a finite set with [...]''). The set $I_N$ is the codeword index set. The random variable $\mathbf{y}^N$ is the source sequence (note that this sequence can be an arbitrary sequence and \textbf{is not} the $N$-fold product of the random variable $\mathbf{y}$ in the mutual information $I(\mathbf{y};\mathbf{z})$). The random variable $\mathbf{z}^N(m)$ for each $m$ is a codeword sequence and each of its coordinate follows the density function $\mu_Z$. The sentence ``Assume that for $m,m'\in I_N$[...]'' means that all codewords are pairwise independent. Then, the last equation of the covering lemma states that, the probability that the sequence $\mathbf{y}^N$ is typical, but the sequence $(\mathbf{y}^N,\mathbf{z}^N(m))$ are not jointly typical for all codewords $\mathbf{z}^N(m),m\in I_N$ is exponentially small in the code length $N$.

Now let us examine the sentence ``Then, there exists a $\mu_{YZ}$-typicality criterion [...]'' which is somewhat subtle. It may seem that from this definition, the typicality criteria $\mathcal{W}_0$ and $\mathcal{V}$ (which are defined based on two sets of typicality requirements $\mathscr{F}_{\mathcal{W}_0}$ and $\mathscr{F}_{\mathcal{V}}$) exist but have not been written explicitly. In fact, they can be defined explicitly, but the construction is quite careful. The explicit constructions of these typicality criterion can be found in \cite[Pg. 12-14]{Jeon_Arx_15}. In this paper, We only need the existence of $\mathcal{W}_0$ and $\mathcal{U}$ typicality criterion, and hence for our use the lemma as stated above suffices.

Next, it may also seem that the generalized covering lemma does not apply to the problem in our paper because the construction of typicality requirements in \cite{Jeon_Arx_15} does not necessarily include the typicality requirements for function $\log q_U(\cdot)$ and $\left\|\cdot\right\|$ in \eqref{typi1} to \eqref{typi4}. In fact, the inequality $\mathcal{W}\le \mathcal{W}_0$ is the key of the generalized covering lemma. It means $\mathcal{W}$ can be any typicality criterion that has stricter typicality requirements than $\mathcal{W}_0$. More specifically, one can incorporate any classical typicality requirements (e.g.~weak or strong typicality) into the typicality criterion $\mathcal{W}$ by adding more $\mu$-integrable functions into the function set $\mathscr{F}$ in $\mathcal{W}=(\mathscr{F},\epsilon,N)$.

\begin{figure}
  \centering
  \includegraphics[scale=0.3]{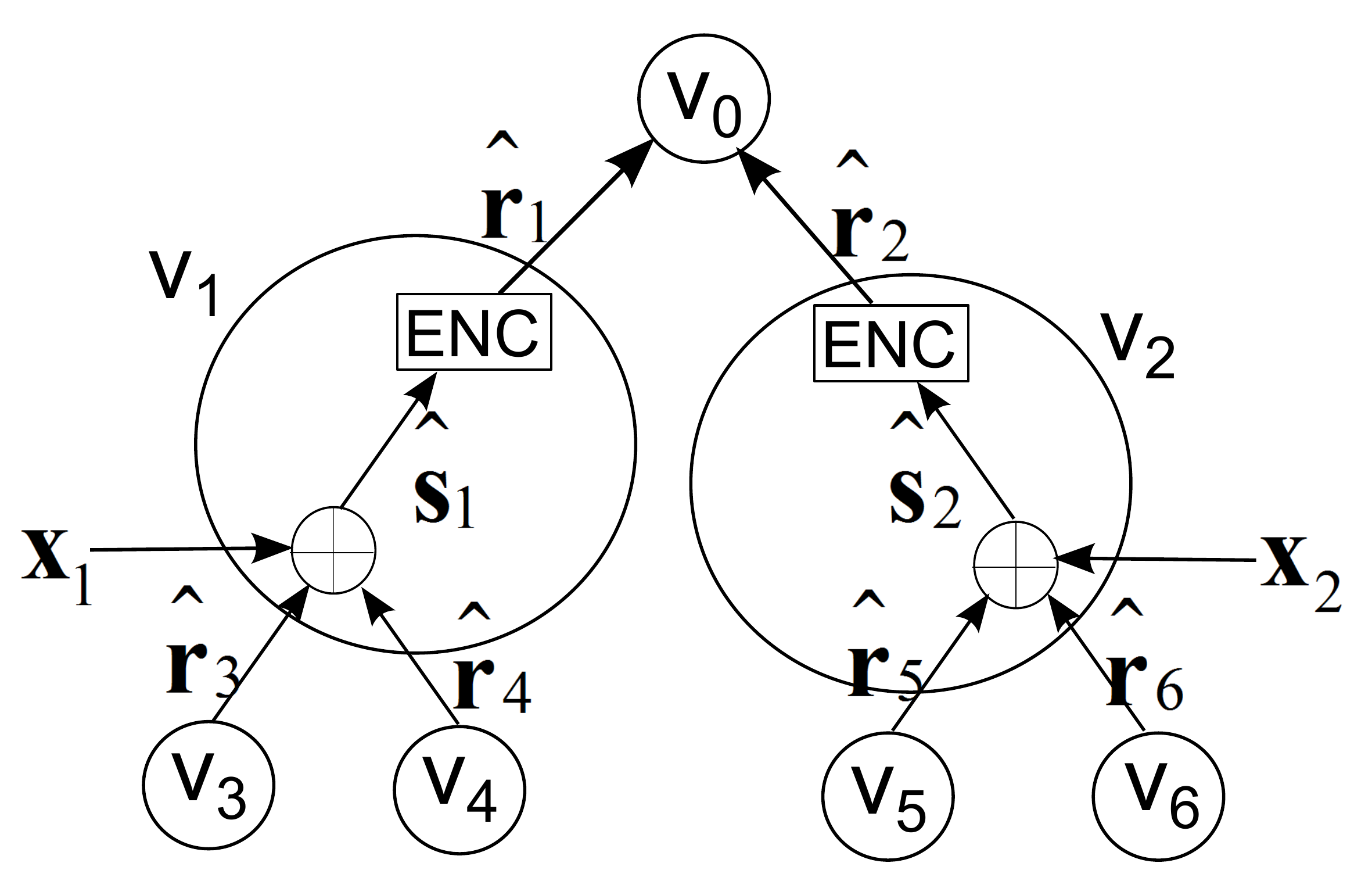}\\
  \caption{This is an illustration of the typicality encoding in a simple two-layer tree network.}\label{typi}
\end{figure}

Now we make explicit the use of this generalized version of covering lemma in our problem. In order to avoid cumbersome notation, we will only look at a specific example shown in Fig. \ref{typi}. At a particular node $v_i$, there are three sequences: the source sequence $\mathbf{x}_i$, the estimate sequence $\hat{\mathbf{s}}_i$ which is the estimate of the partial sum $\mathbf{y}_i$, and the description sequence $\hat{\mathbf{r}}_i$ which is the chosen codeword to be transmitted from node $v_i$ to its parent node. These sequences are defined in Section~\ref{Gaussian_code_alg}. At the root note $v_0$, there is only an estimate sequence $\hat{\mathbf{s}}_0$, which is the estimate of the overall sum. We will use mathematical induction to define $\mu$-typical sets from the root to the leaves of the tree network. This will ensure that the estimate sequence $\hat{\mathbf{s}}_i$ and the description sequence $\hat{\mathbf{r}}_i$ for each node $v_i$ lie inside the typical sets in \eqref{typi1} to \eqref{typi4} with high probability.

At the root node, the estimate sequence $\hat{\mathbf{s}}_0=\hat{\mathbf{r}}_1+\hat{\mathbf{r}}_2$. Therefore, for $\hat{\mathbf{s}}_0$ to be typical with respect to \eqref{typi1}, $\hat{\mathbf{r}}_1$ and $\hat{\mathbf{r}}_2$ have to satisfy the joint typicality criterion that the sum $\hat{\mathbf{r}}_1+\hat{\mathbf{r}}_2\in \mathcal{T}_{U,\epsilon}^N$. To be specific, the two dimensional sequence $(\hat{\mathbf{r}}_1,\hat{\mathbf{r}}_2)$ has to satisfy the typicality requirement $\mathscr{F}_{1,2}$ defined by the ``sum-square function'' $f(r_1,r_2)=(r_1+r_2)^2$ and the sum-$\log q_U$ function $g(r_1,r_2)=\log q_U(r_1+r_2)$, which are
\begin{eqnarray}
|1/N\sum_{i=1}^N f(\hat{r}_1^i,\hat{r}_2^i)-\mathbb{E}_{V_1^\text{TC},V_2^\text{TC}}[f(V_1^\text{TC},V_2^\text{TC})]|<\epsilon\\
|1/N\sum_{i=1}^N g(\hat{r}_1^i,\hat{r}_2^i)-\mathbb{E}_{V_1^\text{TC},V_2^\text{TC}}[g(V_1^\text{TC},V_2^\text{TC})]|<\epsilon,
\end{eqnarray}
where $V_1^\text{TC}$ and $V_2^\text{TC}$ are the test-channel random variables that are used to define the density functions respect to which $\hat{\mathbf{r}}_1$ and $\hat{\mathbf{r}}_2$ are typical.

Now, look at the second layer of the network, which is composed of two nodes $v_1$ and $v_2$. In this layer, we will use the generalized covering lemma. The description sequences $\hat{\mathbf{r}}_1$ and $\hat{\mathbf{r}}_1$ are the output (the chosen codewords $\mathbf{z}^N(m)$) in the generalized covering lemma. Note that although the two sequences $\hat{\mathbf{r}}_1$ and $\hat{\mathbf{r}}_1$ are jointly typical, the codewords at two nodes $v_0$ and $v_1$ can be separately generated because the joint density factorizes as $\phi(V_1^\text{TC},V_2^\text{TC})=\phi(V_1^\text{TC})\phi(V_2^\text{TC})$. The input (the sequence $\mathbf{y}^N$ in the generalized covering lemma) to the second layer are the collections of description sequences and the sources sequences $(\hat{\mathbf{r}}_3,\hat{\mathbf{r}}_4,\hat{\mathbf{r}}_5,\hat{\mathbf{r}}_6,\mathbf{x}_1,\mathbf{x}_2)$. Apart from the typicality requirement $\mathscr{F}_{1,2}$, the output and the input have to satisfy the joint typicality requirements specified in \eqref{typi1} to \eqref{typi4}, which we denote by $\mathscr{G}_{1,2}$.

Then, according to the generalized covering lemma, there exists a typicality criterion $\mathcal{W}_0$ so that we can find two codewords $(\hat{\mathbf{r}}_1,\hat{\mathbf{r}}_2)$ that are jointly typical with the sequence $(\hat{\mathbf{r}}_3,\hat{\mathbf{r}}_4,\hat{\mathbf{r}}_5,\hat{\mathbf{r}}_6,\mathbf{x}_1,\mathbf{x}_2)$. Then, we add the joint typicality requirements $\mathscr{F}_{1,2}\cup \mathscr{G}_{1,2}$ to $\mathcal{W}_0$ to get the new joint-typicality criterion $\mathcal{W}$ (as we mentioned earlier, the covering lemma holds for any typicality criterion $\mathcal{W}\le \mathcal{W}_0$, we can add any typicality requirements that are $\mu$-integrable). Note that the generalized covering lemma requires that the source $(\hat{\mathbf{r}}_3,\hat{\mathbf{r}}_4,\hat{\mathbf{r}}_5,\hat{\mathbf{r}}_6,\mathbf{x}_1,\mathbf{x}_2)$ satisfy an extra typicality criterion $\mathcal{V}$ (see the generalized covering lemma and the explanation). This is very similar to the joint typicality requirement $\mathscr{F}_{1,2}$ that we have used to define the typicality criterion in the second layer of the network, and we can use the same method to define the joint typicality criterion at the third layer.

The general induction method in the tree network goes in the following manner:
\begin{itemize}
\item Specify the output $\mathbf{o}_l$ and input $\mathbf{i}_l$ at each layer $l$;
\item Define the typicality criterion at each layer from the first layer (the root node) to the last layer (the leaves that have the largest distance to the root);
\item At each layer $l$, the definition of the typicality criterion at layer $l-1$ imposes extra typicality criterion $\mathcal{V}_{l-1}$ at the output of layer $l$. Compute the union of the typicality criterion $\mathcal{V}_{l-1}$ and the criterion imposed by \eqref{typi1}-\eqref{typi4} at layer $l$ and apply the generalized covering lemma. This definition process at the layer $l$ will impost extra typicality criterion $\mathcal{V}_l$ to the output of the $(l+1)$-th layer.
\item Repeat the above process until the last layer of the network.
\end{itemize}

Using this definition, the joint typicality requirement at each node in the network is always the weak joint typicality requirement \eqref{joint_typicality} in addition to a finite set of extra typicality requirements. According to the covering lemma, all of these typicality requirements will hold with high probability. This is ensured by the definition of typicality criterion and the definition of typical sequences. When the covering lemma holds, all sequences will be typical with high probability. To be more specific, they will be atypical with exponentially small probability in the code length $N$. We use the generalized covering lemma to ensure that all description sequences $\mathbf{s}_i^N$ and estimate sequences $\mathbf{r}_i^N$ are jointly typical (so that the empirical distance $\frac{1}{N}||\mathbf{s}_i^N-\mathbf{r}_i^N||\approx d_i$ with high probability). The conclusion~\eqref{encoding_error} can thus be obtained. After that, we use the other lemmas in the revised manuscript (Lemma~\ref{var_bounded} to Lemma~\ref{distortion_close_lemma} in our own paper) to compute the expectation of mean-square error at all nodes.

To summarize, we again comment on how the conclusion~\eqref{encoding_error} can be obtained by induction in the tree network (see Remark~\ref{induction_on_tree}). First, on an arbitrary leaf $v_l$, the rate satisfies $R_l>I(U^\text{TC}_l;V^\text{TC}_l)$ (see~\eqref{r_up}). According to the covering lemma, there exists a codeword $\mathbf{c}_l(M_{lPN(l)})$ jointly typical with data $\mathbf{x}_l$ with high probability. This also ensures that, with high probability, the reconstruction $V^\text{TC}_l$ at the parent-node $v_{PN(l)}$ is typical with respect to distribution $\phi_{V^\text{TC}_l}$. On an arbitrary non-leaf node $v_b$, the estimate $\widehat{\mathbf{s}}_b$ defined by~\eqref{U_b_hat} is typical with respect to $\phi_{U^\text{TC}_b}$ with high probability, provided that all descriptions $V^\text{TC}_1,\ldots V^\text{TC}_d$ from the children of $v_b$ are typical with high probability (which is ensured by induction). Since rate satisfies $R_b>I(U^\text{TC}_b;V^\text{TC}_b)$, according to the covering lemma, there exists a codeword $\mathbf{c}_b(M_{b\to a})$ jointly typical with data $\mathbf{x}_b$ and reconstructions $\mathbf{c}_k(M_{k\to b}),k=1,\ldots  d$ with high probability. This also ensures that the codeword $\mathbf{c}_b(M_{b\to a})$ is typical with respect to distribution $\phi_{V^\text{TC}_b}$. Thus, it is clear that equation~\eqref{encoding_error} can be proved using induction in the tree.
\subsection{Proof of Lemma~\ref{var_bounded}}\label{app_var_bounded}
We prove this lemma using induction in the tree (see Remark~\ref{induction_on_tree}). At an arbitrary leaf $v_l$, the estimate $\widehat{\mathbf{s}}_l=w_l\mathbf{x}_l$ satisfies
\begin{equation}
  \mathbb{E}\left[ \frac{1}{N}\left\| \widehat{\mathbf{s}}_l \right\|_{2}^2  \right]=\mathbb{E}\left[ \frac{1}{N}\left\| w_l\mathbf{x}_l \right\|_{2}^2  \right]=w_l^2 ,
\end{equation}
while $\widehat{\sigma}_l^2=w_l^2$, which is the variance of the scalar random variable $U^\text{TC}_l$. Therefore,~\eqref{var_bounded_eqn_1} holds for the leaf $v_b$ with $\varepsilon_{lN}>0$, where $\underset{N\to \infty }{\mathop{\lim }}\,\varepsilon_{lN}=0$. That is
\begin{equation}\label{upp_var_der1}
  \left| \mathbb{E}\left[ \frac{1}{N}\left\| \widehat{\mathbf{s}}_l \right\|_{2}^2  \right]-\widehat{\sigma }_l^2  \right|<\varepsilon_{lN}.
\end{equation}
For the description $\widehat{\mathbf{r}}_l$,
\begin{equation}
\begin{split}
  \mathbb{E}\left[ \frac{1}{N}\left\| \widehat{\mathbf{r}}_l \right\|_{2}^2  \right]=&\Pr ({E_l}=1)\mathbb{E}\left[ \left. \frac{1}{N}\left\| \widehat{\mathbf{r}}_l \right\|_{2}^2  \right|{E_l}=1 \right]
  +\Pr ({E_l}=0)\mathbb{E}\left[ \left. \frac{1}{N}\left\| \widehat{\mathbf{r}}_l \right\|_{2}^2  \right|{E_l}=0 \right],
\end{split}
\end{equation}
where $E_l$ is the indicator random variable of an encoding failure at a leaf $v_l$. Recall that when $E_l=1$, no codeword generated at $v_l$ is jointly typical with the source message $w_l \mathbf{x}_l$ and the $0$-th codeword, which is also a random codeword, is transmitted. When ${E_l}=0$, $\widehat{\mathbf{r}}_l$ is typical with respect to the distribution $\phi_{V^\text{TC}_l}$, and hence
\begin{equation}
  \left| \frac{1}{N}\left\| \widehat{\mathbf{r}}_l \right\|_{2}^2 -(\widehat{\sigma }_l^2 -d_l) \right|<\varepsilon _{lN},
\end{equation}
where $\underset{N\to \infty }{\mathop{\lim }}\,\varepsilon _{lN}=0$. When ${E_l}=1$,
\begin{equation}
\begin{split}
  &\mathbb{E}\left[ \left. \frac{1}{N}\left\| \widehat{\mathbf{r}}_l \right\|_{2}^2  \right|{E_l}=1 \right]=\mathbb{E}\left[ \left. \frac{1}{N}\left\| {{c}_l}(0) \right\|_{2}^2  \right|{E_l}=1 \right]
  =\mathbb{E}\left[ \frac{1}{N}\left\| {{c}_l}(0) \right\|_{2}^2  \right]=\widehat{\sigma }_l^2 -d_l.
\end{split}
\end{equation}
Suppose $\Pr ({E_l}=0)=1-{{\beta }_{lN}},$ where $\underset{N\to \infty }{\mathop{\lim }}\,\beta _{lN}=0$, then
\begin{equation}
\begin{split}
  & \left| \mathbb{E}\left[ \frac{1}{N}\left\| \widehat{\mathbf{r}}_l \right\|_{2}^2  \right]-(\widehat{\sigma}_l^2-d_l)  \right| \\
 & <\Pr ({E_l}=1)\left| \mathbb{E}\left[ \left. \frac{1}{N}\left\| \widehat{\mathbf{r}}_l \right\|_{2}^2  \right|{E_l}=1 \right]-(\widehat{\sigma}_l^2-d_l)  \right|\\
 &{\;\;\;\;}+\Pr ({E_l}=0)\left| \mathbb{E}\left[ \left. \frac{1}{N}\left\| \widehat{\mathbf{r}}_l \right\|_{2}^2  \right|{E_l}=0 \right]-(\widehat{\sigma}_l^2-d_l)  \right| \\
 & ={{\beta }_{lN}}\left| \mathbb{E}\left[ \frac{1}{N}\left\| {{c}_l}(0) \right\|_{2}^2  \right]-(\widehat{\sigma}_l^2-d_l)  \right|+(1-{{\beta }_{lN}})\varepsilon _{lN}\\
 &<\varepsilon _{lN}.
\end{split}
\end{equation}
Therefore,~\eqref{var_bounded_eqn_2} holds for the leaf $v_l$. To prove~\eqref{typical_small_distortion_1} for a leaf $v_l$, we have to use the following fact
\begin{equation}\label{upp_var_der2}
  \Pr ({E_l}=1)\mathbb{E}\left[ \left. \frac{1}{N}\left\| \widehat{\mathbf{s}}_l \right\|_{2}^2  \right|{E_l}=1 \right]<{{\alpha }_{lN}},
\end{equation}
where $\underset{N\to \infty }{\mathop{\lim }}\,{{\alpha }_{lN}}=0$. This can be proved as follows. First, we have that
\begin{equation}
\begin{split}
  &\Pr ({E_l}=1)\mathbb{E}\left[ \left. \frac{1}{N}\left\| \widehat{\mathbf{s}}_l \right\|_{2}^2  \right|{E_l}=1 \right] =\mathbb{E}\left[ \frac{1}{N}\left\| \widehat{\mathbf{s}}_l \right\|_{2}^2  \right]-\Pr ({E_l}=0)\mathbb{E}\left[ \left. \frac{1}{N}\left\| \widehat{\mathbf{s}}_l \right\|_{2}^2  \right|{E_l}=0 \right].
\end{split}
\end{equation}
From~\eqref{upp_var_der1},
\begin{equation}
  \mathbb{E}\left[ \frac{1}{N}\left\| \widehat{\mathbf{s}}_l \right\|_{2}^2  \right]<\widehat{\sigma }_l^2 +\varepsilon _{lN}.
\end{equation}
When ${E_l}=0$, $\widehat{\mathbf{s}}_l$ is typical with respect to the distribution ${p_{{U^\text{TC}_l}}}$, and hence
\begin{equation}
  \mathbb{E}\left[ \left. \frac{1}{N}\left\| \widehat{\mathbf{s}}_l \right\|_{2}^2  \right|{E_l}=0 \right]>\widehat{\sigma }_l^2 -\varepsilon _{lN},
\end{equation}
when $N$ is large enough. Therefore,
\begin{equation}
\begin{split}
  \Pr ({E_l}=1)\mathbb{E}\left[ \left. \frac{1}{N}\left\| \widehat{\mathbf{s}}_l \right\|_{2}^2  \right|{E_l}=1 \right]<&\widehat{\sigma }_l^2 +\varepsilon _{lN}-\text{(}1-{{\beta }_{lN}}\text{)}\left( \widehat{\sigma }_l^2 -\varepsilon _{lN} \right)\\
  =&{{\beta }_{lN}}\widehat{\sigma }_l^2 \text{+(2}-{{\beta }_{lN}}\text{)}\varepsilon _{lN}=:{{\alpha }_{lN}}.
\end{split}
\end{equation}
Then, we prove~\eqref{typical_small_distortion_1} for a leaf $v_l$. We notice that
\begin{equation}\label{dist_decomp}
\begin{split}
\mathbb{E}\left[ \frac{1}{N}\left\|\widehat{\mathbf{r}}_l-\widehat{\mathbf{s}}_l \right\|_{2}^2  \right]
=&\Pr \left( {E_l}=1 \right)\mathbb{E}\left[ \left. \frac{1}{N}\left\| \widehat{\mathbf{r}}_l-\widehat{\mathbf{s}}_l \right\|_{2}^2  \right|{E_l}=1 \right]\\
&+\Pr \left( {E_l}=0 \right)\mathbb{E}\left[ \left. \frac{1}{N}\left\| \widehat{\mathbf{r}}_l-\widehat{\mathbf{s}}_l \right\|_{2}^2  \right|{E_l}=0 \right].
\end{split}
\end{equation}
When ${E_l}=0$, i.e., when the estimate $\widehat{\mathbf{s}}_l$ and the description $\widehat{\mathbf{r}}_l$ are jointly typical and encoding is successful,
\begin{equation}\label{dist_decomp_1}
  \left| \frac{1}{N}\left\| \text{ }\widehat{\mathbf{r}}_l-\widehat{\mathbf{s}}_l \right\|_{2}^2 -{{d}_l} \right|<\frac{{{\varepsilon }_{lN}}}{2},
\end{equation}
where $\underset{N\to \infty }{\mathop{\lim }}\,{{\varepsilon }_{lN}}=0$. When ${E_b}=1$, the transmitted sequence $\widehat{\mathbf{r}}_l=\mathbf{c}_l(0)$ is a predetermined random sequence independent of $\widehat{\mathbf{s}}_l$, and hence $\widehat{\mathbf{r}}_l$ is independent of $\widehat{\mathbf{s}}_l$ conditioned on ${E_b}=1$. Therefore,
\begin{equation}\label{dist_decomp_2}
\begin{split}
&{\;\;\;\;\;}\mathbb{E}\left[ \left. \frac{1}{N}\left\| \widehat{\mathbf{r}}_l-\widehat{\mathbf{s}}_l \right\|_{2}^2  \right|{E_b}=1 \right]-{{d}_l} .\\
&=\mathbb{E}\left[ \left. \frac{1}{N}\left\| \widehat{\mathbf{r}}_l \right\|_{2}^2 +\frac{1}{N}\left\| \widehat{\mathbf{s}}_l \right\|_{2}^2  \right|{E_b}=1 \right]-{{d}_l} \\
&=\mathbb{E}\left[ \left. \frac{1}{N}\left\| \widehat{\mathbf{r}}_l \right\|_{2}^2  \right|{E_b}=1 \right]+\mathbb{E}\left[ \left. \frac{1}{N}\left\| \widehat{\mathbf{s}}_l \right\|_{2}^2  \right|{E_b}=1 \right]-{{d}_l} \\
&\overset{(a)}{\mathop{=}}\,\mathbb{E}\left[ \frac{1}{N}\left\| \mathbf{c}_l(0) \right\|_{2}^2  \right]+\mathbb{E}\left[ \left. \frac{1}{N}\left\| \widehat{\mathbf{s}}_l \right\|_{2}^2  \right|{E_b}=1 \right]-{{d}_l} \\
&=(\widehat{\sigma}_l^2-d_l) +\mathbb{E}\left[ \left. \frac{1}{N}\left\| \widehat{\mathbf{s}}_l \right\|_{2}^2  \right|{E_b}=1 \right]-{{d}_l} \\
&=\widehat{\sigma }_l^2 -2d_l +\mathbb{E}\left[ \left. \frac{1}{N}\left\| \widehat{\mathbf{s}}_l \right\|_{2}^2  \right|{E_b}=1 \right].
\end{split}
\end{equation}
where (a) holds because $\mathbf{c}_l(0)$ is independent of ${E_b}$. Combining~\eqref{dist_decomp}-\eqref{dist_decomp_2} and \eqref{upp_var_der2}, we get
\begin{equation}
\begin{split}
  & \left| \mathbb{E}\left[ \frac{1}{N}\left\| \widehat{\mathbf{r}}_l-\widehat{\mathbf{s}}_l \right\|_{2}^2  \right]-{{d}_l}  \right|\\
  &<\Pr \left( {E_l}=1 \right)\cdot \left[ (\widehat{\sigma }_l^2 -2d_l)+\mathbb{E}\left[ \left. \frac{1}{N}\left\| \widehat{\mathbf{s}}_l \right\|_{2}^2  \right|{E_l}=1 \right] \right]
 +\Pr \left( {E_l}=0 \right){{\varepsilon }_{lN}} \\
 & <{{\beta }_{lN}}(\widehat{\sigma }_l^2 -2d_l)+{{\alpha }_{lN}}+(1-{{\beta }_{lN}}){{\varepsilon }_{lN}}=:{{\eta }_{lN}},
\end{split}
\end{equation}
where $\lim_{N\to\infty}{{\eta }_{lN}}=0$, which can be readily verified from $\lim_{N\to\infty}\alpha_{lN},\beta_{lN},\varepsilon_{lN}=0$. Until now, we have proved~\eqref{var_bounded_eqn_1}, \eqref{var_bounded_eqn_2} and \eqref{typical_small_distortion_1} for a leaf $v_l$.

For a non-leaf node $v_b$, we only prove~\eqref{var_bounded_eqn_1}, because the proof of \eqref{var_bounded_eqn_2} and \eqref{typical_small_distortion_1} is exactly the same as the proof for the case of a leaf, provided that~\eqref{var_bounded_eqn_1} holds. In what follows, we assume that~\eqref{var_bounded_eqn_1}, \eqref{var_bounded_eqn_2} and \eqref{typical_small_distortion_1} hold for all children $v_1, v_2,\ldots v_d$ of a non-leaf node $v_b$. Since at the non-leaf node $v_b$,
\begin{equation}
  \widehat{\mathbf{s}}_b=\sum\limits_{k=1}^{d}{\widehat{\mathbf{r}}_k}+w_b\mathbf{x}_b,
\end{equation}
we have that
\begin{equation}
  \mathbb{E}\left[ \frac{1}{N}\left\| \widehat{\mathbf{s}}_b \right\|_{2}^2  \right]=\sum\limits_{k=1}^{d}{\mathbb{E}\left[ \frac{1}{N}\left\| \widehat{\mathbf{r}}_k \right\|_{2}^2  \right]}+w_b^2 .
\end{equation}
Using the variance relation~\eqref{var_induction} and the fact that~\eqref{var_bounded_eqn_2} holds for all children of $v_b$, we obtain that \eqref{var_bounded_eqn_1} holds for $v_b$.

\subsection{Proof of Lemma~\ref{ent_bounded}}\label{ent_var_bounded}
To prove Lemma~\ref{ent_bounded}, we first prove that the following divergence-bounds hold for all nodes $v_b$:
\begin{equation}\label{div_bounded_eqn_1}
  D\left( \left. {p_{\widehat{\mathbf{s}}_b}} \right\|\phi_{{U^\text{TC}_b}}^N \right)<N{{\gamma }_{b,N}},
\end{equation}
\begin{equation}\label{div_bounded_eqn_2}
  D\left( \left. {p_{\widehat{\mathbf{r}}_b}} \right\|\phi_{{V^\text{TC}_b}}^N \right)<N{{\widetilde{\gamma }}_{b,N}},
\end{equation}
where ${p_{\widehat{\mathbf{s}}_b}}$ and $ {p_{\widehat{\mathbf{r}}_b}}$ are the pdfs of $\widehat{\mathbf{s}}_b$ and $\widehat{\mathbf{r}}_b$, $\phi_{{U^\text{TC}_b}}^N$ and $\phi_{{V^\text{TC}_b}}^N$ are the $N$-fold products of pdfs $\phi_{U^\text{TC}_b}$ and $\phi_{V^\text{TC}_b}$, which are the pdfs of the test-channel random variables $U^\text{TC}_b$ and $V^\text{TC}_b$ that are defined in Section~\ref{test_channel}, and ${{\gamma }_{b,N}}$ and ${{\widetilde{\gamma }}_{b,N}}$ are two small constants such that $\lim_{N\to\infty}{\gamma }_{b,N}=0$ and $\lim_{N\to\infty}\widetilde{\gamma }_{b,N}=0$. In order to prove \eqref{div_bounded_eqn_1} and \eqref{div_bounded_eqn_2} for all nodes, we first prove the following three statements.

Statement 1: Inequality \eqref{div_bounded_eqn_1} holds for all leaves.

Statement 2: If~\eqref{div_bounded_eqn_1} holds at an arbitrary node $v_b$, then~\eqref{div_bounded_eqn_2} also holds at node $v_b$.

Statement 3: If~\eqref{div_bounded_eqn_2} holds at all children of a non-leaf node $v_b$, then~\eqref{div_bounded_eqn_1} holds at $v_b$.

These three statements together can be used to prove~\eqref{div_bounded_eqn_1} and~\eqref{div_bounded_eqn_2} for all nodes in the graph using induction in the tree (see Remark~\ref{induction_on_tree}).

\subsubsection{Proof of Statement 1}
At an arbitrary leaf $v_l$, according to the encoding scheme, the estimate $\widehat{\mathbf{s}}_l=w_l \mathbf{x}_l$ is an $N$-dimensional Gaussian random vector, each entry of which has pdf $\phi_{U^\text{TC}_l}$ (which is the pdf of the test-channel-based random variable $U^\text{TC}_l$). Therefore, at $v_l$, we have $ {p_{\widehat{\mathbf{s}}_l}} =\phi_{{U^\text{TC}_l}}^N$. So the first statement is true, since the KL-divergence is zero.

\subsubsection{Proof of Statement 2}
Denote by ${p_{\widehat{\mathbf{r}}_b\left| \widehat{\mathbf{s}}_b \right.}} $ the conditional distribution of $\widehat{\mathbf{r}}_b$ given $\widehat{\mathbf{s}}_b$, and by $\phi_{{V^\text{TC}_b}\left| {U^\text{TC}_b} \right.}^N $ the $N$-fold product of the conditional distribution $\phi_{{V^\text{TC}_b}\left| {U^\text{TC}_b} \right.}$ of the test-channel-based random variables. Suppose~\eqref{div_bounded_eqn_1} holds at $v_b$, we will prove that \eqref{div_bounded_eqn_2} also holds at $v_b$.
\begin{lemma}\label{div_lemma}
For each node $v_b$,
\begin{equation}
  \frac{1}{N}D\left( \left. {p_{\widehat{\mathbf{r}}_b\left| \widehat{\mathbf{s}}_b \right.}} \right\|\phi_{{V^\text{TC}_b}\left| {U^\text{TC}_b} \right.}^N \right)<\eta_{b,N},
\end{equation}
where $\lim_{N\to \infty}\eta_{b,N}=0$.
\end{lemma}
\begin{IEEEproof}
See Appendix~\ref{pf_div_lemma}.
\end{IEEEproof}
Using the chain rule of KL-divergence, we can expand $D\left( \left. {p_{\widehat{\mathbf{s}}_b,\widehat{\mathbf{r}}_b}} \right\|\phi_{{U^\text{TC}_b},{V^\text{TC}_b}}^N \right)$ in the following two ways:
\begin{equation}
\begin{split}
  &D\left( \left. {p_{\widehat{\mathbf{s}}_b,\widehat{\mathbf{r}}_b}} \right\|\phi_{{U^\text{TC}_b},{V^\text{TC}_b}}^N \right)\\
  =&D\left( \left. {p_{\widehat{\mathbf{s}}_b}} \right\|\phi_{{U^\text{TC}_b}}^N \right)+D\left( \left. {p_{\widehat{\mathbf{r}}_b\left| \widehat{\mathbf{s}}_b \right.}} \right\|\phi_{{V^\text{TC}_b}\left| {U^\text{TC}_b} \right.}^N \right)\\
  =&D\left( \left. {p_{\widehat{\mathbf{r}}_b}} \right\|\phi_{{V^\text{TC}_b}}^N \right)+D\left( \left. {p_{\widehat{\mathbf{s}}_b\left| \widehat{\mathbf{r}}_b \right.}} \right\|\phi_{{U^\text{TC}_b}\left| {V^\text{TC}_b} \right.}^N \right).
\end{split}
\end{equation}
Therefore, using Lemma~\ref{div_lemma} and using the induction assumption that~\eqref{div_bounded_eqn_1} holds at $v_b$, we have that
\begin{equation}
\begin{split}
  D\left( \left. {p_{\widehat{\mathbf{r}}_b}} \right\|\phi_{{V^\text{TC}_b}}^N \right)
  &\le D\left( \left. {p_{\widehat{\mathbf{s}}_b}} \right\|\phi_{{U^\text{TC}_b}}^N \right)+D\left( \left. {p_{\widehat{\mathbf{r}}_b\left| \widehat{\mathbf{s}}_b \right.}} \right\|\phi_{{V^\text{TC}_b}\left| {U^\text{TC}_b} \right.}^N \right)
  \le N\left( {{\widetilde{\eta }}_{b,N}}+{{\gamma }_{b,N}} \right).
\end{split}
\end{equation}
Defining ${\widetilde{\gamma }_{b,N}}={{\widetilde{\eta }}_{b,N}}+{{\gamma }_{b,N}}$, we can show that Statement 2 is true.

\subsubsection{Proof of Statement 3} To prove this statement, we need the following lemma:
\begin{lemma}\label{sum_reduces_KL}
Denote by $\mathbf{x}$ and $\mathbf{y}$ two absolutely continuous and independent random vectors supported in $\mathbb{R}^N$, and denote by $p_\mathbf{x}(\cdot)$ and $p_\mathbf{y}(\cdot)$ the pdfs of $\mathbf{x}$ and $\mathbf{y}$. Denote by $p_{\mathbf{x}+\mathbf{y}}(\cdot)$ the pdf of $\mathbf{x}+\mathbf{y}$. We know that $p_{\mathbf{x}+\mathbf{y}}(\cdot)$ is the convolution of $p_\mathbf{\mathbf{x}}(\cdot)$ and $p_\mathbf{y}(\cdot)$. Then, if there exist two distribution functions $q_{\mathbf{x}'}(\cdot)$ and $q_{\mathbf{y}'}(\cdot)$ of two other independent random variables $\mathbf{x}'$ and $\mathbf{y}'$ such that
\begin{equation}
  D\left(p_\mathbf{x}\|q_{\mathbf{x}'}\right)<\epsilon_1,
\end{equation}
\begin{equation}
  D\left(p_\mathbf{y}\|q_{\mathbf{y}'}\right)<\epsilon_2,
\end{equation}
we have
\begin{equation}
  D\left(p_{\mathbf{x}+\mathbf{y}}\|q_{\mathbf{x}'+\mathbf{y}'}\right)<\epsilon_1+\epsilon_2,
\end{equation}
where $q_{\mathbf{x}'+\mathbf{y}'}$ is the convolution of $q_{\mathbf{x}'}$ and $q_{\mathbf{y}'}$, which is also the pdf of the random variable $\mathbf{x}'+\mathbf{y}'$.
\end{lemma}
\begin{IEEEproof}
Using the chain rule of KL-divergence, we can expand $D(p_{\mathbf{x}+\mathbf{y},\mathbf{x}}\|q_{\mathbf{x}'+\mathbf{y}',\mathbf{x}'})$ in the following two ways:
\begin{equation}\label{D_der1}
\begin{split}
  &D(p_{\mathbf{x}+\mathbf{y},\mathbf{x}}\|q_{\mathbf{x}'+\mathbf{y}',\mathbf{x}'})\\
  =&D(p_{\mathbf{x}+\mathbf{y}}\|q_{\mathbf{x}'+\mathbf{y}'})+D(p_{\mathbf{x}|\mathbf{x}+\mathbf{y}}\|q_{\mathbf{x}'|\mathbf{x}'+\mathbf{y}'})\\
  =&D(p_{\mathbf{x}}\|q_{\mathbf{x}'})+D(p_{\mathbf{x}+\mathbf{y}|\mathbf{x}}\|q_{\mathbf{x}'+\mathbf{y}'|\mathbf{x}'}).
\end{split}
\end{equation}
We denote by $B(x,\delta)$ the $N$-dimensional ball centered at $x$ with volume $\delta$. Then, when $\mathbf{x}$ and $\mathbf{y}$ are independent, for a small constant $\delta$,
\begin{equation}
\begin{split}
&\Pr(\mathbf{x}+\mathbf{y}\in B(x+y,\delta)|\mathbf{x}=x)\\
=&\Pr(\mathbf{y}\in B(y,\delta)|\mathbf{x}=x)\\
=&\Pr(\mathbf{y}\in B(y,\delta)),
\end{split}
\end{equation}
where the conditional probability, such as $\Pr(A|\mathbf{x}=x)$, is defined in the sense of regular conditional probability \cite{jacod2013limit}, which can be written as
\begin{equation}
  \Pr(A|\mathbf{x}=x)=\lim_{m\to\infty}\frac{\Pr(A\cap U_m)}{\Pr(U_m)},
\end{equation}
where $U_1\supset U_2\supset U_3\ldots$ is a sequence of sets such that $\{\mathbf{x}=x\}\subset U_m,\forall m$ and
\begin{equation}
  \lim_{m\to\infty}\text{vol}(U_m)=0.
\end{equation}
The regular conditional probabilities (and densities) exist because the random variables are absolutely continuous and take values in Polish spaces (complete and separable metric spaces).
Therefore, we have that $p_{\mathbf{x}+\mathbf{y}|\mathbf{x}=x}(x+y)=p_{\mathbf{y}}(y)$. Similarly, $q_{\mathbf{x}'+\mathbf{y}'|\mathbf{x}'=x}(x+y)=q_{\mathbf{y}'}(y)$. Therefore,
\begin{equation}
  D(p_{\mathbf{x}+\mathbf{y}|\mathbf{x}=x}\|q_{\mathbf{x}'+\mathbf{y}'|\mathbf{x}'=x})=D(p_{\mathbf{y}}\|q_{\mathbf{y}'}),\forall x.
\end{equation}
Therefore, \eqref{D_der1} changes to
\begin{equation}
\begin{split}
  &D(p_{\mathbf{x}+\mathbf{y}}\|q_{\mathbf{x}'+\mathbf{y}'})+D(p_{\mathbf{x}|\mathbf{x}+\mathbf{y}}\|q_{\mathbf{x}'|\mathbf{x}'+\mathbf{y}'})
  <D(p_{\mathbf{x}}\|q_{\mathbf{x}'})+D(p_{\mathbf{y}}\|q_{\mathbf{y}'}).
\end{split}
\end{equation}
Noticing that $D(p_{\mathbf{x}|\mathbf{x}+\mathbf{y}}\|q_{\mathbf{x}'|\mathbf{x}'+\mathbf{y}'})>0$, we have
\begin{equation}
  D(p_{\mathbf{x}+\mathbf{y}}\|q_{\mathbf{x}'+\mathbf{y}'})<D(p_{\mathbf{x}}\|q_{\mathbf{x}'})+D(p_{\mathbf{y}}\|q_{\mathbf{y}'}),
\end{equation}
which concludes the proof.
\end{IEEEproof}
Now we prove Statement 3. Based on the induction assumption, suppose that for all child nodes $v_i$ of $v_b$, $1\le i\le d$,
\begin{equation}
  \frac{1}{N}D\left( \left. {p_{\widehat{\mathbf{r}}_i}} \right\|\phi_{{V^\text{TC}_i}}^N \right)<{{\widetilde{\gamma }}_{i,N}}.
\end{equation}
Considering
\begin{itemize}
  \item Gaussian random codes:
  \begin{equation}
    \widehat{\mathbf{s}}_b=\sum\limits_{i=1}^{d} \widehat{\mathbf{r}}_i+w_b\mathbf{x}_b,
  \end{equation}
  \item Test-channel Random Variables:
  \begin{equation}
    {U^\text{TC}_b}=\sum\limits_{i=1}^{d}{{V^\text{TC}_i}}+w_bX_b,
  \end{equation}
\end{itemize}
(see equation~\eqref{U_b_hat} and \eqref{non_leaf_estimator}) and using Lemma~\ref{sum_reduces_KL}, we have that
\begin{equation}
\begin{split}
\frac{1}{N}D\left( \left. {p_{\widehat{\mathbf{s}}_b}} \right\|\phi_{{U^\text{TC}_b}}^N \right)<&\sum_{i=1}^d \frac{1}{N} D\left( \left. {p_{\widehat{\mathbf{r}}_i}} \right\|\phi_{{V^\text{TC}_i}}^N \right)
<\sum_{i=1}^d{{\widetilde{\gamma }}_{i,N}}=:{{{\gamma }}_{b,N}},
\end{split}
\end{equation}
which concludes the proof of Statement 3.

\subsubsection{Using Statement 1-3 to Prove Lemma~\ref{ent_bounded}} We only provide the proof for \eqref{ent_bounded_eqn_1} (the first inequality in Lemma~\ref{ent_bounded}) using the divergence bound \eqref{div_bounded_eqn_1} because the proof for \eqref{ent_bounded_eqn_2} using \eqref{div_bounded_eqn_2} is exactly the same.

To simplify notation, we use $p(\cdot)$ and $q(\cdot)$ to denote ${p_{\widehat{\mathbf{s}}_b}}(\cdot)$ and $\phi_{{U^\text{TC}_b}}^N(\cdot)$. Then, by definition, we have that
\begin{equation}\label{q_form_2}
 q({{x}^N})=\frac{1}{{{\left( \sqrt{2\pi }{{{\widehat{\sigma }}}_b} \right)}^N}}\exp \left( -\frac{\left\| {{x}^N} \right\|_{2}^2 }{2\widehat{\sigma }_b^2 } \right).
\end{equation}
and
\begin{equation}\label{q_entropy_2}
  h(q)=\frac{N}{2}{\log_2 }2\pi e\widehat{\sigma}_b^2.
\end{equation}

The difference between $h(p)$ and $h(q)$ is
\begin{equation}\label{entropy_to_MMSE_2}
\begin{split}
  & h(p)-h(q)=-\int_{x\in R^N}{p\log pdx}+\int_{x\in R^N}{q\log qdx} \\
 & =-\int_{x\in R^N}{p\log \frac{p}{q}dx}+\int_{x\in R^N}{(q-p)\log qdx} \\
 & \overset{(a)}{=}-D\left( p||q \right)+{\log_2 }e\int_{x\in R^N}{(q-p)\left( -\frac{{\left\|{x}\right\|_2^2 }}{2\widehat{\sigma}_b^2 } \right)dx} \\
 & =-D\left( p||q \right)+\frac{{\log_2 }e}{2\widehat{\sigma}_b^2 }\mathbb{E}\left[ \left\|\widehat{\mathbf{s}}_b\right\|_2^2 -N\widehat{\sigma}_b^2 \right],
\end{split}
\end{equation}
{where we used~\eqref{q_form_2} in step (a)}. The first term of the RHS can be bounded by the divergence bound~\eqref{div_bounded_eqn_1} and the second term of the RHS can be bounded by Lemma~\ref{var_bounded}:
\begin{equation}\label{der11_2}
  \left|\mathbb{E}\left[ \left\|\widehat{\mathbf{s}}_b\right\|_2^2  \right]-N\widehat{\sigma}_b^2\right|<N\varepsilon_N,
\end{equation}
where $\lim_{N\to \infty} \varepsilon_N=0$.
Therefore, {combining~\eqref{div_bounded_eqn_1} and \eqref{q_entropy_2}-\eqref{der11_2}}, we get
\begin{equation}\label{diff_in_entro_2}
\begin{split}
&h(p)\ge h(q)-D\left( p||q \right)+\frac{{\log_2 }e}{2\widehat{\sigma}_b^2 }\mathbb{E}\left[ \left\|\widehat{\mathbf{s}}_b\right\|_2^2 -N\widehat{\sigma}_b^2 \right]\\
&>\frac{N}{2}{\log_2 }2\pi e\widehat{\sigma}_b^2-N\gamma_{b,N}-\frac{{\log_2 }e}{2\widehat{\sigma}_b^2 }\varepsilon_N N.
\end{split}
\end{equation}
By defining $\beta_N=\max_{1\le b\le n}\gamma_{b,N}+\frac{{\log_2 }e}{2\widehat{\sigma}_b^2 }\varepsilon_N$, we conclude that \eqref{ent_bounded_eqn_1} is true.
\subsection{Proof of Lemma~\ref{distortion_close_lemma}}\label{pf_distortion}
\setcounter{subsubsection}{0}
To simplify notation, for an arbitrary node $v_b$ and its parent node $v_a=v_{\text{PN}(b)}$ define
\begin{align}
  & \widehat{\mathbf{s}}_b^{*}=\widehat{\mathbf{y}}^\text{mmse}_{\mathcal{S}_b,b}, \\
 & \widehat{\mathbf{r}}_b^{*}=\widehat{\mathbf{y}}^\text{mmse}_{\mathcal{S}_b,\text{PN}(b)}.
\end{align}
Therefore, $\widehat{\mathbf{s}}_b^{*}$ is the MMSE estimate of the partial sum $\mathbf{y}_{\mathcal{S}_b}$ at the node $v_b$, while $\widehat{\mathbf{r}}_b^{*}$ is the MMSE estimate of the same variable, but at the parent-node ${v_{\text{PN}(b)}}$. In order to relate the Gaussian-code-based distortion and the MMSE-based distortion, we will prove that, the estimates based on the Gaussian code, i.e., the estimate $\widehat{\mathbf{s}}_b$ and the description $\widehat{\mathbf{r}}_b$, are very close to the MMSE estimates $\widehat{\mathbf{s}}_b^{*}$ and $\widehat{\mathbf{r}}_b^{*}$ in the sense of mean-square error\footnote{Note that according to the intuitive explanation on test channels (see Remark~\ref{remark5}), the estimates based on the Gaussian code and the MMSE estimations are indeed equal to each other when Gaussian test channels can be physically established.}. We prove that as long as $N$ is finite but sufficiently large, the gap between these two types of estimators can be arbitrarily small. Define
\begin{align}
  & {\Delta_b^\text{Tx}}=\mathbb{E}\left[ \frac{1}{N}\left\| \widehat{\mathbf{s}}_b-\widehat{\mathbf{s}}_b^{*} \right\|_{2}^2  \right], \\
 & {\Delta_b^\text{Rx}}=\mathbb{E}\left[ \frac{1}{N}\left\| \widehat{\mathbf{r}}_b-\widehat{\mathbf{r}}_b^{*} \right\|_{2}^2  \right].
\end{align}
We will prove that ${\Delta_b^\text{Tx}}\to 0$ and $\Delta_b^\text{Rx}\to 0$ when $N\to \infty$. In particular, we will prove the following three statements:

Statement 1: For an arbitrary leaf $v_l$,
\begin{equation}
  \Delta_l^\text{Tx}=0.
\end{equation}

Statement 2: For an arbitrary non-leaf node $v_b$ and its $d$ children $v_1,\ldots v_d$ (see Fig.~\ref{cut_set_figure}),
\begin{equation}\label{Delta_relation_1_1}
\sqrt{\Delta_b^\text{Tx}}\le\sum\limits_{k=1}^{d}\sqrt{\Delta_k^\text{Rx}}.
\end{equation}

Statement 3: For an arbitrary node $v_b$,
\begin{equation}
\sqrt{\Delta_b^\text{Rx}}\le \sqrt{\theta_N}+\sqrt{\Delta_b^\text{Tx}},
\end{equation}
where $\lim_{N\to \infty} \theta_N=0$.
\subsubsection{Proof of Statement 1}
For a leaf $v_l$, the random-coding-based estimate is $\widehat{\mathbf{s}}_l=w_l \mathbf{x}_l$, which is exactly the same as the MMSE estimate $\widehat{\mathbf{s}}_l^*$, since $\mathbf{x}_l$ is known to $v_l$. Therefore, $\Delta_l^\text{Tx}=0$.

\subsubsection{Proof of Statement 2}
For a non-leaf node $v_b$ and its children, we have that (see \eqref{U_b_hat})
\begin{equation}\label{dis_der1}
  \widehat{\mathbf{s}}_b=\sum\limits_{k=1}^{d}{\mathbf{c}_k(M_{k\to b})}+w_b\mathbf{x}_b=\sum\limits_{k=1}^{d}{\widehat{\mathbf{r}}_b}+w_b\mathbf{x}_b.
\end{equation}
Since the partial sum $\mathbf{y}_{\mathcal{S}_b}=\sum\limits_{k=1}^{d}{\mathbf{y}_{\mathcal{S}_k}}+w_b\mathbf{x}_b$, we have that
\begin{equation}\label{dis_der2}
\begin{split}
  \widehat{\mathbf{s}}_b^{*}=&\mathbb{E}\left[ \mathbf{y}_{\mathcal{S}_b}\left|I_b\right. \right]=\sum\limits_{k=1}^{d}{\mathbb{E}\left[ \mathbf{y}_{\mathcal{S}_k}\left|I_b\right. \right]}+w_b\mathbf{x}_b
  =\sum\limits_{k=1}^{d}{\widehat{\mathbf{r}}_k^{*}}+w_b\mathbf{x}_b.
\end{split}
\end{equation}
Thus, combining~\eqref{dis_der1} and~\eqref{dis_der2}, we get
\begin{equation}\label{Delta_relation_1}
  {\Delta_b^\text{Tx}}=\left[ \left\| \widehat{\mathbf{s}}_b-\widehat{\mathbf{s}}_b^{*} \right\|_{2}^2  \right]=\sum\limits_{k=1}^{d}{\mathbb{E}\left[ \left\| \widehat{\mathbf{r}}_k-\widehat{\mathbf{r}}_k^{*} \right\|_{2}^2  \right]}=\sum\limits_{k=1}^{d}{\Delta_k^\text{Rx}},
\end{equation}
which can be further relaxed by
\begin{equation}\label{Delta_relation_1_change}
   \sqrt{\Delta_b^\text{Tx}}<\sum\limits_{k=1}^{d}\sqrt{\Delta_k^\text{Rx}}.
\end{equation}
\subsubsection{Proof of Statement 3}
Note that by~\eqref {typical_small_distortion_1}, we have
\begin{equation}\label{dis_der3}
\mathbb{E}\left[ \frac{1}{N}\left\| \widehat{\mathbf{s}}_b-\widehat{\mathbf{r}}_b \right\|_{2}^2  \right]\le d_b +{{\varepsilon }_{N}}.
\end{equation}
Define $\text{Dist}_b=\mathbb{E}_{\mathcal{C}_b}\left[ \frac{1}{N}\left\| \mathbb{E}_{\mathcal{C}_b}\left[ \widehat{\mathbf{s}}_b\left|\widehat{\mathbf{r}}_b\right. \right]-\widehat{\mathbf{s}}_b \right\|_{2}^2  \right]$. We will prove that $\text{Dist}_b$ is approximately greater than $d_b $ (the explicit form is in \eqref{dis_der4}), which means that even the MMSE estimate $\mathbb{E}_{\mathcal{C}_b}\left[ \widehat{\mathbf{s}}_b\left|\widehat{\mathbf{r}}_b\right. \right]$ cannot provide a much better description (in the sense of mean-square error) of $\widehat{\mathbf{s}}_b$ than the typicality-based estimate $\widehat{\mathbf{r}}_b$. Notice that the MMSE estimate $\mathbb{E}_{\mathcal{C}_b}\left[ \widehat{\mathbf{s}}_b|\widehat{\mathbf{r}}_b\right] $ here should be defined for the chosen codebook $\mathcal{C}_b$ at $v_b$, since the receiver $v_a$ also knows the codebook. The outer $\mathbb{E}$ in $\text{Dist}_b=\mathbb{E}_{\mathcal{C}_b}\left[ \frac{1}{N}\left\| \mathbb{E}_{\mathcal{C}_b}\left[ \widehat{\mathbf{s}}_b\left|\widehat{\mathbf{r}}_b\right. \right]-\widehat{\mathbf{s}}_b \right\|_{2}^2  \right]$ is also conditioned on a given codebook $\mathcal{C}_b$ at node $v_b$. From~\eqref{r_up} we have that
\begin{equation}
\begin{split}
\frac{N}{2}\log \frac{\widehat{\sigma}_b^2}{d_b}+N\delta_N
=&NR_b\\
\overset{(a)}{\ge}&I(\widehat{\mathbf{s}}_b;\widehat{\mathbf{r}}_b)\\
\overset{(b)}{\ge}&I\left(\widehat{\mathbf{s}}_b;\mathbb{E}_{\mathcal{C}_b}\left[ \widehat{\mathbf{s}}_b\left|\widehat{\mathbf{r}}_b\right. \right]\right)\\
=&h(\widehat{\mathbf{s}}_b)-h\left(\widehat{\mathbf{s}}_b|\mathbb{E}_{\mathcal{C}_b}\left[ \widehat{\mathbf{s}}_b\left|\widehat{\mathbf{r}}_b\right. \right]\right)\\
=&h(\widehat{\mathbf{s}}_b)-h\left(\widehat{\mathbf{s}}_b-\mathbb{E}_{\mathcal{C}_b}\left[ \widehat{\mathbf{s}}_b\left|\widehat{\mathbf{r}}_b\right. \right]|\mathbb{E}_{\mathcal{C}_b}\left[ \widehat{\mathbf{s}}_b\left|\widehat{\mathbf{r}}_b\right. \right]\right)\\
\ge&h(\widehat{\mathbf{s}}_b)-h\left(\widehat{\mathbf{s}}_b-\mathbb{E}_{\mathcal{C}_b}\left[ \widehat{\mathbf{s}}_b\left|\widehat{\mathbf{r}}_b\right. \right]\right)\\
\overset{(c)}{>}&\frac{N}{2}\log_2 2\pi e \widehat{\sigma}_b^2-N\beta_N-\frac{N}{2}\log_2 2\pi e \text{Dist}_b,
\end{split}
\end{equation}
where (a) follows from the cut set bound, (b) follows from the data processing inequality, and (c) follows from Lemma \ref{ent_bounded}. Notice that although the codebook $\mathcal{C}_b$ is fixed, other codebooks are not fixed, so the random vector $h(\widehat{\mathbf{s}}_b)$ still satisfies Lemma \ref{ent_bounded}. Therefore,
\begin{equation}\label{dis_der4}
\text{Dist}_b>2^{-\delta_N-\beta_N}d_b=(1-\epsilon_N)d_b,
\end{equation}
where $\lim_{N\to\infty} \epsilon_N=0$. \textcolor{black}{Since the inequality \eqref{dis_der4} holds for any given codebook $\mathcal{C}_b$, \eqref{dis_der4} also holds for the entire random codebook ensemble, in which case the outside $\mathbb{E}$ is again taken over the random codebook generation (which is in alignment with the definitions of other mean-square distortions in other parts of this section and all other sections).} Combining~\eqref{dis_der3} and \eqref{dis_der4} and the orthogonality principle
\[\left(\mathbb{E}\left[ \widehat{\mathbf{s}}_b\left|\widehat{\mathbf{r}}_b \right.\right]-\widehat{\mathbf{s}}_b\right)\bot \left(\widehat{\mathbf{r}}_b-\mathbb{E}\left[ \widehat{\mathbf{s}}_b\left|\widehat{\mathbf{r}}_b \right.\right]\right),\]
we get
\begin{equation}\label{dis_der5}
\begin{split}
  \mathbb{E}\left[ \frac{1}{N}\left\| \mathbb{E}\left[ \widehat{\mathbf{s}}_b\left|\widehat{\mathbf{r}}_b\right. \right]-\widehat{\mathbf{r}}_b \right\|_{2}^2  \right]
  \le& d_b+{{\varepsilon }_{N}}-(1-\epsilon_N)d_b
  ={{\varepsilon }_{N}}+\epsilon_Nd_b =:\theta_N,
\end{split}
\end{equation}
where $\lim_{N\to\infty}\theta_N=0$. Further, we have that
\begin{equation}
\begin{split}
\widehat{\mathbf{r}}_b^{*}&=\mathbb{E}\left[ \mathbf{y}_{\mathcal{S}_b}\left|{{I}_{\text{PN}(b)}} \right.\right]=\mathbb{E}\left[ \mathbf{y}_{\mathcal{S}_b}\left|\widehat{\mathbf{r}}_b \right.\right]
\overset{(a)}{\mathop{=}}\,\mathbb{E}\left[ \mathbb{E}\left[ \mathbf{y}_{\mathcal{S}_b}|I_b \right]\left|\widehat{\mathbf{r}}_b \right.\right]=\mathbb{E}\left[ \widehat{\mathbf{s}}_b^{*}\left|\widehat{\mathbf{r}}_b \right.\right],
\end{split}
\end{equation}
where the equality (a) follows from the iterative expectation principle and the fact that $\widehat{\mathbf{r}}_b$ is a function of $I_b$. Therefore
\begin{equation}\label{dis_der6}
\begin{split}
    & \mathbb{E}\left[ \left\| \mathbb{E}\left[ \widehat{\mathbf{s}}_b\left|\widehat{\mathbf{r}}_b\right. \right]-\widehat{\mathbf{r}}_b^{*} \right\|_{2}^2  \right]=\mathbb{E}\left[ \left\| \mathbb{E}\left[ \widehat{\mathbf{s}}_b-\widehat{\mathbf{s}}_b^{*}\left|\widehat{\mathbf{r}}_b \right.\right] \right\|_{2}^2  \right] \\
 & \overset{(a)}{\mathop{\le }}\,\mathbb{E}\left[ \mathbb{E}\left[ \left\| \widehat{\mathbf{s}}_b-\widehat{\mathbf{s}}_b^{*} \right\|_{2}^2 \left|\widehat{\mathbf{r}}_b \right.\right] \right]=\mathbb{E}\left[ \left\| \widehat{\mathbf{s}}_b-\widehat{\mathbf{s}}_b^{*} \right\|_{2}^2  \right]=N{\Delta_b^\text{Tx}},
 \end{split}
\end{equation}
where inequality (a) follows from the Jensen's inequality. Thus, combining~\eqref{dis_der5} and \eqref{dis_der6} and using the triangle inequality, we get
\begin{equation}\label{Delta_relation_2}
\begin{split}
 \sqrt{{\Delta_b^\text{Rx}}}=&\sqrt{\mathbb{E}\left[ \frac{1}{N}\left\| \widehat{\mathbf{r}}_b^{*}-\widehat{\mathbf{r}}_b \right\|_{2}^2  \right]} \\
  \le &\sqrt{\mathbb{E}\left[ \frac{1}{N}\left\| \widehat{\mathbf{r}}_b^{*}-\mathbb{E}\left[ \widehat{\mathbf{s}}_b\left|\widehat{\mathbf{r}}_b \right.\right] \right\|_{2}^2  \right]}
 +\sqrt{\mathbb{E}\left[ \frac{1}{N}\left\| \widehat{\mathbf{r}}_b-\mathbb{E}\left[ \widehat{\mathbf{s}}_b\left|\widehat{\mathbf{r}}_b \right.\right] \right\|_{2}^2  \right]} \\
  \le& \sqrt{{{\theta }_{N}}}+\sqrt{{\Delta_b^\text{Tx}}}.
\end{split}
\end{equation}
\subsubsection{Using Statement 1-3 to Prove Lemma~\ref{distortion_close_lemma}}
Using the three statements and using induction on the tree, we have that
\begin{equation}\label{distortion_close_key}
  \underset{1\le b\le n}{\mathop{\sup }}\,\sqrt{{\Delta_b^\text{Tx}}}\le \underset{1\le b\le n}{\mathop{\sup }}\,\sqrt{{\Delta_b^\text{Rx}}}\le n\sqrt{{{\theta }_{N}}}.
\end{equation}
Thus, the conclusion~\eqref{distortion_close} can be obtained by combining the orthogonality principle
\begin{equation}
\begin{split}
  \mathbb{E}\left[ \frac{1}{N}\left\| \widehat{\mathbf{r}}_b^{*}-\widehat{\mathbf{s}}_b^{*} \right\|_{2}^2  \right]
  =&\mathbb{E}\left[ \frac{1}{N}\left\| \widehat{\mathbf{r}}_b^{*}-\mathbf{y}_{\mathcal{S}_b} \right\|_{2}^2  \right]-\mathbb{E}\left[ \frac{1}{N}\left\| \widehat{\mathbf{s}}_b^{*}-\mathbf{y}_{\mathcal{S}_b} \right\|_{2}^2  \right]
  ={D_b^\text{Rx}}-D_b^\text{Tx}
\end{split}
\end{equation}
and the triangle inequality, which is
\begin{equation}
\begin{split}
   \sqrt{\mathbb{E}\left[ \frac{1}{N}\left\| \widehat{\mathbf{r}}_b^{*}-\widehat{\mathbf{s}}_b^{*} \right\|_{2}^2  \right]}&\le \sqrt{\mathbb{E}\left[ \frac{1}{N}\left\| \widehat{\mathbf{s}}_b-\widehat{\mathbf{s}}_b^{*} \right\|_{2}^2  \right]}+\sqrt{\mathbb{E}\left[ \frac{1}{N}\left\| \widehat{\mathbf{r}}_b-\widehat{\mathbf{r}}_b^{*} \right\|_{2}^2  \right]}+\sqrt{\mathbb{E}\left[ \frac{1}{N}\left\| \widehat{\mathbf{r}}_b-\widehat{\mathbf{s}}_b \right\|_{2}^2  \right]} \\
 & \le \sqrt{d_b +{{\varepsilon }_{N}}}+2n\sqrt{{{\theta }_{N}}},
 \end{split}
\end{equation}
and
\begin{equation}
\begin{split}
   \sqrt{\mathbb{E}\left[ \frac{1}{N}\left\| \widehat{\mathbf{r}}_b^{*}-\widehat{\mathbf{s}}_b^{*} \right\|_{2}^2  \right]}&\ge \sqrt{\mathbb{E}\left[ \frac{1}{N}\left\| \widehat{\mathbf{s}}_b-\widehat{\mathbf{s}}_b^{*} \right\|_{2}^2  \right]}
  -\sqrt{\mathbb{E}\left[ \frac{1}{N}\left\| \widehat{\mathbf{r}}_b-\widehat{\mathbf{r}}_b^{*} \right\|_{2}^2  \right]}-\sqrt{\mathbb{E}\left[ \frac{1}{N}\left\| \widehat{\mathbf{r}}_b-\widehat{\mathbf{s}}_b \right\|_{2}^2  \right]} \\
 & \ge \sqrt{d_b -{{\varepsilon }_{N}}}-2n\sqrt{{{\theta }_{N}}}.
 \end{split}
\end{equation}

\subsection{Proof of Lemma~\ref{div_lemma}}\label{pf_div_lemma}
\setcounter{subsubsection}{0}

We use ${s^N}\in \mathbb{R}^N$ to denote one sample of the random vector $\widehat{\mathbf{s}}_b$, and use ${r^N} \in \mathbb{R}^N$ to denote one sample of the codeword (description) $\widehat{\mathbf{r}}_b$. We will show that the KL-divergence $D\left( \left. {p_{\widehat{\mathbf{r}}_b\left| \widehat{\mathbf{s}}_b \right.}} \right\|\phi_{{V^\text{TC}_b}\left| {U^\text{TC}_b} \right.}^N \right)$ is small. We will prove this statement using two steps:
\begin{itemize}
  \item When the estimate $\widehat{\mathbf{s}}_b=s^N$ is typical, $\frac{1}{N}D\left( \left. {p_{\widehat{\mathbf{r}}_b\left| \widehat{\mathbf{s}}_b \right.}} \right\|\phi_{{V^\text{TC}_b}\left| {U^\text{TC}_b} \right.}^N \right)$ is small.
  \item When the estimate $\widehat{\mathbf{s}}_b=s^N$ is not typical, $\frac{1}{N}D\left( \left. {p_{\widehat{\mathbf{r}}_b\left| \widehat{\mathbf{s}}_b \right.}} \right\|\phi_{{V^\text{TC}_b}\left| {U^\text{TC}_b} \right.}^N \right)$ is bounded.
\end{itemize}
\begin{remark}
These two steps only provide the intuition underlying the two major parts of the proof. The proof itself is rigorous.
\end{remark}

Denote by $\mathcal{T}_\epsilon^N$ the set of all $N$-length sequences ${s^N}$ that is typical with respect to ${\phi_U^\text{TC}}$. Denote by $\mathcal{J}_{\epsilon }^{2N}$ the set of all $2N$-length sequences $\left( {s^N},{r^N} \right)$ that are jointly typical with respect to ${\phi}_{U^\text{TC},V^\text{TC}}$. Denote by $\mathcal{T}_\epsilon^N(s^N)$ the set of sequences ${r^N}$ that are jointly typical with the typical sequence ${s^N}$. Notice that we define typical sets as the distortion typical set (weak typical set with Euclidean distortion) in \cite[Sec. 10.5]{Cover_Wiley_06}. Denote by $B({{x}^N},v)$ the $N$-dimensional ball centered at ${{x}^N}$ and with volume $v$. We simplify notation and omit the subscripts of all pdfs and use notation $p(\cdot )$ and $q(\cdot )$ to respectively denote typical-codes-based pdfs and test-channel-based pdfs. Note that the support set of the $N$-fold product pdf $\phi_{{V^\text{TC}_b}\left| {U^\text{TC}_b} \right.}^N$ is the entire $\mathbb{R}^N$ and it has no singular point and it does not vanish everywhere. Therefore, the ratio $\frac{p(\cdot)}{q(\cdot)}$ is always properly defined\footnote{\textcolor{black}{Here, `properly defined' means that $\frac{0}{0}$ or $\frac{\infty}{\infty}$ will not happen. This property only requires that the $N$-fold product pdf $\phi_{{V^\text{TC}_b}\left| {U^\text{TC}_b} \right.}^N$ is properly defined. In fact, based on the randomness of the generation of the codewords, it may be possible to prove a stronger result that the pdf ${p_{\widehat{\mathbf{r}}_b\left| \widehat{\mathbf{s}}_b \right.}}$ does not have any singular point as well.}}.

\subsubsection{Proof of the first statement}
when the estimate (source) ${s^N}$ is a typical sequence, i.e., when ${s^N}\in \mathcal{T}_{\epsilon }^N$,
\begin{equation}\label{div_dec}
\begin{split}
  D\left( p\left( {r^N}\left| {s^N} \right. \right)\left\| q\left( {r^N}\left| {s^N} \right. \right) \right. \right)
  =&\int_{{{\mathbb{R}}^N}}{p\left( {r^N}\left| {s^N} \right. \right)\log \frac{p\left( {r^N}\left| {s^N} \right. \right)}{q\left( {r^N}\left| {s^N} \right. \right)}}d{r^N}\\
  =&\int_{\mathcal{T}_{\epsilon }^N({s^N})}{p\left( {r^N}\left| {s^N} \right. \right)\log \frac{p\left( {r^N}\left| {s^N} \right. \right)}{q\left( {r^N}\left| {s^N} \right. \right)}}d{r^N}\\
  &+\int_{{{\mathbb{R}}^N}\backslash \mathcal{T}_{\epsilon }^N({s^N})}{p\left( {r^N}\left| {s^N} \right. \right)\log \frac{p\left( {r^N}\left| {s^N} \right. \right)}{q\left( {r^N}\left| {s^N} \right. \right)}}d{r^N}.
\end{split}
\end{equation}
We look at the first term on the RHS of~\eqref{div_dec}. When ${r^N}\in \mathcal{T}_{\epsilon }^N\left( {s^N} \right)$, since the sent codeword $\widehat{\mathbf{r}}_b$ is chosen to be an arbitrary codeword in $\mathcal{C}_b\setminus \{\mathbf{c}_b(0)\}$ that is jointly typical with $\widehat{\mathbf{s}}_b$ (see Section~\ref{Gaussian_code_alg}), there are two possible cases when the sent codeword is close\footnote{Since we compute the pdf in a continuous space, we have to compute the probability that the sent codeword is close to $r^N$ and then compute the limit when the ``distance'' approaches zero (see $\lim_{v\to 0}$ in \eqref{vto0}).} to ${r^N}$: there is at least one codeword in $\mathcal{C}_b\setminus \{\mathbf{c}_b(0)\}$ that is within the ball $B({r^N},v)$, or no codeword in $\mathcal{C}_b\setminus \{\mathbf{c}_b(0)\}$ is within $B({r^N},v)$ but the codeword $\{\mathbf{c}_b(0)\}$ (which is only sent when an error happens) is within $B({r^N},v)$. Therefore, when ${r^N}\in \mathcal{T}_{\epsilon }^N\left( {s^N} \right)$, we have that
\begin{equation}\label{vto0}
\begin{split}
  p\left( {r^N}\left| {s^N} \right. \right)
  &=\underset{v\to 0 }{\mathop{\lim }}\,\frac{1}{v}\Pr \left(\widehat{\mathbf{r}}_b\in B({r^N},v)\left| \widehat{\mathbf{s}}_b={s^N} \right. \right)\\
  & <\underset{v\to 0 }{\mathop{\lim }}\,\frac{1}{v}\Pr \left(\exists c^N\in\mathcal{C}_b\setminus \{\mathbf{c}_b(0)\}, \right.\\
  &{\;\;\;\;\;\;\;\;\;\;\;\;\;\;\;\;}\left.\text{ s.t. } c^N\in B({r^N},v)\left| \widehat{\mathbf{s}}_b=s^N \right. \right) \\
 & +\underset{v\to 0 }{\mathop{\lim }}\,\frac{1}{v}\Pr \left({{c}_b}(0)\in B({r^N},v)\left| \widehat{\mathbf{s}}_b=s^N \right. \right) \\
 &=\underset{v\to 0 }{\mathop{\lim }}\,\frac{1}{v}\left\{ 1-{{\left[ 1-q\left( {r^N} \right)v \right]}^{{{2}^{N{R_b}}}}}+q\left( {r^N} \right)v \right\}\\
 &=\underset{v\to 0 }{\mathop{\lim }}\,\frac{1}{v}\left\{ q\left( {r^N} \right) {2}^{N{R_b}}v+o(v)+q\left( {r^N} \right)v \right\}\\
 &=\left( {{2}^{N{R_b}}}+1 \right)q\left( {r^N} \right)\\
 &< {{2}^{N{R_b}+1}} q\left( {r^N} \right).
\end{split}
\end{equation}
\textcolor{black}{Here the conditional probability is also defined in the sense of regular conditional probability.} Also notice that in this case, since $\left( {s^N},{r^N} \right)\in \mathcal{J}_{\epsilon }^{2N}$, due to the weak typicality, we have\footnote{Notice that this typicality is defined for random variables with continuous alphabets, the details of which are provided in \cite[Sec. 8.2]{Cover_Wiley_06}.}
\begin{equation}\label{typ1}
  {{2}^{-N\left( h\left( {V^\text{TC}_b} \right)+\varepsilon_N  \right)}}\le q\left( {r^N} \right)\le {{2}^{-N\left( h\left( {V^\text{TC}_b} \right)-\varepsilon_N  \right)}},
\end{equation}
\begin{equation}\label{typ2}
  {{2}^{-N\left( h\left( {U^\text{TC}_b} \right)+\varepsilon_N  \right)}}\le q\left( {s^N} \right)\le {{2}^{-N\left( h\left( {U^\text{TC}_b} \right)-\varepsilon_N  \right)}},
\end{equation}
\begin{equation}\label{typ3}
  {{2}^{-N\left( h\left( {U^\text{TC}_b},{V^\text{TC}_b} \right)+\varepsilon_N  \right)}}\le q\left( {s^N},{r^N} \right)\le {{2}^{-N\left( h\left( {U^\text{TC}_b},{V^\text{TC}_b} \right)-\varepsilon_N  \right)}},
\end{equation}
where $\lim_{N\to \infty}\varepsilon_N=0$. Therefore,
\begin{equation}\label{div_der_1}
\begin{split}
  \int_{\mathcal{T}_{\epsilon }^N({s^N})}{p\left( {r^N}\left| {s^N} \right. \right)\log \frac{p\left( {r^N}\left| {s^N} \right. \right)}{q\left( {r^N}\left| {s^N} \right. \right)}}d{r^N}
  \overset{(a)}{<}&\int_{\mathcal{T}_{\epsilon }^N({s^N})}{p\left( {r^N}\left| {s^N} \right. \right)\log \frac{{{2}^{N{R_b}+1}}q\left( {r^N} \right)}{q\left( {r^N}\left| {s^N} \right. \right)}}d{r^N}\\
  =&\int_{\mathcal{T}_{\epsilon }^N({s^N})}{p\left( {r^N}\left| {s^N} \right. \right)\log \frac{ {{2}^{N{R_b}+1}} q\left( {r^N} \right)q\left( {s^N} \right)}{q\left( {s^N},{r^N} \right)}}d{r^N}\\
  \overset{(b)}{<}&N\left( {R_b}+\frac{1}{N}-I\left( {U^\text{TC}_b};{V^\text{TC}_b} \right)+3\varepsilon_N  \right)\\
  &\cdot\int_{\mathcal{T}_{\epsilon }^N({s^N})}{p\left( {r^N}\left| {s^N} \right. \right)}d{r^N}\\
  <&N\left( {R_b}+\frac{1}{N}-I\left( {U^\text{TC}_b};{V^\text{TC}_b} \right)+3\varepsilon_N  \right)\\
  \overset{(c)}{=}&N\left( {{\delta }_{N}}+\frac{1}{N}+3\varepsilon_N  \right),
\end{split}
\end{equation}
where step (a) follows from \eqref{vto0}, step (b) holds because when $r^N\in{\mathcal{T}_{\epsilon }^N({s^N})}$, \eqref{typ1}-\eqref{typ3} hold, and because
\begin{equation}
\begin{split}
\log \frac{{{2}^{N{R_b}+1}}q\left( {r^N} \right)q\left( {s^N} \right)}{q\left( {s^N},{r^N} \right)}
\le& \log \frac{{{2}^{N{R_b}+1}}{{2}^{-N\left( h\left( {V^\text{TC}_b} \right)-{{\varepsilon }_{N}} \right)}}{{2}^{-N\left( h\left( {U^\text{TC}_b} \right)-{{\varepsilon }_{N}} \right)}}}{{{2}^{-N\left( h\left( {U^\text{TC}_b},{V^\text{TC}_b} \right)+{{\varepsilon }_{N}} \right)}}}\\
=&N{R_b}+1-N\left( h\left( {V^\text{TC}_b} \right)+h\left( {U^\text{TC}_b} \right)-h\left( {U^\text{TC}_b},{V^\text{TC}_b} \right)-3\varepsilon_N  \right)\\
=&N{R_b}+1-N\left( I\left( {U^\text{TC}_b};{V^\text{TC}_b} \right)-3\varepsilon_N  \right)\\
=&N\left({R_b}+\frac{1}{N}- I\left( {U^\text{TC}_b};{V^\text{TC}_b} \right)+3\varepsilon_N \right),
\end{split}
\end{equation}
and ${{\delta }_{N}}$ in step (c) is defined in \eqref{r_up}, which says that $R_b=I(U^\text{TC}_b;V^\text{TC}_b)+\delta_N$.

Then, we look at the second term on the RHS of~\eqref{div_dec}. When the estimate (source) ${s^N}$ is a typical sequence but ${r^N}\notin \mathcal{T}_{\epsilon }^N\left( {s^N} \right)$, we have that
\begin{equation}
\begin{split}
  &p\left( {r^N}\left| {s^N} \right. \right)\\
  \overset{(a)}{=}&\lim_{v\to 0}\frac{1}{v}\Pr \left( \text{no codeword}\in\mathcal{C}_b\setminus\{\mathbf{c}_b(0)\} \text{ is jointly typical with }{s^N} \right)\cdot\Pr \left(\mathbf{c}_b(0)\in B({r^N},v)\right)\\
  =&\Pr \left( \text{no codeword}\in\mathcal{C}_b\setminus\{\mathbf{c}_b(0)\} \text{ is jointly typical with }{s^N} \right)\cdot\lim_{v\to 0}\frac{1}{v}\Pr \left(\mathbf{c}_b(0)\in B({r^N},v)\right)\\
  =&\Pr \left( \text{no codeword}\in\mathcal{C}_b\setminus\{\mathbf{c}_b(0)\} \text{ is jointly typical with }{s^N} \right)\cdot q\left( {r^N} \right),
\end{split}
\end{equation}
where (a) holds because the only case to obtain a codeword ${r^N}\notin \mathcal{T}_{\epsilon }^N\left( {s^N} \right)$ is when no codeword in the code ${{\mathcal{C}}_b}\backslash \left\{ {\mathbf{c}_b}(0) \right\}=\{\mathbf{c}_b(w):\ w\in \{1,2,\ldots {{2}^{N{R_i}}}\}\}$ is ${r^N}$ but the first codeword ${\mathbf{c}_b}(0)$ (only sent when an encoding error happens) is ${r^N}$. Define
\begin{equation}
  {{\lambda }_{b,N}}=\Pr \left( \text{no codeword}\in\mathcal{C}_b\setminus\{\mathbf{c}_b(0)\} \text{ is jointly typical with } s^N\right).
\end{equation}
Then, we have that $p\left( {r^N}\left| {s^N} \right. \right)={{\lambda }_{b,N}}q\left( {r^N} \right)$. From the covering lemma (see Lemma~\ref{covering_lemma}), we have that $\underset{N\to \infty }{\mathop{\lim }}\,{{\lambda }_{b,N}}=0$. Then, the second term on the RHS of~\eqref{div_dec} can be upper-bounded by
\begin{equation}\label{div_der_2}
\begin{split}
  \int_{{{\mathbb{R}}^N}\backslash \mathcal{T}_{\epsilon }^N({s^N})}{p\left( {r^N}\left| {s^N} \right. \right)\log \frac{p\left( {r^N}\left| {s^N} \right. \right)}{q\left( {r^N}\left| {s^N} \right. \right)}d{r^N}}
  =&\int_{{{\mathbb{R}}^N}\backslash \mathcal{T}_{\epsilon }^N({s^N})}{{{\lambda }_{b,N}}q\left( {r^N} \right)\log \frac{{{\lambda }_{b,N}}q\left( {r^N} \right)}{q\left( {r^N}\left| {s^N} \right. \right)}d{r^N}}\\
  \overset{(a)}{<}&\int_{{{\mathbb{R}}^N}\backslash \mathcal{T}_{\epsilon }^N({s^N})}{{{\lambda }_{b,N}}q\left( {r^N} \right)\log \frac{q\left( {r^N} \right)}{q\left( {r^N}\left| {s^N} \right. \right)}d{r^N}}\\
  =&{{\lambda }_{b,N}}\int_{{{\mathbb{R}}^N}\backslash \mathcal{T}_{\epsilon }^N({s^N})}{q\left( {r^N} \right)\log \frac{q\left( {r^N} \right)}{q\left( {r^N}\left| {s^N} \right. \right)}d{r^N}},
\end{split}
\end{equation}
where step (a) holds because ${{\lambda }_{b,N}}<1$. We respectively bound the above integral within two integral regions. First, we notice that
\begin{equation}\label{upp_der_1}
\begin{split}
\int_{\mathcal{T}_{\epsilon }^N({s^N})}{q\left( {r^N} \right)\log \frac{q\left( {r^N} \right)}{q\left( {r^N}\left| {s^N} \right. \right)}d{r^N}}
=&\int_{\mathcal{T}_{\epsilon }^N({s^N})}{q\left( {r^N} \right)\log \frac{q\left( {r^N} \right)q\left( {s^N} \right)}{q\left( {s^N},{r^N} \right)}d{r^N}}\\
\ge &\int_{\mathcal{T}_{\epsilon }^N({s^N})}{q\left( {r^N} \right)\log \frac{{{2}^{-N\left( h\left( {V^\text{TC}_b} \right)+{{\varepsilon }_{N}} \right)}}{{2}^{-N\left( h\left( {U^\text{TC}_b} \right)+{{\varepsilon }_{N}} \right)}}}{{{2}^{-N\left( h\left( {U^\text{TC}_b},{V^\text{TC}_b} \right)-{{\varepsilon }_{N}} \right)}}}d{r^N}}\\
=&-N\left( I\left( {V^\text{TC}_b};{U^\text{TC}_b} \right)+3{{\varepsilon }_{N}} \right)\int_{\mathcal{T}_{\epsilon }^N({s^N})}{q\left( {r^N} \right)d{r^N}}\\
>&-N\left( I\left( {V^\text{TC}_b};{U^\text{TC}_b} \right)+3{{\varepsilon }_{N}} \right)\\
=&-N\left( \frac{1}{2}\log \frac{\widehat{\sigma}_b^2}{d_b}+{{\delta }_{N}}+3{{\varepsilon }_{N}} \right).
\end{split}
\end{equation}
Then, we notice that
\begin{equation}\label{upp_der_2}
\begin{split}
  \int_{{{\mathbb{R}}^N}}{q\left( {r^N} \right)\log \frac{q\left( {r^N} \right)}{q\left( {r^N}\left| {s^N} \right. \right)}d{r^N}}
  =&D\left( q\left( {r^N} \right)\left\| q\left( {r^N}\left| {s^N} \right. \right) \right. \right)\\
  \overset{(a)}{=}&\sum\limits_{i=1}^ND\left( q\left( r_i \right)\left\| q\left( r_i\left| s_i \right. \right) \right. \right)\\
  =&\sum\limits_{i=1}^N{D\left( \mathcal{N}\left( 0,(\widehat{\sigma }_b^2-d_b)  \right)\left\| \mathcal{N}\left( \frac{\widehat{\sigma }_b^2s_i}{\widehat{\sigma }_b^2-d_b},\left(1-\frac{d_b}{\widehat{\sigma }_b^2}\right)d_b  \right) \right. \right)}\\
  \overset{(b)}{=}&\sum\limits_{i=1}^N{{{c}_{b,1}}+{{c}_{b,2}}{s_i}^2 }
  ={{c}_{b,1}}N+{{c}_{b,2}}\left\| {s^N} \right\|_{2}^2 ,
\end{split}
\end{equation}
where (a) holds because the typicality-based pdf $q(\cdot)$ can be decomposed into the product of $N$ identical pdfs such that each identical pdf corresponds to the pdf of each entry of the corresponding $N$-length vector, and (b) follows from the formula of KL-divergence between two Gaussian random variables (see~\eqref{Gaussian_div}):
\begin{equation}
\begin{split}
&{D\left( \mathcal{N}\left( 0,(1-d_b)\widehat{\sigma }_b^2  \right)\left\| \mathcal{N}\left( \frac{\widehat{\sigma }_b^2s_i}{\widehat{\sigma }_b^2-d_b},\left(1-\frac{d_b}{\widehat{\sigma }_b^2}\right)d_b  \right) \right. \right)}\\
  =&\log \frac{\widehat{\sigma }_b^2(\widehat{\sigma }_b^2 -d_b)}{d_b(\widehat{\sigma }_b^2 -d_b)}-1+\frac{\widehat{\sigma }_b^2(\widehat{\sigma }_b^2 -d_b)}{d_b(\widehat{\sigma }_b^2 -d_b)}+\frac{1}{(1-d_b/\widehat{\sigma }_b^2)d_b }{{\left( \frac{s_i}{1-d_b/\widehat{\sigma }_b^2} \right)}^2 }\\
  =&\log\frac{\widehat{\sigma }_b^2}{d_b}-1+\frac{\widehat{\sigma }_b^2}{d_b}+\frac{s_i^2}{(1-d_b/\widehat{\sigma }_b^2)^3d_b }\\
  =:&\sum\limits_{i=1}^N{{{c}_{b,1}}+{{c}_{b,2}}s_i^2 }.
\end{split}
\end{equation}
Thus, combining \eqref{div_der_2}-\eqref{upp_der_2}, we have that, the second term on the RHS of~\eqref{div_dec} can be upper-bounded by
\begin{equation}\label{div_der_2_2}
\begin{split}
 &\int_{{{\mathbb{R}}^N}\backslash \mathcal{T}_{\epsilon }^N({s^N})}{p\left( {r^N}\left| {s^N} \right. \right)\log \frac{p\left( {r^N}\left| {s^N} \right. \right)}{q\left( {r^N}\left| {s^N} \right. \right)}d{r^N}}\\
 &{\le} \lambda_{b,N} \left({{c}_{b,1}}N+{{c}_{b,2}}\left\| {s^N} \right\|_{2}^2  +N\left( \frac{1}{2}\log \frac{\widehat{\sigma}_b^2}{d_b}+{{\delta }_{N}}+3{{\varepsilon }_{N}} \right)\right),
\end{split}
\end{equation}
where the inequality follows by adding up the RHSs of~\eqref{upp_der_1} and \eqref{upp_der_2}. Also note that ${{c}_{b,2}}\ge 0$ (otherwise we can upper-bound ${{c}_{b,2}}$ with $\max (0,{{c}_{b,2}})$). Therefore, combining~\eqref{div_der_1} and~\eqref{div_der_2_2}, we get that, when ${s^N}\in \mathcal{T}_{\epsilon }^N$,
\begin{equation}
\begin{split}
  D\left( p\left( {r^N}\left| {s^N} \right. \right)\left\| q\left( {r^N}\left| {s^N} \right. \right) \right. \right)  <&N\left( {{\delta }_{N}}+\frac{1}{N}+3\varepsilon_N  \right)+{{\lambda }_{b,N}}\left( {{c}_{b,1}}N+{{c}_{b,2}}\left\| {s^N} \right\|_{2}^2  \right)\\
  &+\lambda_{b,N}N\left( \frac{1}{2}\log \frac{\widehat{\sigma}_b^2}{d_b}+{{\delta }_{N}}+3{{\varepsilon }_{N}} \right).
\end{split}
\end{equation}
Define
\begin{equation}
\begin{split}
  \zeta_{b,N}=& \left({{\delta }_{N}}+\frac{1}{N}+3\varepsilon_N\right) + \lambda_{b,N}{{c}_{b,1}}
  +\lambda_{b,N}\left(\frac{1}{2}\log \frac{\widehat{\sigma}_b^2}{d_b}+{{\delta }_{N}}+3{{\varepsilon }_{N}}\right).
\end{split}
\end{equation}
Then,
\begin{equation}
  D\left( p\left( {r^N}\left| {s^N} \right. \right)\left\| q\left( {r^N}\left| {s^N} \right. \right) \right. \right)<\zeta_{b,N} N+\lambda_{b,N}c_{b,2}\left\| {s^N} \right\|_{2}^2  ,
\end{equation}
where $\lim_{N\to \infty}\zeta_{b,N}=0$, because $c_{b,1},\frac{1}{2}\log\frac{1}{d_b}<\infty$ and ${{\delta }_{N}},\frac{1}{N},\varepsilon_N,\lambda_{b,N}\to 0$.

\subsubsection{Proof of the second statement} When ${s^N}$ is not a typical sequence, i.e., when ${s^N}\notin \mathcal{T}_{\epsilon }^N$. In this case,
\begin{equation}
  p\left( {r^N}\left| {s^N} \right. \right)=q\left( {r^N} \right),
\end{equation}
because the encoding automatically fails (even without checking the existence of a codeword) when the estimate (source) $\widehat{\mathbf{s}}_b$ is not typical itself, and we directly send $\mathbf{c}_b(0)$. Therefore, when ${s^N}\notin \mathcal{T}_{\epsilon }^N$,
\begin{equation}
\begin{split}
  D\left( p\left( {r^N}\left| {s^N} \right. \right)\left\| q\left( {r^N}\left| {s^N} \right. \right) \right. \right)
=&D\left( q\left( {r^N} \right)\left\| q\left( {r^N}\left| {s^N} \right. \right) \right. \right)\\
  =&\sum\limits_{i=1}^N{D\left( \mathcal{N}\left( 0,\widehat{\sigma }_b^2 -d_b \right)\left\| \mathcal{N}\left( \frac{\widehat{\sigma }_b^2s_i}{\widehat{\sigma }_b^2-d_b},\left(1-\frac{d_b}{\widehat{\sigma }_b^2}\right)d_b  \right) \right. \right)}\\
  =&{{c}_{b,1}}N+{{c}_{b,2}}\left\| {s^N} \right\|_{2}^2 .
\end{split}
\end{equation}
Here, we only need the fact that $c_{b,1}+c_{b,2}\frac{1}{N}\left\| {s^N} \right\|_{2}^2 $ is bounded to complete the remaining proof.

\subsubsection{Using the two statements to prove Lemma~\ref{div_lemma}}
Finally, we can upper-bound the KL-divergence $D\left( \left. {p_{\widehat{\mathbf{r}}_b\left| \widehat{\mathbf{s}}_b \right.}} \right\|\phi_{{V^\text{TC}_b}\left| {U^\text{TC}_b} \right.}^N \right)$ using the following integral:
\begin{equation}
\begin{split}
  D\left( \left. {p_{\widehat{\mathbf{r}}_b\left| \widehat{\mathbf{s}}_b \right.}} \right\|\phi_{{V^\text{TC}_b}\left| {U^\text{TC}_b} \right.}^N \right)
  =&\int_{{{\mathbb{R}}^N}}{p\left( {s^N} \right)D\left( p\left( {r^N}\left| {s^N} \right. \right)\left\| q\left( {r^N}\left| {s^N} \right. \right) \right. \right)d{s^N}}\\
  =&\left( \int_{{{\mathbb{R}}^N}\backslash \mathcal{T}_{\epsilon }^N}{+\int_{\mathcal{T}_{\epsilon }^N}{{}}} \right)p\left( {s^N} \right)D\left( p\left( {r^N}\left| {s^N} \right. \right)\left\| q\left( {r^N}\left| {s^N} \right. \right) \right. \right)d{s^N}\\
  <&\int_{{{\mathbb{R}}^N}\backslash \mathcal{T}_{\epsilon }^N}{p\left( {s^N} \right)\left[ {{c}_{b,1}}N+{{c}_{b,2}}\left\| {s^N} \right\|_{2}^2  \right]d{s^N}}\\
  &+\int_{\mathcal{T}_{\epsilon }^N}p\left( {s^N} \right)\left[\zeta_{b,N}N+{{\lambda }_{b,N}}{{c}_{b,2}}\left\| {s^N} \right\|_{2}^2 \right] d{s^N}\\
  <&\int_{{{\mathbb{R}}^N}\backslash \mathcal{T}_{\epsilon }^N}{p\left( {s^N} \right)\left[ {{c}_{b,1}}N+{{c}_{b,2}}\left\| {s^N} \right\|_{2}^2  \right]d{s^N}}\\
  &+\int_{{{\mathbb{R}}^N}}p\left( {s^N} \right)\left[\zeta_{b,N}N+{{\lambda }_{b,N}}{{c}_{b,2}}\left\| {s^N} \right\|_{2}^2 \right]d{s^N}\\
  &=\left( 1-\Pr\left( \mathcal{T}_{\epsilon }^N \right)  \right){{c}_{b,1}}N+{{c}_{b,2}}\int_{{{\mathbb{R}}^N}\backslash \mathcal{T}_{\epsilon }^N}{p\left( {s^N} \right)\left\| {s^N} \right\|_{2}^2 d{s^N}}\\
  &+ \zeta_{b,N}N + {\lambda _{b,N}}{c_{b,2}}\mathbb{E}\left[\left\| {\widehat{\mathbf{s}}_b} \right\|_2^2\right],
\end{split}
\end{equation}
where $1-\Pr \left( \mathcal{T}_{\epsilon }^N \right)$, ${{\lambda }_{b,N}},\zeta_{b,N} \xrightarrow{N\to \infty }0$, and $\frac{1}{N}\mathbb{E}\left[\left\| \widehat{\mathbf{s}}_b \right\|_{2}^2 \right]<\infty $ (see \eqref{var_bounded_eqn_1}). Therefore, to prove that
\begin{equation}
  \frac{1}{N}D\left( \left. {p_{\widehat{\mathbf{r}}_b\left| \widehat{\mathbf{s}}_b \right.}} \right\|\phi_{{V^\text{TC}_b}\left| {U^\text{TC}_b} \right.}^N \right)<{{\widetilde{\eta }}_{b,N}},
\end{equation}
for some constant ${{\widetilde{\eta }}_{b,N}}$ such that $\lim_{N\to\infty}{{\widetilde{\eta }}_{b,N}}\to 0$, we only need to show that
\begin{equation}
  \lim_{N\to\infty}\frac{1}{N}\int_{{{\mathbb{R}}^N}\backslash \mathcal{T}_{\epsilon }^N}{p\left( {s^N} \right)\left\| {s^N} \right\|_{2}^2 d{s^N}}=0.
\end{equation}
Note that based on the induction in~\eqref{upp_var_der2}, we already know that (recall that $E_l=1$ means that the encoding at node $v_l$ is not successful)
\begin{equation}
\begin{split}
  {{\alpha }_{lN}}N>&\Pr ({E_l}=1)\mathbb{E}\left[ \left. \left\| \widehat{\mathbf{s}}_l \right\|_{2}^2  \right|{E_l}=1 \right]\\
  =&\iint_{{{\mathbb{R}}^{2N}}\backslash \mathcal{J}_{\epsilon }^{2N}}{p\left( {s^N} \right)\left\| {s^N} \right\|_{2}^2 }d{s^N}d{r^N}\\
  \overset{(a)}{\mathop{>}}&\,\int_{{{\mathbb{R}}^N}}{d{r^N}\int_{{{\mathbb{R}}^N}\backslash \mathcal{T}_{\epsilon }^N}{p\left( {s^N} \right)\left\| {s^N} \right\|_{2}^2 d{s^N}}}\\
  =&\int_{{{\mathbb{R}}^N}\backslash \mathcal{T}_{\epsilon }^N}{p\left( {s^N} \right)\left\| {s^N} \right\|_{2}^2 d{s^N}},
\end{split}
\end{equation}
where step (a) follows from the fact that when ${s^N}$ is not typical, the pair $\left( {s^N},{r^N} \right)$ is not jointly-typical, which means that integral region ${{\mathbb{R}}^{2N}}\backslash \mathcal{J}_{\epsilon }^{2N}$ (the pair is not typical) contains the region$\left( {{\mathbb{R}}^N}\backslash \mathcal{T}_{\epsilon }^N \right)\times {{\mathbb{R}}^N}$(${s^N}$ is not typical). Therefore, we conclude that
\begin{equation}
  \frac{1}{N}D\left( \left. {p_{\widehat{\mathbf{r}}_b\left| \widehat{\mathbf{s}}_b \right.}} \right\|\phi_{{V^\text{TC}_b}\left| {U^\text{TC}_b} \right.}^N \right)<{{\widetilde{\eta }}_{b,N}},
\end{equation}
for some constant ${{\widetilde{\eta }}_{b,N}}\to 0$.
%From Lemma~\ref{distortion_close_lemma}, we have that
%\begin{equation}
%  \sqrt{{{d}_i}\widehat{\sigma}_i^2 -{\varepsilon }_{N}}-{{\eta }_{N}}\le\sqrt{D_i^\text{Inc}}\le \sqrt{{{d}_i}\widehat{\sigma}_i^2 +{\varepsilon }_{N}}+{{\eta }_{N}},
%\end{equation}
%where $\widehat{\sigma}_i^2 $, the variance of $U^\text{TC}_i$, can be calculated explicitly and $\underset{N\to \infty }{\mathop{\lim }}\,{{\eta }_{N}}=0$ and $\underset{N\to \infty }{\mathop{\lim }}\,{{\varepsilon }_{N}}=0$. Thus, we can set ${d}_i\widehat{\sigma}_i^2 =C,\forall i$, where $C$ is a constant. When the code length $N\to\infty$
%\begin{equation}
%  C=\lim_{N\to\infty} {d}_i\widehat{\sigma}_i^2 =\lim_{N\to\infty}D_i^\text{Inc},\forall i.
%\end{equation}
%Combining with the distortion accumulation result $D_0=\sum\limits_{i=1}^n{{D_i^\text{Inc}}}$, we get
%\begin{equation}\label{222}
%  \lim_{N\to\infty} D_i^\text{Inc}=D_0/n.
%\end{equation}
%Therefore,~\eqref{minimize_upper_bound} can be obtained by plugging~\eqref{222} into~\eqref{Gaussian_code_rate}.

\section{Proofs for Section \ref{consensus_sec}}
\subsection{Proof of Theorem~\ref{Distortion_consensus_add_up}}\label{consensus_add_up_proof}
We consider a general case in Fig.~\ref{cut_set_figure}, where the set $\mathcal{S}$ represents ${\mathcal{S}_{b\to a}}$. Using exactly the same arguments from~\eqref{MMSE_b} to~\eqref{orthogonal}, we obtain
\begin{equation}
 \left({\widehat{\mathbf{y}}^\text{mmse}_{\mathcal{S}_{b\to a},b}}-{\widehat{\mathbf{y}}^\text{mmse}_{\mathcal{S}_{b\to a},a}}\right)\bot \left({\widehat{\mathbf{y}}^\text{mmse}_{\mathcal{S}_{b\to a},b}}-{\mathbf{y}_{\mathcal{S}_{b\to a}}}\right).
\end{equation}
Therefore, using Pythagoras theorem, we get
\begin{equation}\label{Pyth_consensus}
  {D_{i\to j}^\text{Rx}}={D_{i\to j}^\text{Tx}}+D_{i\to j}^\text{Inc}.
\end{equation}
From the definition of an MMSE estimate, we have that
\begin{equation}
\begin{split}
  {\widehat{\mathbf{y}}^\text{mmse}_{\mathcal{S}_{b\to a},b}}=&\mathbb{E}\left[ {\mathbf{y}_{\mathcal{S}_{b\to a}}}|I_b \right]
  =\mathbb{E}\left[ \sum\limits_{k=1}^{d} \mathbf{y}_{{\mathcal{S}_{k\to b}}}+w_b{\mathbf{x}_b}\left|I_b\right. \right]
  =\sum\limits_{k=1}^{d}\widehat{\mathbf{y}}^\text{mmse}_{{\mathcal{S}_{k\to b}},b}+w_b{\mathbf{x}_b}.
\end{split}
\end{equation}
Therefore
\begin{equation}
\begin{split}
  {D_{b\to a}^\text{Tx}}=&\mathbb{E}\left[ {{\left( {\mathbf{y}_\mathcal{S}}-{\widehat{\mathbf{y}}^\text{mmse}_{\mathcal{S}_{b\to a},b}} \right)}^2 } \right]
  =\sum\limits_{k=1}^{d}{\mathbb{E}\left[ {{\left( \mathbf{y}_{\mathcal{S}_{k\to b},b}-\widehat{\mathbf{y}}^\text{mmse}_{\mathcal{S}_{k\to b},b} \right)}^2 } \right]}
  =\sum\limits_{k=1}^{d}{{D_{k\to b}^\text{Rx}}+D_{k\to b}^\text{Inc}}.
\end{split}
\end{equation}
Using induction on the edge set ${{\overrightarrow{\mathcal{T}}}_{k}}$ of the directed tree towards the root $v_k$, we get~\eqref{Total_Distortion_consensus}.
\subsection{Proof of Theorem~\ref{main_thm_2}}\label{consensus_lower_bound_proof}
The main part is to show that in Fig.~\ref{cut_set_figure}
\begin{equation}\label{consensus_one_link}
  {R_{b\to a}}\ge \frac{1}{2}{\log_2 }\frac{\sigma _{\mathcal{S}_{b\to a}}^2 }{D_{b\to a}^\text{Inc}}-\mathcal{O}\left((D_{b\to a}^\text{Tx})^{1/2}\right),
\end{equation}
which is a counterpart of \eqref{good_bound}. As long as~\eqref{consensus_one_link} holds, the outer bound in Theorem~\ref{main_thm_2} can be obtained by summing~\eqref{consensus_one_link} over all links.

The proof of~\eqref{consensus_one_link} can be obtained similarly as in the proof of~\eqref{good_bound}. We know that the set $\mathcal{S}$ in Fig.~\ref{cut_set_figure} represents ${\mathcal{S}_{b\to a}}\subset\mathcal{V}$. Then, using the same derivations in~\eqref{cut_set_bound_ineq}, we get
\begin{equation}\label{cut_set_consensus}
  N{R_{b\to a}}\ge h({\widehat{\mathbf{y}}^\text{mmse}_{\mathcal{S}_{b\to a},b}})-\frac{N}{2}{\log_2 }2\pi eD_{b\to a}^\text{Inc}.
\end{equation}
Using Lemma~\ref{Transportation_Inequality_lmm} and the same derivations in~\eqref{entropy_to_MMSE} and \eqref{der11}, we get
\begin{equation}
\begin{split}
   h({\widehat{\mathbf{y}}^\text{mmse}_{\mathcal{S}_{b\to a},b}})-h({\mathbf{y}_{\mathcal{S}_{b\to a}}})
  &=-D\left( p||q \right)+\frac{{\log_2 }e}{2\sigma _{\mathcal{S}_{b\to a}}^2 }\mathbb{E}\left[ \left\|\widehat{\mathbf{y}}^\text{mmse}_{\mathcal{S}_{b\to a},b}\right\|_2^2 -\left\|\mathbf{y}_{\mathcal{S}_{b\to a}}\right\|_2^2  \right] \\
 & \ge -\frac{N{D_{b\to a}^\text{Tx}}}{2w_b^2 }-\frac{N{\log_2 }e}{2\sigma _{\mathcal{S}_{b\to a}}^2 }\sqrt{2{D_{b\to a}^\text{Tx}}\left( 4\sigma _{\mathcal{S}_{b\to a}}^2 +{D_{b\to a}^\text{Tx}} \right)}\label{der12},
\end{split}
\end{equation}
where $p(\cdot)$ and $q(\cdot)$ are the pdfs of ${\widehat{\mathbf{y}}^\text{mmse}_{\mathcal{S}_{b\to a},b}}$ and ${\mathbf{y}_{\mathcal{S}_{b\to a}}}$ respectively.
Combining~\eqref{cut_set_consensus},~\eqref{der12} and the fact that $h(\mathbf{y}_\mathcal{S})=\frac{1}{2}{\log_2 }2\pi e \sigma _{\mathcal{S}}^2 $, we get
\begin{equation}
\begin{split}
   {R_{b\to a}}\ge &\frac{1}{2}{\log_2 }\frac{\sigma _{\mathcal{S}_{b\to a}}^2 }{D_{b\to a}^\text{Inc}}-\frac{{D_{b\to a}^\text{Tx}}}{2w_b^2 }
  -\frac{{\log_2 }e}{2\sigma _{\mathcal{S}_{b\to a}}^2 }\sqrt{2{D_{b\to a}^\text{Tx}}\left( 4\sigma _{\mathcal{S}_{b\to a}}^2 +{D_{b\to a}^\text{Tx}} \right)} \\
  =&\frac{1}{2}{\log_2 }\frac{\sigma _{\mathcal{S}_{b\to a}}^2 }{D_{b\to a}^\text{Inc}}-\mathcal{O}\left((D_{b\to a}^\text{Tx})^{1/2}\right).
\end{split}
\end{equation}
This completes the proof.

\subsection{Proof of Theorem~\ref{Gaussian_code_thm_2}}\label{Gaussian_thm2_proof}
In this proof, we provide an achievable scheme for the Gaussian network consensus problem. We basically generalize the scheme for linear function computation in Section~\ref{Gaussian_sec} to the network consensus problem. Therefore, we will first use Gaussian test channels to define some distribution functions that we will use in this section. Then, we will provide the encoding and decoding procedures for the Gaussian random codes. Finally, we will prove that this scheme achieves the sum rate inner bound~\eqref{consensus_upper_bound}.

Recall that at each node $v_i$, $\mathbf{y}_{{\mathcal{S}}_{i\to j}}$ denotes the partial weighted sum of all data at all descendants of $v_i$ when the node $v_j$ is viewed as the parent node of $v_i$. Denote by $\widehat{\mathbf{s}}_{i\to j}$ the estimate of the partial sum $\mathbf{y}_{{\mathcal{S}}_{i\to j}}$. Denote by $\widehat{\mathbf{r}}_{i\to j}$ the description of $\widehat{\mathbf{s}}_{i\to j}$ that is sent by $v_i$ to $v_j$. The formal definition of the estimates and descriptions will be provided in the encoding and decoding procedures. Following the same procedures in Section~\ref{Gaussian_sec}, we first define some distribution functions using Gaussian test channels. These distribution functions will be defined such that the estimates $\widehat{\mathbf{s}}_{i\to j}$ and descriptions $\widehat{\mathbf{r}}_{i\to j}$ are typical with respect to them.

At each link $v_i\to v_j$, we define two scalar random variables $U^\text{TC}_{i\to j}$ and $V^\text{TC}_{i\to j}$. Define $\widehat{\sigma}_{i\to j}^2$ as the variance of $U^\text{TC}_{i\to j}$. When $U^\text{TC}_{i\to j}$ is given, $V^\text{TC}_{i\to j}$ is defined by the Gaussian test channel
\begin{equation}\label{Gaussian_test_consensus}
  U^\text{TC}_{i\to j}=V^\text{TC}_{i\to j}+Z_{i\to j},
\end{equation}
where $Z_{i\to j}\sim\mathcal{N}(0,d_{i\to j})$ is independent of ${V^\text{TC}_{i\to j}}$ and $d_{i\to j}$ is the distortion parameter, which can be tuned.

For any arbitrary leaf $v_l$, define
\begin{equation}\label{leaf_estimator_consensus}
  {U^\text{TC}_{l\to n(l)}}=w_lX_l,
\end{equation}
where $X_l$ denotes a random variable that has the same distribution as each entry of $\mathbf{x}_l$, and ${v_{n(l)}}$ denotes the only neighbor of the node $v_l$. For an arbitrary non-leaf node $v_b$ and an arbitrary neighbor $v_a\in \mathcal{N}(v_b)$ as shown in Fig.~\ref{cut_set_figure}, define
\begin{equation}\label{non_leaf_estimator_consensus}
  {U^\text{TC}_{b\to a}}=\sum\limits_{v_k\in \mathcal{N}(v_b)\setminus \{{v_a}\}}{V^\text{TC}_{k\to b}}+w_bX_b,
\end{equation}
where $X_b$ denotes a random variable that has the same distribution as each entry of $\mathbf{x}_b$. Since the network is a tree, all descriptions $V^\text{TC}_{k\to b}$ at different neighbors $v_k$ of $v_b$ are independent of each other. Therefore,
\begin{equation}\label{var_induction_consensus}
  \widehat{\sigma }_{b\to a}^2 =\sum\limits_{k=1}^{d}{(\widehat{\sigma }_{k\to b}^2 -{d}_{k\to b})}+w_b^2 .
\end{equation}
Define $\phi_{U^\text{TC}_{i\to j}}$ and $\phi_{V^\text{TC}_{i\to j}}$ as distribution functions of $U^\text{TC}_{i\to j}$ and $V^\text{TC}_{i\to j}$. We also use joint pdfs, where the meanings are always clear from the context. Note that Gaussian test channels and the calculations in~\eqref{leaf_estimator_consensus} and~\eqref{non_leaf_estimator_consensus} are all linear. Therefore, all pdfs $\phi_{U^\text{TC}_{i\to j}}$ and $\phi_{V^\text{TC}_{i\to j}}$ are Gaussian. Moreover, the pdfs $\phi_{U^\text{TC}_{i\to j}}$ and $\phi_{V^\text{TC}_{i\to j}}$ are tunable by changing the normalized distortions $d_{i\to j}$.
\begin{remark}
The random variable $U^\text{TC}_{i\to j}$ can be viewed intuitively as the estimate at the node $v_i$ of the partial weighted sum $\mathbf{y}_{{\mathcal{S}}_{i\to j}}$ when test-channels can be physically established, while $V^\text{TC}_{i\to j}$ can be viewed as the description of $U^\text{TC}_{i\to j}$.
\end{remark}
Before the computation starts, each node $v_i$ generates $d(v_i)$ random codebooks $\mathcal{C}_{i\to j}=\{\mathbf{c}_{i\to j}(w):\; w \in \{0,1,\ldots 2^{NR_{i\to j}}\}\},\forall j$ s.t. $v_j\in \mathcal{N}(v_i)$, where each codeword is generated i.i.d. according to distribution $\phi_{V^\text{TC}_{i\to j}}$. The rate is chosen such that
\begin{equation}\label{r_up_consensus}
 R_{i\to j}=I(U^\text{TC}_{i\to j};V^\text{TC}_{i\to j})+\delta_N=\frac{1}{2}\log \frac{\widehat{\sigma}_{i\to j}^2}{d_{i\to j}}+\delta_N,
\end{equation}
where $U^\text{TC}_{i\to j}$ and $V^\text{TC}_{i\to j}$ are respectively the `estimate' scalar random variable and the `description' scalar random variable, and $\underset{N\to \infty }{\mathop{\lim }}\delta_N=0$. Thus, the formula of the sum rate $R$ in \eqref{total_rate_consensus} can be proved by summing up the rates on all links in the network.

The codebook $\mathcal{C}_{i\to j}$ is revealed to the node $v_j$. During the computation, as shown in Fig~\ref{cut_set_figure}, each node $v_b$, upon receiving description indexes $M_{1b}, M_{2b},\ldots M_{db}$ from the $d$ neighbors $v_1,\ldots v_d$ except the neighbor $v_a$, decodes these descriptions, computes the sum of them and the data vector generated at $v_b$
\begin{equation}\label{U_b_hat_consensus}
  \widehat{\mathbf{s}}_{b\to a}=\sum\limits_{k=1}^{d}{\mathbf{c}_{k\to b}(M_{k\to b})}+w_b\mathbf{x}_b,
\end{equation}
and re-encodes $\widehat{\mathbf{s}}_{b\to a}$ into a new description index $M_{b\to a}\in \{1,2,\ldots 2^{NR_{b\to a}}\}$ and sends the description index to the neighbor $v_a$ with $NR_{b\to a}$ bits. We denote the reconstructed description by $\widehat{\mathbf{r}}_{b\to a}=\mathbf{c}_{b\to a}(M_{b\to a})$. The decoding and encoding at the node $v_b$ are defined as follows.
\begin{itemize}
  \item \textbf{Decoding: }In each codebook $\mathcal{C}_{k\to b},\forall k\text{ s.t. }v_k\in \mathcal{N}(v_b)$, use the codeword $\mathbf{c}_{k\to b}(M_{k\to b})$ as the description $\widehat{\mathbf{r}}_{k\to b}$. If $v_b$ has obtained all descriptions from all neighbors, it computes the sum of all descriptions and its own data as the estimate of $\mathbf{y}$:
      \begin{equation}
        \widehat{\mathbf{y}}_b=\sum\limits_{v_k\in\mathcal{N}(v_b)}\widehat{\mathbf{r}}_{k\to b}+w_b\mathbf{x}_b.
      \end{equation}
  \item \textbf{Encoding: }For each neighbor $v_a\in \mathcal{N}(v_b)$, find the codeword $\mathbf{c}_{b\to a}(M_{b\to a}) \in \mathcal{C}_{b\to a}\setminus \{\mathbf{c}_{b\to a}(0)\}$ such that the two sequences $\widehat{\mathbf{s}}_{b\to a}=\sum\limits_{k=1}^{d}{\mathbf{c}_{k\to b}(M_{k\to b})}+w_b\mathbf{x}_b$ and $\widehat{\mathbf{r}}_{b\to a}=\mathbf{c}_{b\to a}(M_{b\to a})$ are jointly typical with respect to the distribution $\phi_{U^\text{TC}_{b\to a},V^\text{TC}_{b\to a}}$. If there are more than one codewords that satisfy this condition, arbitrarily choose one of them. However, if $\widehat{\mathbf{s}}_{b\to a}$ is not typical with respect to the distribution $\phi_{U^\text{TC}_{b\to a}}$, or if there is no codeword in $\mathcal{C}_{b\to a}\setminus \{\mathbf{c}_{b\to a}(0)\}$ that satisfies the joint typicality condition, send description index $M_{b\to a}=0$.
\end{itemize}
Similar to the linear function computation case, the encoding step for network consensus may fail, because the estimate $\widehat{\mathbf{s}}_{b\to a}=\sum\limits_{k=1}^{d}{\mathbf{c}_{k\to b}(M_{k\to b})}+w_b\mathbf{x}_b$ may not be a typical sequence respect to pdf $\phi_{U^\text{TC}_{b\to a}}$, or there may not exist codewords in $\mathcal{C}_{b\to a}$ that satisfy the typicality requirement. In this case, the description index $M_{b\to a}=0$ is sent and this description is decoded to a predetermined random sequence $\mathbf{c}_{b\to a}(0)$ on the receiver side.

\begin{lemma}[Covering Lemma for Network Consensus]\label{covering_lemma_consensus}
Denote by $E_{i\to j}=1$ the event that the encoding of the estimate $\widehat{\mathbf{s}}_{i\to j}$ at the node $v_i$ is not successful. Then
\begin{equation}\label{encoding_error_consensus}
 \underset{N\to \infty }{\mathop{\lim }}\,\underset{(i,j)\in\mathcal{E}}{\mathop{\sup }}\,\Pr ({E_{i\to j}}=1)=0,
\end{equation}
where $\mathcal{E}$ denotes all links in the tree network $\mathcal{G}=(\mathcal{V}.\mathcal{E})$ ($(i,j)$ and $(j,i)$ are viewed as two links in the undirected graph $\mathcal{G}$), and the probability is taken over random data sampling and random codebook generation.
\end{lemma}
\begin{IEEEproof}
The proof of this lemma is almost the same as the proof for linear function computing case (see Appendix~\ref{pf_covering}). This is because the distributed computation algorithm used in this section can be viewed as a group of $n=|\mathcal{V}|$ linear function computations in $n$ different directed trees $\vec{\mathcal{T}}_k,1\le k\le n$ towards $n$ different roots (see definition of $\vec{\mathcal{T}}_k$ below equation \eqref{RD_op_problem_consensus}). Therefore, we can use the conditional typicality lemma and mathematical induction on each directed tree to obtain the conclusion.
\end{IEEEproof}
\begin{remark}
\textcolor{black}{The proofs for network consensus are also based on the induction on the tree (see Remark~\ref{induction_on_tree}), except that we may often want to prove that some property $P$ holds at all links $v_b\to v_a$ in the tree network. Firstly, we prove that $P$ holds for all links $v_l\to v_n(l)$, where $v_l$ is a leaf and $v_{n(l)}$ is the only neighbor of $v_l$. Secondly, we prove that, for an arbitrary node $v_b$ with $d+1$ neighbors, denoted by $v_1,v_2,\ldots v_d$ and a special neighbor $v_a$, if $P$ holds for all links $v_1\to v_b,v_2\to v_b,\ldots v_d\to v_b$, then the property holds for the link $v_b\to v_a$. It is obvious that these two arguments lead to the conclusion that $P$ holds for all links in the tree network.}
\end{remark}
Lemma~\ref{covering_lemma_consensus} states that the estimate $\widehat{\mathbf{s}}_{b\to a}$ and the description $\widehat{\mathbf{r}}_{b\to a}$ are jointly typical with high probability for all links $v_b\to v_a$ in the tree network. The following Lemma~\ref{var_bounded_consensus} and Lemma~\ref{ent_bounded_consensus} are counterparts of Lemma~\ref{var_bounded} and Lemma~\ref{ent_bounded} in the linear function computation problem.

\begin{lemma}\label{var_bounded_consensus}
For an arbitrary link $v_b\to v_a$, the description $\widehat{\mathbf{r}}_{b\to a}=\mathbf{c}_{b\to a}({{M}_{ba}})$ and the estimate $\widehat{\mathbf{s}}_{b\to a}$ satisfy
\begin{equation}\label{var_bounded_eqn_1_consensus}
  \left|\mathbb{E}\left[\frac{1}{N}\left\|\widehat{\mathbf{s}}_{b\to a}\right\|_2^2\right]-\widehat{\sigma}_{b\to a}^2\right|<\varepsilon_N,
\end{equation}
\begin{equation}\label{var_bounded_eqn_2_consensus}
  \left|\mathbb{E}\left[\frac{1}{N}\left\|\widehat{\mathbf{r}}_{b\to a}\right\|_2^2\right]-(\widehat{\sigma}_{b\to a}^2-d_{b\to a})\right|<\varepsilon_N,
\end{equation}
\begin{equation}\label{typical_small_distortion_1_consensus}
  \left|\mathbb{E}\left[ \frac{1}{N}{{\left\| \widehat{\mathbf{r}}_{b\to a}-\widehat{\mathbf{s}}_{b\to a} \right\|}_2^2 } \right]-{{d}_{b\to a}} \right|<\varepsilon_N,
\end{equation}
where $\lim_{N\to \infty}\varepsilon_N=0$.
\end{lemma}
\begin{IEEEproof}
Similar with the proof of Lemma~\ref{covering_lemma_consensus}, the proof of this lemma can be derived similarly as the proof for the linear function computation case (see Appendix~\ref{app_var_bounded}), because the proof for the linear function computation case is mathematical induction in the tree network, while the network consensus computation scheme in this section can be viewed as a group of linear function computations on $n$ different directed trees.
\end{IEEEproof}

\begin{lemma}\label{ent_bounded_consensus}
For an arbitrary link $v_b\to v_a$, the description $\widehat{\mathbf{r}}_{b\to a}=\mathbf{c}_{b\to a}({{M}_{ba}})$ and the estimate $\widehat{\mathbf{s}}_{b\to a}$ satisfy
\begin{equation}\label{ent_bounded_eqn_1_consensus}
 h(\widehat{\mathbf{s}}_{b\to a})>\frac{N}{2}\log_{2}2\pi e \widehat{\sigma}^2_{b\to a}-N\beta_N,
\end{equation}
\begin{equation}\label{ent_bounded_eqn_2_consensus}
  h(\widehat{\mathbf{r}}_{b\to a})>\frac{N}{2}\log_{2}2\pi e (\widehat{\sigma}^2_{b\to a}-d_{b\to a})-N\beta_N,
\end{equation}
where $\lim_{N\to \infty}\beta_N=0$.
\end{lemma}
\begin{IEEEproof}
One can use the same argument as the one used in the proof of Lemma~\ref{var_bounded_consensus}.
\end{IEEEproof}

The following lemma characterizes the relationship between the Gaussian-code-based distortion $d_{i\to j}$ (normalized distortion) and the MMSE-based distortion $D_{i\to j}^\text{Tx}$ {for the Gaussian code.}
\begin{lemma}\label{distortion_close_lemma_consensus}
For an arbitrary link $v_i\to v_j$
\begin{equation}\label{distortion_close_consensus}
\begin{split}
\sqrt{{{d}_{i\to j}} -{\varepsilon }_{N}}-{{\eta }_{N}}&\le\sqrt{{D_{i\to j}^\text{Rx}}-D_{i\to j}^\text{Tx}}
\le \sqrt{{{d}_{i\to j}}+{\varepsilon }_{N}}+{{\eta }_{N}},
\end{split}
\end{equation}
where $\underset{N\to \infty }{\mathop{\lim }}\,{{\eta }_{N}}=0$ and $\varepsilon_N$ is the same as in~\eqref{typical_small_distortion_1_consensus}.
\end{lemma}
\begin{IEEEproof}
The proof of this lemma essentially follows the same procedures with the ones in the proof for linear function computation in the Appendix~\ref{pf_distortion}. We only provide the sketch of the proof. First, define
\begin{align}
  & \widehat{\mathbf{s}}_{i\to j}^{*}=\widehat{\mathbf{y}}^\text{mmse}_{\mathcal{S}_{i\to j},i}, \\
 & \widehat{\mathbf{r}}_{i\to j}^{*}=\widehat{\mathbf{y}}^\text{mmse}_{\mathcal{S}_{i\to j},j}.
\end{align}
Therefore, $\widehat{\mathbf{s}}_{i\to j}^{*}$ is the MMSE estimate of the partial weighted sum $\mathbf{y}_{{\mathcal{S}}_{i\to j}}$ at node $v_i$, while $\widehat{\mathbf{r}}_{i\to j}^{*}$ is the MMSE estimate of the same weighted sum at node $v_j$. Define
\begin{align}
  & {\Delta_{i\to j}^\text{Tx}}=\mathbb{E}\left[ \frac{1}{N}\left\| \widehat{\mathbf{s}}_{i\to j}-\widehat{\mathbf{s}}_{i\to j}^{*} \right\|_{2}^2  \right], \\
 & {\Delta_{i\to j}^\text{Rx}}=\mathbb{E}\left[ \frac{1}{N}\left\| \widehat{\mathbf{r}}_{i\to j}-\widehat{\mathbf{r}}_{i\to j}^{*} \right\|_{2}^2  \right].
\end{align}
We will prove that ${\Delta_{i\to j}^\text{Tx}}\to 0$ and $\Delta_{i\to j}^\text{Rx}\to 0$ when $N\to \infty$.

Using the same derivations with equation~\eqref{dis_der1} to~\eqref{Delta_relation_1}, we get
\begin{equation}\label{Delta_relation_1_consensus}
  {\Delta_{b\to a}^\text{Tx}}=\sum\limits_{k=1}^{d}{\Delta_{k\to b}^\text{Rx}},
\end{equation}
for an arbitrary link $v_b\to v_a$ and the neighborhood structure $\mathcal{N}(v_b)=\{v_1,\ldots v_d\}\cup \{v_a\}$ (see Figure~\ref{cut_set_figure}). Using the same derivations with equation~\eqref{dis_der3} to~\eqref{Delta_relation_2}, we get
\begin{equation}\label{Delta_relation_2_consensus}
 \sqrt{\Delta_{b\to a}^\text{Rx}}\le \sqrt{{{\theta }_{N}}}+\sqrt{{\Delta_{b\to a}^\text{Tx}}},
\end{equation}
for an arbitrary link $v_b\to v_a$ and the constant $\lim_{N\to\infty}\theta_N=0$. Using induction on $n$ different directed tree networks, we get
\begin{equation}\label{distortion_close_key_consensus}
  \underset{(i,j)\in\mathcal{E}}{\mathop{\sup }}\,\sqrt{{\Delta_{i\to j}^\text{Tx}}}\le \underset{(i,j)\in\mathcal{E}}{\mathop{\sup }}\,\sqrt{{\Delta_{i\to j}^\text{Rx}}}\le n\sqrt{{{\theta }_{N}}}.
\end{equation}
Using the triangle inequality, we get
\begin{equation}
  \sqrt{\mathbb{E}\left[ \frac{1}{N}\left\| \widehat{\mathbf{r}}_{i\to j}^{*}-\widehat{\mathbf{s}}_{i\to j}^{*} \right\|_{2}^2  \right]}\le \sqrt{{{d}_{i\to j}}+{{\varepsilon }_{N}}}+2n\sqrt{{{\theta }_{N}}},
\end{equation}
and
\begin{equation}
  \sqrt{\mathbb{E}\left[ \frac{1}{N}\left\| \widehat{\mathbf{r}}_{i\to j}^{*}-\widehat{\mathbf{s}}_{i\to j}^{*} \right\|_{2}^2  \right]}\ge \sqrt{{{d}_{i\to j}} -{{\varepsilon }_{N}}}-2n\sqrt{{{\theta }_{N}}},
\end{equation}
which conclude the proof.
\end{IEEEproof}
Using the same procedures from~\eqref{333} to \eqref{555}, one can prove that the overall distortion at one node, averaged over the random code ensemble satisfies
\begin{equation}
  D_i^\text{Total}\le \sum\limits_{(i,j)\in {{\overrightarrow{\mathcal{T}}}_{k}}}{d_{i\to j}}+\epsilon_N.
\end{equation}
Summing the above equations over all directed trees in the network, we have that~\eqref{distortion_upper_bound_consensus} holds for the overall distortion averaged over the random code ensemble. Therefore, we can at least find one code for which \eqref{distortion_upper_bound_consensus} holds.

\bibliographystyle{ieeetr}
\bibliography{rough}

%\section*{Acknowledgment}



%\bibliographystyle{IEEEtran}

\end{document}